\documentclass[12pt,a4paper,twoside]{article}

\usepackage{amscd}
\usepackage{amsfonts}
\usepackage{amsmath}
\usepackage{amssymb}
\usepackage{authblk}
\usepackage{cancel}
\usepackage{enumitem}
\usepackage{graphicx}
\usepackage{ifthen}
\usepackage{mathdots}
\usepackage{mathtools}
\usepackage{pict2e}
\usepackage{subdepth}
\usepackage{stmaryrd}
\usepackage{tcolorbox}
\usepackage{theorem}
\usepackage{tikz-cd}
\usepackage{tikz}
\usetikzlibrary{tikzmark}
\newtheorem{thm}{Theorem}

\newtheorem{definition}[thm]{Definition}
\newtheorem{lemma}[thm]{Lemma}
\newtheorem{proposition}[thm]{Proposition}
{\theorembodyfont{\upshape}\newtheorem{remark}[thm]{\it Remark}}
{\theorembodyfont{\upshape}\newtheorem{example}[thm]{\it Example}}
\newcommand{\R}{{\mathord{\mathbb R}}}
\newcommand{\Z}{{\mathord{\mathbb Z}}}
\newcommand{\N}{{\mathord{\mathbb N}}}
\newcommand{\C}{{\mathord{\mathbb C}}}

\newcommand{\mB}{\mathcal B}
\newcommand{\mC}{\mathcal C}
\newcommand{\mF}{\mathcal F}
\newcommand{\mG}{\mathcal G}
\newcommand{\mH}{\mathcal H}
\newcommand{\mJ}{\mathcal J}
\newcommand{\mK}{\mathcal K}
\newcommand{\mL}{\mathcal L}
\newcommand{\mM}{\mathcal M}
\newcommand{\mS}{\mathcal S}
\newcommand{\mU}{\mathcal U}
\newcommand{\mV}{\mathcal V}
\newcommand{\mW}{\mathcal W}

\newcommand{\clo}{{\rm clo}}
\newcommand{\ran}{{\rm ran\,}}		
\newcommand{\sign}{{\rm sign}}
\newcommand{\spa}{{\rm span\,}}

\newcommand{\sAut}{{{}^\ast\hspace{-0.5mm}\rm Aut}}

\newcommand{\sIso}{{{}^\ast\rm Iso}}
\newcommand{\Hom}{{\rm Hom}}

\newcommand{\sHom}{{{}^\ast\rm Hom}}

\newcommand{\sMon}{{{}^\ast\rm Mon}}
\newcommand{\Cov}{{\rm Cov}}

\newcommand{\Rep}{{\rm Rep}}
\newcommand{\card}{{\rm card}}
\newcommand{\rd}{\hspace{-0.5mm}{\rm d}}
\newcommand{\diag}{{\rm diag}}
\newcommand{\fA}{{\mathfrak A}}
\newcommand{\fB}{{\mathfrak B}}
\newcommand{\fC}{{\mathfrak C}}
\newcommand{\Fin}{{\rm Fin}}
\newcommand{\Bs}{$\text{B}^\ast$\hspace{-0.8mm}-} 
\newcommand{\Cs}{$\text{C}^\ast$\hspace{-0.8mm}-} 
\newcommand{\Ws}{$\text{W}^\ast$\hspace{-0.8mm}-} 
\newcommand{\str}{${}^\ast$\hspace{-0.4mm}-} 
\newcommand{\kp}{\kappa}
\newcommand{\lm}{\lambda}
\newcommand{\Lm}{\Lambda}
\newcommand{\Gm}{\Gamma}
\newcommand{\veps}{{\varepsilon}}

\newcommand{\spec}{{\rm spec}}

\newcommand{\vi}{\varphi}
\newcommand{\wt}{\widetilde}
\newcommand{\wh}{\widehat}
\newcommand{\bd}{\begin{definition}}
\newcommand{\ed}{\end{definition}\vspace{1mm}}
\newcommand{\bt}{\begin{thm}}
\newcommand{\et}{\end{thm}\vspace{2mm}}
\newcommand{\bc}{\begin{corollary}}
\newcommand{\ec}{\end{corollary}\vspace{2mm}}
\newcommand{\bl}{\begin{lemma}}
\newcommand{\el}{\end{lemma}\vspace{2mm}}
\newcommand{\bp}{\begin{proposition}}
\newcommand{\ep}{\end{proposition}\vspace{2mm}}
\newcommand{\bx}{\begin{example}}
\newcommand{\ex}{\end{example}}
\newcommand{\br}{\begin{remark}}
\newcommand{\er}{\end{remark}}
\newcommand{\bprf}{\noindent{\it Proof.}\hspace{2mm}}
\newcommand{\eprf}{\hfill $\Box$ \vspace{5mm}}
\newcommand{\ie}{i.e.}
\def\bas#1\eas{\begin{align*}#1\end{align*}}
\def\ba#1\ea{\begin{align}#1\end{align}}
\newcommand{\bn}{\begin{enumerate}}
\newcommand{\en}{\end{enumerate}}
\newcommand{\M}[1] {\C^{{#1}\times{#1}}}
\newcommand{\nm}[2] {\C^{{#1}\times{#2}}}

\newcommand{\eN}[1] {{\|{#1}\|}_1}
\newcommand{\n}[1] {\|{#1}\|}
\newcommand{\Ln}[1] {{\|{#1}\|}_\Lambda}
\newcommand{\num}[2]{\llbracket{#1},{#2}\rrbracket}

\newcommand{\Pol}{{\rm Pol}}
\newcommand{\tA}{\widetilde\fA}
\newcommand{\hA}{\widehat\fA}
\newcommand{\ii}{{\rm i}}
\renewcommand{\l}{\langle}
\renewcommand{\r}{\rangle}

\title{On Araki's extension of the Jordan-Wigner transformation}
\author{Walter H. Aschbacher\footnote{walter.aschbacher@univ-tln.fr}}
\affil{Universit\'e de Toulon, Aix Marseille Univ, CNRS, CPT, Toulon, France}

\begin{document}
\pagestyle{myheadings}
\markboth{Walter H. Aschbacher}{On Araki's extension of the Jordan-Wigner transformation}

\date{}
\maketitle

\begin{center}
{\it Dedicated to the memory of Huzihiro Araki}
\end{center}

\vspace{2mm}

\begin{abstract}
In his seminal paper \cite{Ara84}, Araki introduced an elegant extension of the Jordan-Wigner transformation which establishes a precise connection between quantum spin systems and Fermi lattice gases in one dimension in the so-called infinite system idealization of quantum statistical mechanics. His extension allows in particular for the rigorous study of numerous aspects of the prominent XY chain over the two sided infinite discrete line without having to resort to a thermodynamic limit procedure at an intermediate or at the final stage. We rigorously review and elaborate this extension from scratch which makes the paper rather self-contained. In the course of the construction, we also present a simple and concrete realization of Araki's crossed product extension.
\end{abstract}



\section{Introduction}

The rigorous study, from first  principles, of the ubiquitous open quantum systems is of fundamental importance for a deepened understanding of their thermodynamic properties in and out of equilibrium. Since, by definition, an open quantum system has a very large number of degrees of freedom and since the finite accuracy of any feasible experiment does not allow an empirical distinction between an infinite system and a finite system with sufficiently many degrees of freedom, a powerful strategy consists in approximating (in a somewhat reversed sense) the actual finite system by an idealized one with infinitely many degrees of freedom (see \cite{Pri} for an extensive discussion of this idealization and its implications). Furthermore, it is conceptually more appealing and often mathematically more rigorous to treat the idealized system from the outset in a framework designed for infinite systems rather than to take the thermodynamic limit at an intermediate or at the final stage.

One of the most important axiomatic frameworks for the study of such idealized  infinite systems is the so-called algebraic approach to quantum mechanics based on operator algebras. Indeed, after having been heavily used from as early as the 1960s on, in particular for the quantum statistical description of quantum systems in thermal equilibrium (see, for example, \cite{Emch, Sew, BR}), the benefits of this framework have again started to unfold more recently in the physically much more general situation of open quantum systems out of equilibrium (see, for example, \cite{AJP}).

A large class of models which has been widely used in quantum statistical mechanics  is the class of quantum spin models whose first representatives, the so-called (Lenz-) Ising and Heisenberg models, were respectively introduced in 1920 in \cite{Lenz} (and analyzed in 1925 in \cite{Isi}) and in 1928 in \cite{Hei} in order to describe magnetic properties of crystalline solids. An important special instance of the Heisenberg model in one dimension is the so-called XY chain whose Hamiltonian density has the form
\ba
\label{XYDensity}
(1+\gamma)\,\sigma_1^{(x)}
\sigma_1^{(x+1)}+(1-\gamma)\,\sigma_2^{(x)}\sigma_2^{(x+1)},
\ea
where $\gamma\in\R$ stands for the anisotropy and the superscripts of the Pauli matrices denote the sites $x\in\Z$ of the chain. The isotropic version of \eqref{XYDensity} (\ie, $\gamma=0$) was studied in 1950 in \cite{Nambu} and the more general anisotropic version (\ie, $\gamma\in[-1,1]$) was introduced in 1961 in \cite{LSM1961} (where also the name ''XY model'' was given to \eqref{XYDensity}). In 1962 in \cite{Kat}, \eqref{XYDensity} was supplemented by an additional external magnetic field of the form $\lm \sigma_3^{(x)}$, where $\lm\in\R$ denotes its strength (see also \cite{Nie} from 1967). Already in 1969, a first physical realization of the XY chain has been identified (see, for example, \cite{CSP1969}). The impact of the XY chain on the experimental, numerical, theoretical, and mathematical research activity in the field of low-dimensional magnetic systems is ongoing ever since (see, for example,  \cite{MK2004}). 

The detailed analysis of the XY chain carried out in \cite{LSM1961} relied on the classical version of the Jordan-Wigner transformation from 1928 (see \cite{JW}) which defines a fermionic annihilation operator $a_x$ on site $x$ of the finite system with sites $\{1,\ldots, n\}$ by
\ba
\label{oJW}
S_x\sigma_-^{(x)},
\ea
where $S_x:=\sigma_3^{(1)}\ldots\sigma_3^{(x-1)}$ and $\sigma_-^{(x)}:=(\sigma_1^{(x)}-\ii\sigma_2^{(x)})/2$. This transformation allowed the authors of \cite{LSM1961} to establish an equivalence between the finite XY chain and a lattice gas of fermions and to explicitly compute various physical quantities of interest because \eqref{oJW} transforms \eqref{XYDensity} into a quadratic form in the fermionic creation and annihilation operators, \ie, \eqref{XYDensity} becomes (up to a global prefactor) 
\ba
\label{FDensity}
a_x^\ast a_{x+1} + a_{x+1}^\ast a_x
+\gamma (a_x^\ast a_{x+1}^\ast +a_{x+1}a_x).
\ea

In order to study the XY chain in its infinite system idealization whose configuration space is the entire discrete line $\Z$, we have to anchor the Jordan-Wigner transformation \eqref{oJW} at minus infinity. If  we formally define such an anchor $T$ by 
\ba
\label{limex}
\lim_{y\to-\infty}\sigma_3^{(y)}\ldots\sigma_3^{(0)}, 
\ea
we find that $TS_x$ corresponds to $\lim_{y\to-\infty}\sigma_3^{(y)}\ldots\sigma_3^{(x-1)}$ for all sites $x$, where we redefined $S_x$ as $\sigma_3^{(1)}\ldots\sigma_3^{(x-1)}$ if $x\ge2$, as $1$ if $x=1$, and as $\sigma_3^{(x)}\ldots\sigma_3^{(0)}$ if $x\le 0$. Moreover, a formal computation also yields the properties $T^2=1$, $T^\ast=T$, and $T\sigma_\kp^{(x)}=\Theta'(\sigma_\kp^{(x)})T$ for all directions $\kp$ of the spins, where $\Theta'$ describes the rotation around the 3-axis by an angle of $\pi$ of the observables on the nonpositive sites $x$ (and leaving the observables unchanged on the positive sites $x$). Hence, we could replace \eqref{oJW}, for all sites $x$, by 
\ba
\label{AJWFermion}
TS_x\sigma_-^{(x)},
\ea
but, unfortunately, the limit \eqref{limex} does not exist within the spin algebra (see Remark \ref{rem:formalT} below). 

That's why, in 1984 in \cite{Ara84}, Araki introduces an elegant extension of the Jordan-Wigner transformation providing us with a rigorous construction of such an anchor element $T$ outside the spin algebra (and which still allows for the transformation of \eqref{XYDensity} into \eqref{FDensity}). On many occasions, I have been asked by the audience to explain in more detail Araki's construction which is based on a so-called crossed product built out of a \Cs dynamical system. The main motivation of the present paper is to serve this purpose. At the same time, it seems useful to seize the opportunity to somewhat deepen the frequently rather sketchy presentations of the construction of the infinite spin algebra starting from the local observable algebras. In this way, the complete construction is carried out from scratch in a rigorous and pedestrian manner and makes the paper rather self-contained. Finally, in the course of the construction, I present a simple and concrete realization of Araki's crossed product which, to the best of my knowledge, has not been used in the literature so far.

\vspace{5mm}

The subsequent sections contain the following.

\vspace{2mm}

{\it Section \ref{sec:Local} (Local observable algebras)}\, 
We use the algebraic tensor product and its universal property in order to construct the local observable algebras. Moreover, with the help of the classical Kronecker product, we provide a simple identification of any local observable algebra with a full matrix algebra and use the spectral norm on the latter to induce a \Cs norm on the former.

\vspace{1mm}

{\it Section \ref{sec:infinite} (Infinite tensor product)}\, 
We first briefly discuss  the general construction of the so-called inductive limit before specializing the general case to the concrete case at hand. In the concrete case (which we elaborate), the inductive limit is based on the net of local observable algebras and a family of what we call isotonies. The resulting object, the infinite tensor product  (also called uniformly hyperfinite or Glimm algebra), plays the role of the observable algebra over the infinitely extended configuration space $\Z$. 

\vspace{1mm}

{\it Section \ref{sec:crossed} (Crossed product extension)}\, 
We again first briefly discuss  the general construction of the crossed product built out of a \Cs dynamical system before specializing the general case to the case of the group $\Z_2$ (but without making a special choice for the \Cs algebra and the group of \str automorphisms yet). In order to set the precise framework for the construction of the latter, we recall the notion of a Hilbert \Cs module. Within this framework, we construct an extension of the \Cs algebra of the \Cs dynamical system and identify it with the $\Z_2$-crossed product.

\vspace{1mm}

{\it Section \ref{sec:JW} (Jordan-Wigner transformation)}\, 
Based on the preceding sections, we give the precise definitions of the ingredients of \eqref{AJWFermion} leading to a concrete realization of \eqref{AJWFermion}. Moreover, we construct the canonical anticommutation relation (CAR) subalgebra of the $\Z_2$-crossed product extension of the spin algebra. For all the constructed algebras, we introduce an important decomposition into even and odd parts and show that the even part of the spin algebra coincides with the even part of the CAR algebra. Finally, we use the local structure of the CAR algebra to construct a \str isomorphism between the the spin algebra and the CAR algebra but we show that no \str isomorphism exists which preserves the spin structure.

\vspace{1mm}

{\it Appendix \ref{app:Cstar} (\Cs completion)}\, 
For the convenience of the reader, we briefly recall the main definitions and some of the basic facts about \Cs algebras used in the foregoing sections. Moreover, we introduce the so-called \Cs completion and prove some of its properties. In all of the foregoing sections, we use the content of this appendix without necessarily referring to it each time.

\section{Local observable algebras}
\label{sec:Local}

In this section, we make use of the Kronecker product and the algebraic tensor product in order to define the local observable algebras, \ie,  the spin one-half algebras over finite subsets of the infinitely extended configuration space $\Z$. The local observable algebras are used for the construction of the infinite tensor product in the next section.

In the following, let $\N:=\{1,2,\ldots\}$ and $\N_0:=\N\cup\{0\}$. For all $n, m\in\N$, we denote by $\nm{n}{m}$ the set of all complex $n\times m$ matrices and we set $\C^n:=\nm{n}{1}$. Moreover, $[a_{ij}]_{i\in\num{1}{n}, j\in\num{1}{m}}$ stands for the matrix in $\nm{n}{m}$ with entries $a_{ij}\in\C$ for all $i\in\num{1}{n}$ and all $j\in\num{1}{m}$,
where, for all $x,y\in\Z$ with $x\le y$, we set
\ba
\label{Setxy}
\num{x}{y}
:=\begin{cases}
\{x,x+1,\ldots,y\}, & x<y,\\
\hfill\{x\}, & x=y.
\end{cases}
\ea
For all $A\in\nm{n}{m}$, we denote by $A^T$,  $\bar A$, and $A^\ast:=\bar A^T$ the transpose, the complex conjugate, and the conjugate transpose (adjoint) of $A$, respectively. 

Moreover, all the general vector spaces are assumed to be complex and all the norms are denoted by $\|\cdot\|$ unless there are several norms on the same vector space.

For all $n\in\N$, we first want to make $\M{n}$ into a unital \Cs algebra. To this end, we equip $\M{n}$ with the usual matrix addition $\M{n}\times\M{n}\to\M{n}$, the usual scalar multiplication $\C\times\M{n}\to\M{n}$, the usual matrix multiplication $\M{n}\times\M{n}\to\M{n}$, and the involution $\M{n}\to\M{n}$ which associates to any matrix $A\in\M{n}$ its conjugate transpose $A^\ast\in\M{n}$. Equipped with these operations, $\M{n}$ becomes a \str algebra. Moreover, we know that there exists a unique \Cs norm on $\M{n}$, the so-called spectral norm $\|\cdot\|:\M{n}\to\R$ defined, for all $A\in\M{n}$, by
\ba
\label{specnorm}
\|A\|
:=\max_{\substack{z\in\C^n\\ \|z\|=1}}\hspace{0.5mm} \|Az\|,
\ea
where the (complex) Euclidean norm on $\C^n$ is defined as usual by $\|z\|^2:=\sum_{i\in\num{1}{n}}|z_i|^2$ for all $z=[z_1,\ldots,z_n]^T\in\C^n$ (see before \eqref{piIso}). Hence, $\M{n}$ becomes a unital \Cs algebra with identity $1_n:=\diag[1,\ldots,1]$, where $\diag[\lm_1,\ldots,\lm_n]\in\M{n}$ stands for the diagonal matrix with diagonal entries $\lm_1,\ldots,\lm_n\in\C$ (and the additive identity of $\M{n}$, \ie, the null matrix, is sometimes denoted by $0_n$ if needed).

We next define the following  (non internal) binary function. 

\bd[Kronecker product]
\label{def:kron}
Let $n,m\in\N$. The Kronecker product (for $n$ and $m$) $\M{n}\times\M{m}\to\M{nm}$ is defined, for all $A=[a_{ij}]_{i,j\in\num{1}{n}}\in\M{n}$ and all $B=[b_{kl}]_{k,l\in\num{1}{m}}\in\M{m}$, by
\ba
\label{KP}
A\oslash B
:=\begin{bmatrix}
a_{11} B & \ldots & a_{1n} B\\
\vdots & & \vdots\\
a_{n1} B & \ldots & a_{nn} B
\end{bmatrix},
\ea
\ie, $(A\oslash B)_{k+(i-1)m, l+(j-1)m}=a_{ij}b_{kl}$ for all $i,j\in\num{1}{n}$ and all $k,l\in\num{1}{m}$ (and we do not display the dependence on $n, m$ on the left hand side of \eqref{KP}).
\ed

We next collect the properties of the Kronecker product which are used in the following.

\bl[Kronecker product]
\label{lem:kron}
Let $n,m\in\N$. The Kronecker product (for $n$ and $m$) has the following properties:
\bn[label=(\alph*), ref={\it (\alph*)}]
\setlength{\itemsep}{0mm}
\item 
\label{kron:bilin}
It is bilinear, has the mixed-product property, \ie, $(A\oslash B)(C\oslash D)=(AB)\oslash (CD)$ for all $A, C\in\M{n}$ and all $B, D\in\M{m}$, and it preserves the involution, \ie, $(A\oslash B)^\ast=A^\ast\oslash B^\ast$ for all $A\in\M{n}$ and all $B\in\M{m}$.

\item
\label{kron:base}
If $\{E_i\}_{i\in\num{1}{n^2}}\subseteq\M{n}$ and $\{F_j\}_{j\in\num{1}{m^2}}\subseteq\M{m}$ are bases of $\M{n}$ and $\M{m}$, respectively, then, $\{E_i\oslash F_j\}_{i\in\num{1}{n^2}, \hspace{0.5mm}j\in\num{1}{m^2}}$ is a basis of $\M{nm}$.
 
 \item
 \label{kron:assoc}
If $p\in\N$, the Kronecker product (for $n$ and $m$, for $nm$ and $p$, for $n$ and $mp$, and for $m$ and $p$, respectively) is associative in the sense that $(A\oslash B)\oslash C=A\oslash (B\oslash C)$ for all $A\in\M{n}$, all $B\in\M{m}$, and all $C\in\M{p}$ (and, hence, we can write $A\oslash B\oslash C$).

\item
\label{kron:cross}
The spectral norm has the so-called cross norm property, \ie, $\|A\oslash B\|=\|A\| \|B\|$ for all $A\in\M{n}$ and all $B\in\M{m}$.
\en
\el
 
\bprf
See, for example, \cite{HJ2}.  
\eprf

\br
\label{rem:inj}
Let $n, m\in\N$, $A\in\M{n}$, and $B\in\M{m}$. If 
\ba
\label{Kron-1}
A\oslash B
=0, 
\ea
we get from \eqref{KP} that $(A,B)\in (\{0\}\times\M{m})\cup (\M{n}\times \{0\})$. Moreover, if $A\neq 0$, $B\neq 0$, $C\in\M{n}$, $D\in\M{m}$, and 
\ba
\label{Kron-2}
A\oslash B
=C\oslash D, 
\ea
 \eqref{KP} implies that there exists $\lambda\in\C\setminus\{0\}$ such that $C=\lambda A$ and $D=B/\lambda$. For both properties \eqref{Kron-1} and \eqref{Kron-2}, Lemma \ref{lem:kron} \ref{kron:assoc} yields direct generalizations to more than two factors.
\er

In the following, if $n\in\N$ and if $\mV_1,\ldots,\mV_n$ and $\mU$ are any vector spaces, we denote by $L_n(\mV_1,\ldots,\mV_n; \mU)$ the set of all $n$-multilinear maps $\mV_1\times\ldots\times\mV_n\to\mU$ (defined for all $(v_1,\ldots,v_n)\in\mV_1\times\ldots\times\mV_n$). If $n=1$, we write $L(\mV,\mU):=L_1(\mV;\mU)$ and $L(\mV):=L(\mV,\mV)$. Analogously, $\bar L_n(\mV_1,\ldots,\mV_n;\mU)$ stands for the corresponding multi-antilinear maps. 

\begin{figure}
\begin{center}
\begin{tikzcd}
&&\mV/\mW
\arrow[dd, "\textstyle S"]\\
& \mV
\arrow[ur,"\textstyle p", two heads, start anchor={[xshift=-0.5mm, yshift=-1.5mm]}, end anchor={[xshift=2mm, yshift=0mm]}]
\arrow[dr, "\textstyle T"', start anchor={[xshift=-0.5mm, yshift=1.5mm]}, end anchor={[xshift=0mm, yshift=-1mm]}]\\
&& \mU
\end{tikzcd}
\end{center}
\caption{Existence and uniqueness of $S\in L(\mV/\mW,\mU)$ for given $\mU$ and given $T\in L(\mV,\mU)$ with $\mW\subseteq\ker(T)$.}
\label{fig:univ0}
\end{figure}
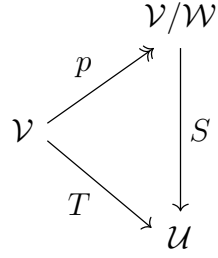

In order to define the local observable algebras, we make use of the algebraic tensor product of vector spaces whose definition we want to recall next. Let $\mV$ be any vector space, $\mW$ a vector subspace of $\mV$, and let $\{(v,v')\in \mV\times\mV\,|\, v'-v\in\mW\}$ be the usual equivalence relation on $\mV$ which defines the set
\ba
\label{V/W}
\mV/\mW
:=\{[v]\,|\,v\in\mV\}
\ea
of equivalence classes $[v]:=\{v'\in\mV\,|\, v'-v\in\mW\}$ for all $v\in\mV$. Equipped with the addition $\mV/\mW\times \mV/\mW\to\mV/\mW$ and the scalar multiplication  $\C\times \mV/\mW\to\mV/\mW$ (well-)defined, for all  $v, w\in\mV$ and all $\lambda\in\C$, by $[v]+[w]:=[v+w]$ and $\lambda[v]:=[\lambda v]$, the set $\mV/\mW$ becomes a vector space called the quotient space of $\mV$ by $\mW$. Moreover, $\mV/\mW$ possesses the following (universal) property:  for all vector spaces $\mU$ and all $T\in L(\mV,\mU)$ satisfying 
\ba
\label{WkerT}
\mW\subseteq\ker(T), 
\ea
where $\ker(T)$ stands for the kernel of $T$, there exists a unique $S\in L(\mV/\mW,\mU)$ such that $T=S\circ p$, where the so-called quotient map $p\in L(\mV,\mV/\mW)$ is defined by $pv:=[v]$ for all $v\in\mV$ (recall that $0\to\mW\hookrightarrow\mV\stackrel{p}{\to}\mV/\mW\to 0$ is a short exact sequence, where the arrow with a hook stands for the canonical inclusion map), see Figure \ref{fig:univ0} (a two-headed arrow represents a surjection).

We next use Figure \ref{fig:univ0} as follows (see, for example, \cite{Lang}). Let $n\in\N$ with $n\ge 2$, let $\mV_1, \ldots,\mV_n$ be vector spaces, and set $X:=\mV_1\times\ldots\times\mV_n$ (considered as a set only). The so-called free vector space $\mF(X)$ over $X$ (also called the formal linear combinations of elements of $X$) is defined to be the set ($\card$ stands for cardinality)
\ba
\mF(X)
:=\{\vi:X\to\C\,|\,\card(\vi^{-1}(\C\setminus\{0\}))\in\N_0\},
\ea
equipped with the addition $\mF(X)\times\mF(X)\to\mF(X)$ and the scalar multiplication $\C\times\mF(X)\to\mF(X)$ defined, for all $\vi,\psi\in\mF(X)$, all $x\in X$, and all $\lambda\in\C$, by $(\vi+\psi)(x):=\vi(x)+\psi(x)$ and $(\lm\vi)(x):=\lm\vi(x)$, respectively, making $\mF(X)$ into a vector space. Moreover, we know that the map $\delta:X\to\mF(X)$ defined, for all $x,y\in X$, by
\ba
\label{delta}
\delta(x)(y)
:=\begin{cases}
1, & y=x,\\
0, & y\neq x,
\end{cases}
\ea
is injective, has the property that $\ran(\delta)$ is linearly independent (where $\ran(\delta)$ stands for the range of $\delta$), and satisfies 
\ba
\label{SpanRan}
\spa(\ran(\delta))
=\mF(X), 
\ea
where $\spa(\mM)$ stands for the (finite) linear span of a subset $\mM\subseteq\mV$ of a vector space $\mV$. Finally, we define the vector subspace $\mF_0(X)$ of $\mF(X)$ by
\ba
\mF_0(X)
:=\spa(\{&\delta(v_1,\ldots,v_i+w_i,\ldots,v_n)-\delta(v_1,\ldots,v_i,\ldots,v_n)-\delta(v_1,\ldots,w_i,\ldots,v_n), \nonumber\\
&\delta(v_1,\ldots,\lm v_i,\ldots,v_n)-\lm \delta(v_1,\ldots,v_n)\,|\, i\in\num{1}{n}, v_i, w_i\in V_i, \lm\in\C\}), 
\ea
and we apply the upper branch of Figure \ref{fig:univ0} to $\mV:=\mF(X)$ and $\mW:=\mF_0(X)$ as follows.

\bd[Algebraic tensor product]
\label{def:atp}
Let $n\in\N$ with $n\ge 2$, let $\mV_1, \ldots, \mV_n$ be vector spaces, and set $X:=\mV_1\times\ldots\times\mV_n$.  The algebraic tensor product of $\mV_1, \ldots,\mV_n$  is defined by 
\ba
\label{tensor1}
\mV_1\odot\ldots\odot\mV_n
:=\mF(X)/\mF_0(X).
\ea
Moreover, for all $v_i\in\mV_i$ with $i\in\num{1}{n}$, the simple tensors are defined by
\ba
\label{tensor2}
v_1\otimes\ldots\otimes \hspace{1mm}v_n
:=p(\delta(v_1,\ldots,v_n)).
\ea
\ed

\begin{figure}
\begin{center}
\begin{tikzcd}
	& 
	& \mV_1\odot\ldots\odot \mV_n \arrow[dd, "\textstyle S_f"]\\
\mV_1\times\ldots\times\mV_n \arrow[r, "\textstyle\delta", tail, shift left=0mm] \arrow[rrd, "\textstyle f"', start anchor={[xshift=0mm, yshift=0.5mm]}, end anchor={[xshift=0mm, yshift=0mm]}] 
	& \mF(\mV_1\times\ldots\times\mV_n) \arrow[ur, "\textstyle p", two heads, start anchor={[xshift=7mm, yshift=-3mm]}, end anchor={[xshift=6mm, yshift=-1mm]}] \arrow[dr, "\textstyle T_f", start anchor={[xshift=7mm, yshift=3mm]}, end anchor={[xshift=0mm, yshift=1mm]}]
	& \\
	& 
	& \mU
\end{tikzcd}
\end{center}
\caption{Existence and uniqueness of  $S_f\in L(\mV_1\odot\ldots\odot\mV_n,\mU)$ for given $\mU$ and $f\in L_n(\mV_1,\ldots,\mV_n;\mU)$.}
\label{fig:universal}
\end{figure}
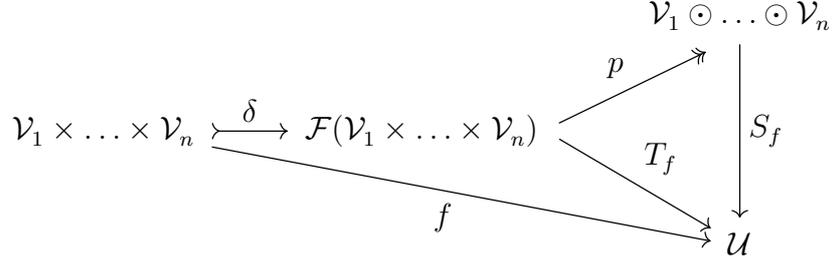

In order to prove various features of \eqref{tensor1} and \eqref{tensor2}, we also use the lower branch of Figure \ref{fig:univ0}, \ie, the universal property of quotient spaces. To this end, let $\mU$ be any vector space and $f\in L_n(\mV_1,\ldots,\mV_n;\mU)$ (with $n\ge 2$). Then, we know that the operator $T_f\in L(\mF(X),\mU)$, defined by $T_f(\delta(x)):=f(x)$ for all $x\in X$ (and by linear extension to the whole of $\mF(X)$), has the property that $\mF_0(X)\subseteq\ker(T_f)$. Hence, due to Figure \ref{fig:univ0}, there exists a unique $S_f\in L(\mV_1\odot\ldots\odot\mV_n,\mU)$ such that, for all $v_i\in\mV_i$ with $i\in\num{1}{n}$, 
\ba
\label{Sfv}
S_f(v_1\otimes\ldots\otimes \hspace{1mm}v_n)
=f(v_1,\ldots,v_n),
\ea
see Figure \ref{fig:universal} (an arrow with a tail represents an injection).

\br
In Figure \ref{fig:universal}, let $n=1$ and set $\mV:=\mV_1$. Moreover, let $\mU:=\mV$ and let $f\in L(\mV)$ be the identity map. Then, $S_f\in L(\mF(\mV)/\mF_0(\mV),\mV)$ is a bijection. In order to verify this claim, we first note that $S_f$ is clearly surjective since, due to the commutativity of Figure \ref{fig:universal}, we have $v=f(v)=S_f(p(\delta(v)))$ for all $v\in\mV$. Next, let us show that $\ker(T_f)\subseteq\mF_0(\mV)$ (recall \eqref{WkerT}). Due to \eqref{SpanRan}, for all $\vi\in\mF(\mV)$, there exists $m\in\N$ and,  for all $i\in\num{1}{m}$, there exist $\alpha_i\in\C$ and pairwise distinct $v_i\in\mV$ such that $\vi=\sum_{i\in\num{1}{m}}\alpha_i\delta(v_i)$. Hence, if $\vi\in\ker(T_f)$, Figure \ref{fig:universal} yields $\sum_{i\in\num{1}{m}}\alpha_iv_i=0$. If $\{v_1,\ldots,v_m\}$ is a linearly independent set, we are done. Otherwise, there exists $i\in\num{1}{m}$ such that $\alpha_i\neq 0$. If $m=1$, we have $v_1=0$ and $\vi=\alpha_1\delta(0)\in\mF_0(\mV)$. If $m=2$, we suppose that $\alpha_2\neq 0$ and write $v_2=\lambda v_1$, where $\lambda:=-\alpha_1/\alpha_2$. Hence, we get $\vi=\alpha_2\{\delta(\lm v_1)-\lm\delta(v_1)\}\in\mF_0(\mV)$. If $m\ge 3$, we suppose that $\alpha_m\neq0$ and write $v_m=\sum_{i\in\num{1}{m-1}}(-\alpha_i/\alpha_m)v_i$. Since $\delta(\sum_{i\in\num{1}{N}}\lm_iv_i)=\delta(\lm_{N}v_{N})+\{\delta(\sum_{i\in\num{1}{N}}\lm_iv_i)-\delta(\sum_{i\in\num{1}{N-1}}\lm_iv_i)-\delta(\lm_{N}v_{N})\}+\delta(\sum_{i\in\num{1}{N-1}}\lm_iv_i)$, where $N:=m-1$ and $\lm_i:=-\alpha_i/\alpha_m$ for all $i\in\num{1}{N}$, we recursively get 
\ba
\label{kerF0}
\vi
&=\alpha_m\hspace{-2mm}\sum_{i\in\num{1}{m-1}}\hspace{-2mm}\big\{\delta(\lm_iv_i)-\lm_i\delta(v_i)\big\}\nonumber\\
&+\alpha_m\hspace{-2mm}\sum_{j\in\num{2}{m-1}}\hspace{-2mm}\big\{\delta\big(\sum\nolimits_{i\in\num{1}{j}}\lm_iv_i\big)-\delta\big(\sum\nolimits_{i\in\num{1}{j-1}}\lm_iv_i\big)-\delta(\lm_jv_j)\big\},
\ea
\ie, $\vi\in\mF_0(\mV)$ as desired. 
We now want to show that $S_f$ is injective. Due to Figure \ref{fig:universal}, if $\eta\in\mF(\mV)/\mF_0(\mV)$, there exists $\vi\in\mF(\mV)$ such that $\eta=p\vi$. Hence, if $\eta\in\ker(S_f)$, we have $0=S_f\eta=T_f\vi$ and \eqref{kerF0} yields $\vi\in\mF_0(\mV)$. Since $\ker(p)=\mF_0(\mV)$, we get $\eta=0$.
\er

In the following, for all vector spaces $\mV$ and $\mW$, a map $\mV\to\mW$ is called a vector space isomorphism if it belongs to $L(\mV,\mW)$ and if it is a bijection.

We next collect some properties of the algebraic tensor product. 

\bl[Algebraic tensor product]
\label{lem:odot}
Let $n\in\N$ with $n\ge 2$ and let $\mV_1,\ldots,\mV_n$ be vector spaces. Then, the algebraic tensor product has the following properties:
\bn[label=(\alph*), ref={\it (\alph*)}]
\setlength{\itemsep}{0mm}
\item 
\label{odot:ml}
The map $\mV_1\times\ldots\times\mV_n\ni (v_1,\ldots,v_n)\mapsto v_1\otimes\ldots\otimes \hspace{1mm}v_n\in\mV_1\odot\ldots\odot\mV_n$ is $n$-multilinear (and called the tensor product of $v_i\in\mV_i$ with $i\in\num{1}{n}$).

\item 
\label{odot:bas}
Let $n_1, n_2\in\N$ and let $\{v_i\}_{i\in\num{1}{n_1}}\subseteq\mV_1$ and $\{w_j\}_{j\in\num{1}{n_2}}\subseteq\mV_2$ be bases of $\mV_1$ and $\mV_2$, respectively. Then, $\{v_i\otimes w_j\}_{i\in\num{1}{n_1}, \hspace{0.5mm}j\in\num{1}{n_2}}$ is a basis of $\mV_1\odot\mV_2$.

\item 
\label{odot:comm}
There exists a unique vector space isomorphism $\mV_1\odot\mV_2\to\mV_2\odot\mV_1$ such that $v_1\otimes v_2\mapsto v_2\otimes v_1$ for all $v_i\in\mV_i$ with $i\in\num{1}{2}$.

\item 
\label{odot:assoc}
There exists a unique vector space isomorphism $(\mV_1\odot\mV_2)\odot\mV_3\to\mV_1\odot (\mV_2\odot\mV_3)$ 
 such that $(v_1\otimes v_2)\otimes v_3\mapsto v_1\otimes (v_2\otimes v_3)$ for all $v_i\in\mV_i$ with $i\in\num{1}{3}$.  Moreover, for all $n\ge 3$, there exists a unique  bijection in $L((\mV_1\odot\ldots\odot\mV_{n-1})\odot\mV_n,\mV_1\odot\ldots\odot\mV_n)$ such that $(v_1\otimes\ldots\otimes v_{n-1})\otimes v_n\mapsto v_1\otimes\ldots\otimes v_n$ for all $v_i\in\mV_i$ with $i\in\num{1}{n}$.
\en
\el

\bprf
See, for example, \cite{Lang}. 
\eprf

\br
\label{rem:st}
Due to Lemma \ref{lem:odot} \ref{odot:ml} and the properties of \eqref{delta}, every element of the algebraic tensor product can be written as a finite sum of simple tensors. Of course, Lemma \ref{lem:odot} \ref{odot:ml} and \ref{odot:bas} yield the same conclusion.
\er

\br
\label{rem:tpvanish}
Let $n=2$ and let $v_1\in\mV_1$ and $v_2\in\mV_2$ be such that $v_1\otimes v_2=0$. Then, Figure \ref{fig:universal} implies that $f(v_1,v_2)=0$ for all vector spaces $\mU$ and all $f\in L_2(\mV_1,\mV_2;\mU)$. 
\er

\br
Let $n\in\N$, let $\mV_1,\ldots,\mV_n$ be fixed vector spaces, and let us consider the category whose objects are elements of $L_n(\mV_1,\ldots,\mV_n;\mU)$ for (the proper class of) all vector spaces $\mU$ and whose morphisms from $f\in L_n(\mV_1,\ldots,\mV_n;\mU)$ to $f'\in L_n(\mV_1,\ldots,\mV_n;\mU')$ are operators $T\in L(\mU,\mU')$ such that $f'(x)=T f(x)$ for all $x\in\mV_1\times\ldots\times\mV_n$. Hence, the tensor product $p\circ\delta\in L_n(\mV_1,\ldots,\mV_n; \mV_1\odot\ldots\odot\mV_n)$ is a universally repelling object in this category.
\er

\br
Due to Lemma \ref{lem:kron} \ref{kron:assoc}, the multiple Kronecker product is constructed by iteration whereas the multiple algebraic tensor product is defined directly (and the iterated algebraic tensor products are isomorphic, see Lemma \ref{lem:odot} \ref{odot:assoc}).
\er

Next, we want to bring the configuration space $\Z$ into play. To this end, we denote the set of all finite subsets of the configuration space by
\ba
\label{FinZ}
\Fin(\Z)
:=\{\Lambda\subseteq\Z\,|\,\card(\Lambda)\in\N\},
\ea
and we equip it with a direction (\ie, with a upward directed partial ordering) defined by set inclusion. Moreover,  for all $x\in\Z$, let $\fA_{x}$ be a unital \Cs algebra with identity $1_{\{x\}}$ and suppose that there exists
\ba
\label{copy}
\xi_x\in\sIso(\fA_x, \M{2}),
\ea
\ie,  suppose that $\fA_x$ is a ''copy'' of $\M{2}$ for all sites $x\in\Z$.

\bd[Local tensor product]
\label{def:ltp}
Let $\Lambda=\{x_1,\ldots,x_n\}\in\Fin(\Z)$ for some $n\in\N$ be such that $x_1<\ldots<x_n$ if $n\ge 2$. The vector space 
\ba
\label{AL}
\fA_\Lambda
:=\begin{cases}
\hfill \fA_{x_1}, & n=1,\\
\fA_{x_1}\odot\ldots\odot \fA_{x_n}, & n\ge 2,
\end{cases}
\ea
is called the local tensor product (over $\Lambda$).
\ed

\br
Due to Lemma \ref{lem:odot} \ref{odot:comm}, $\fA_{x_{\pi(1)}}\odot\ldots\odot \fA_{x_{\pi(n)}}$ and $\fA_{x_1}\odot\ldots\odot \fA_{x_n}$ for all $n\ge 2$ are (vector space) isomomorphic for all permutations $\pi\in\mS_n$ (where, for all $n\in\N$, we denote by $\mS_n$ the permutation group over the set $\{1,\ldots,n\}$).
\er

For the right hand side of \eqref{NormLambda} below, we recall that the (unique) \Cs norm on $\M{n}$ is the spectral norm given in \eqref{specnorm}.

We next make  the local tensor product into a  \Cs algebra (as mentioned in the Introduction, see Appendix \ref{app:Cstar} for the main definitions and some of the basic facts about \Cs algebras used in the following).

\bl[Local observable algebras]
\label{lem:ltp}
Let $\Lambda=\{x_1,\ldots,x_n\}\in\Fin(\Z)$ for some $n\ge 2$ be such that $x_1<\ldots<x_n$ and let $\fA_\Lambda$ be the local tensor product over $\Lambda$. Then:
\bn[label=(\alph*), ref={\it (\alph*)}]
\setlength{\itemsep}{0mm}
\item 
\label{ltp:star}
There exists a unique multiplication $\fA_\Lambda\times\fA_\Lambda\to\fA_\Lambda$ and a unique involution $\fA_\Lambda\to\fA_\Lambda$ satisfying, for all $A_i, B_i\in\fA_{x_i}$ with $i\in\num{1}{n}$,
\ba
\label{ALmProd}
(A_1\otimes\ldots\otimes A_n, B_1\otimes\ldots\otimes B_n)
&\mapsto (A_1B_1)\otimes\ldots\otimes(A_nB_n),\\
\label{ALmInvo}
A_1\otimes\ldots\otimes A_n
&\mapsto A_1^\ast\otimes\ldots\otimes A_n^\ast,
\ea
respectively, making $\fA_\Lambda$ into a $\ast$-algebra.

\item
\label{ltp:iso}
There exists a unique $\xi_\Lm\in\sIso(\fA_\Lambda, \M{2^n})$ satisfying,  for all $A_i\in\fA_{x_i}$ with $i\in\num{1}{n}$, 
\ba
\label{KTiso}
\xi_\Lm(A_1\otimes\ldots\otimes A_n)
=\xi_{x_1}(A_1)\oslash\ldots\oslash\xi_{x_n}(A_n).
\ea

\item
\label{ltp:Cstar}
The map $\Ln{\cdot}:\fA_\Lambda\to\R$ defined, for all $A\in\fA_\Lambda$, by
\ba
\label{NormLambda}
\Ln{A}
:=\|\xi_\Lm(A)\|,
\ea
is the unique \Cs norm on the $\ast$-algebra $\fA_\Lambda$. Hence, $\fA_\Lambda$ is a unital \Cs algebra which is called the  local observable algebra (over $\Lm$).
\en
\el

\br
\label{rem:op}
Let $\fA$ be a unital \Cs algebra, $\mV$ a vector space, and let $\vi\in L(\mV,\fA)$ be a  
vector space isomorphism. Then, equipped with the multiplication $\mV\times\mV\to\mV$, the involution $\mV\to\mV$, and the \Cs norm $\mV\to\R$ defined, for all $v,w\in\mV$, by $(v,w)\mapsto\vi^{-1}(\vi(v)\vi(w))$, by $v\mapsto\vi^{-1}([\vi(v)]^\ast)$, and  by $v\mapsto\|\vi(v)\|$, respectively, the vector space $\mV$ becomes a \Cs algebra. Moreover, it is a unital \Cs algebra with identity $1_\mV:=\vi^{-1}(1_\fA)$, where $1_\fA$ is the identity of $\fA$.
\er

In the following, we repeatedly apply the identification from Remark \ref{rem:op} to various vector spaces which are vector space isomorphic to $\fA_\Lm$ through Lemma \ref{lem:odot} \ref{odot:comm} and \ref{odot:assoc} (without mentioning it explicitly each time). 

\vspace{5mm}

\bprf
\ref{ltp:star}\,
In order to define the desired multiplication which we denote by $M:\fA_\Lambda\times\fA_\Lambda\to\fA_\Lambda$, we use Figure \ref{fig:universal} twice. First, we specify the upper branch, \ie, $\mV_1, \ldots,\mV_n$, by $\mV_i:=\fA_{x_i}$ for all $i\in\num{1}{n}$, and the lower branch, \ie, $\mU$ and $f\in L_n(\mV_1,\ldots,\mV_n;\mU)$, by $\mU:=\fA_\Lambda$  and, for fixed $A_i\in\fA_{x_i}$ with $i\in\num{1}{n}$, by the map $f:=g_{A_1,\ldots,A_n}: \fA_{x_1}\times\ldots\times\fA_{x_n}\to\fA_\Lambda$ defined by $g_{A_1,\ldots,A_n}(B_1,\ldots,B_n):=(A_1B_1)\otimes\ldots\otimes(A_nB_n)$ for all $B_i\in\fA_{x_i}$ with $i\in\num{1}{n}$. Due to  Lemma \ref{lem:odot} \ref{odot:ml}, we have $g_{A_1,\ldots,A_n}\in L_n(\fA_{x_1},\ldots,\fA_{x_n};\fA_\Lambda)$ and Figure \ref{fig:universal} thus yields $S_{g_{A_1,\ldots,A_n}}\in L(\fA_\Lambda)$ satisfying, for all $B_i\in\fA_{x_i}$ with $i\in\num{1}{n}$,
\ba
S_{g_{A_1,\ldots,A_n}}(B_1\otimes\ldots\otimes B_n)
=(A_1B_1)\otimes\ldots\otimes(A_nB_n).
\ea
Second, we again use Figure \ref{fig:universal} but, this time, for $\mV_i:=\fA_{x_i}$ for all $i\in\num{1}{n}$, for $\mU:=L(\fA_\Lambda)$, and for the map $f:\fA_{x_1}\times\ldots\times\fA_{x_n}\to L(\fA_\Lambda)$ defined, for all $A_i\in\fA_{x_i}$ with $i\in\num{1}{n}$, by 
\ba
\label{f2nd}
f(A_1,\ldots,A_n)
:=S_{g_{A_1,\ldots,A_n}}.
\ea
Due to Lemma \ref{lem:odot} \ref{odot:ml} and Remark \ref{rem:st}, \eqref{f2nd} satisfies $f\in L_n(\fA_{x_1},\ldots,\fA_{x_n};L(\fA_\Lambda))$. Hence, Figure \ref{fig:universal} yields $S_f\in L(\fA_\Lambda, L(\fA_\Lambda))$. Moreover, with the help of the usual vector space isomorphism $\vi: L(\fA_\Lambda, L(\fA_\Lambda))\to L_2(\fA_\Lambda, \fA_\Lambda;\fA_\Lambda)$ given by $\vi(S)(A,B):=S(A)B$ for all $A,B\in\fA_\Lambda$ (with inverse $\vi^{-1}_g(A) B=g(A,B)$ for all $g\in L_2(\fA_\Lambda, \fA_\Lambda;\fA_\Lambda)$ and all $A, B\in\fA_\Lambda$), we define the map $M:\fA_\Lambda\times\fA_\Lambda\to\fA_\Lambda$ by $M:=\vi(S_f)\in L_2(\fA_\Lambda, \fA_\Lambda;\fA_\Lambda)$. Hence, for all  $A_i, B_i\in\fA_{x_i}$ with $i\in\num{1}{n}$,
\ba
\label{Mprop}
M(A_1\otimes\ldots\otimes A_n,B_1\otimes\ldots\otimes B_n)
&=S_f(A_1\otimes\ldots\otimes A_n)(B_1\otimes\ldots\otimes B_n)\nonumber\\
&=f(A_1,\ldots,A_n)(B_1\otimes\ldots\otimes B_n)\nonumber\\
&=S_{g_{A_1,\ldots,A_n}}(B_1\otimes\ldots\otimes B_n)\nonumber\\
&=(A_1B_1)\otimes\ldots\otimes(A_nB_n),
\ea
and, due to the uniqueness property of the vertical branch of Figure \ref{fig:universal}, $M$ is the unique map in $L_2(\fA_\Lambda, \fA_\Lambda;\fA_\Lambda)$ satisfying \eqref{Mprop}. Finally, due to Remark \ref{rem:st} and the associativity of the multiplications on the \Cs algebras $\fA_{x_i}$ with $i\in\num{1}{n}$, the map $M$ is associative, too. In order to define the involution which we denote by $I:\fA_\Lambda\to\fA_\Lambda$, we proceed similarly. But, since an involution is antihomogeneous (and additive), we have to modify Figure \ref{fig:univ0} by considering $T\in\bar L(\mV,\mU)$ with $\mW\in\ker (T)$ (where $\ker$ is defined for $T\in\bar L(\mV,\mU)$ as it is for the linear case). Defining $S[v]:=Tv$ for all $v\in\mV$ (using the same notations as after \eqref{WkerT}), we get that $S$ is well-defined and antilinear, that it satisfies $T=S\circ p$ and that it is unique. As to the resulting modification of Figure \ref{fig:universal}, we pick $f\in \bar L_n(\mV_1,\ldots,\mV_n;\mU)$ which defines $T_f\in\bar L(F(X),\mU)$ by $T_f\delta_x:=f(x)$ for all $x\in X$ (using the same notations as before \eqref{Sfv}) and, now, by antilinear extension to the whole of $F(X)$. Moreover, $T_f$ again satisfies $F_0(X)\subseteq\ker(T_f)$. The modification of Figure \ref{fig:univ0} then yields the unique $S_f\in\bar L(\mV_1\odot\ldots\odot\mV_n,\mU)$ such that $S_f(v_1\otimes\ldots\otimes \hspace{1mm}v_n)=f(v_1,\ldots,v_n)$ for all $v_i\in\mV_i$ with $i\in\num{1}{n}$. 
We next specify the upper branch of the modification of Figure \ref{fig:universal} by $\mV_i:=\fA_{x_i}$ for all $i\in\num{1}{n}$ and the lower branch by $\mU:=\fA_\Lambda$ and the map $f:\fA_{x_1}\times\ldots\times\fA_{x_n}\to \mU$ defined by $f(A_1,\ldots,A_n):=A_1^\ast\otimes\ldots\otimes A_n^\ast$ for all $A_i\in\fA_{x_i}$ with $i\in\num{1}{n}$. Due to Lemma \ref{lem:odot} \ref{odot:ml}, we have $f\in \bar L_n(\fA_{x_1},\ldots,\fA_{x_n};\fA_\Lambda)$, and using the modification of Figure \ref{fig:universal}, we define the map $I:\fA_\Lambda\to\fA_\Lambda$ by $I
:=S_f\in \bar L(\fA_\Lambda,\fA_\Lambda)$. Hence, for all $A_i\in\fA_{x_i}$ with $i\in\num{1}{n}$,
\ba
\label{Iprop}
I(A_1\otimes\ldots\otimes A_n)
=A_1^\ast\otimes\ldots\otimes A_n^\ast,
\ea
and, due to the uniqueness property of the vertical branch of the modification of Figure \ref{fig:universal}, $I$ is the unique map in $\bar L(\fA_\Lambda,\fA_\Lambda)$ satisfying \eqref{Iprop}. Finally,  due to Remark \ref{rem:st} and since, for all $i\in\num{1}{n}$, the antilinear operators ${}^\ast:\fA_{x_i}\to\fA_{x_i}$ are involutive and antidistributive (see Appendix \ref{app:Cstar}), the same holds for the map $I$.

\ref{ltp:iso}\, 
In order to define the desired \str isomorphism, we again use Figure \ref{fig:universal} whose upper and lower branches are respectively specified by $\mV_i:=\fA_{x_i}$ for all $i\in\num{1}{n}$, and by $\mU:=\M{2^n}$ and the map $f:\fA_{x_1}\times\ldots\times\fA_{x_n}\to\mU$ defined by $f(A_1,\ldots,A_n):=\xi_{x_1}(A_1)\oslash\ldots\oslash \xi_{x_n}(A_n)$ for all $A_i\in\fA_{x_i}$ with $i\in\num{1}{n}$. Due to  Lemma \ref{lem:kron} \ref{kron:bilin} and \ref{kron:assoc} and since $\xi_x\in\sIso(\fA_x, \M{2})$ for all $x\in\Z$, we have $f\in L_n(\fA_{x_1},\ldots,\fA_{x_n}; \M{2^n})$. Using Figure \ref{fig:universal}, we define the map $\xi_\Lm:\fA_\Lambda\to\M{2^n}$ by $\xi_\Lm:=S_f\in L(\fA_\Lambda,  \M{2^n})$. Hence, for all $A_i\in\fA_{x_i}$ with $i\in\num{1}{n}$,
\ba
\label{piprop}
\xi_\Lm(A_1\otimes\ldots\otimes A_n)
=\xi_{x_1}(A_1)\oslash\ldots\oslash\xi_{x_n}(A_n),
\ea
and, due to the uniqueness property of the vertical branch of Figure \ref{fig:universal}, $\xi_\Lm$ is the unique map in $L(\fA_\Lambda, \M{2^n})$ satisfying \eqref{piprop}. Moreover, due to Remark \ref{rem:st} and using, in particular, Lemma \ref{lem:kron} \ref{kron:bilin}, we get $\xi_\Lm\in\sHom(\fA_\Lambda,  \M{2^n})$. Finally, we have to show that $\xi_\Lm$ is bijective. To this end, let $\{E_i\}_{i\in\num{1}{4}}$ be a basis of $\M{2}$. Then, $\{\xi_x^{-1}(E_i)\}_{i\in\num{1}{4}}$ is a basis of $\fA_x$ for all $x\in\Z$ since $\xi_x\in\sIso(\fA_x, \M{2})$ for all $x\in\Z$. Due to Lemma \ref{lem:kron} \ref{kron:base} and Lemma \ref{lem:odot} \ref{odot:bas}, respectively, $\{E_{i_1}\oslash\ldots\oslash E_{i_n}\}_{i_1,\ldots, i_n\in\num{1}{4}}$ is a basis of $\M{2^n}$ and $\{\xi_{x_1}^{-1}(E_{i_1})\otimes\ldots\otimes \xi_{x_n}^{-1}(E_{i_n})\}_{i_1,\ldots, i_n\in\num{1}{4}}$ is a basis of $\fA_\Lambda$. Moreover, $\xi_\Lm$ satisfies $\xi_\Lm(\xi_{x_1}^{-1}(E_{i_1})\otimes\ldots\otimes \xi_{x_n}^{-1}(E_{i_n}))=E_{i_1}\oslash\ldots\oslash E_{i_n}$ for all $i_1,\ldots, i_n\in\num{1}{4}$.

\ref{ltp:Cstar}\,
Since the spectral norm \eqref{specnorm} is a \Cs norm on $\M{2^n}$ and since $\xi_\Lm\in\sIso(\fA_\Lambda, \M{2^n})$, the map ${\n{\cdot}}_\Lambda:\fA_\Lambda\to\R$ is a \Cs norm on the \str algebra $\fA_\Lambda$. Moreover, we know that a \Cs norm on a $\ast$-algebra is unique. Finally, $\fA_\Lambda$ has the identity $1_\Lambda:=\xi_\Lm^{-1}(1_{2^n})=1_{\{x_1\}}\otimes\ldots\otimes 1_{\{x_n\}}$.
\eprf

\section{Infinite tensor product}
\label{sec:infinite}

In this section, we construct the infinite tensor product of the local \Cs algebras from Section \ref{sec:Local}. Playing the role of the observable algebra over the infinitely extended configuration space $\Z$, the infinite tensor product is the central object upon which the infinite system approach is based. Since the infinite tensor product is an example of a so-called inductive limit, let us first briefly recall the ingredients needed for the general construction of the latter (see, for example, \cite{Sakai} [or \cite{Emch}]). Subsequently, we specialize in detail the general case to the concrete case at hand.

One starts off with a net of unital \Cs algebras $(\fA_\alpha)_{\alpha\in\mJ}$, where $\mJ$ is a directed set of indices whose (upward directed) partial ordering is denoted by $\preceq$ and where,  for all $\alpha\in\mJ$, the \Cs norm on $\fA_\alpha$ is written as $\n{\cdot}_\alpha$. Moreover, let us assume that, for all $\alpha, \alpha'\in\mJ$ with $\alpha\preceq\alpha'$, there exists a map $\vi_{\alpha',\alpha}\in\sMon(\fA_\alpha,\fA_{\alpha'})$ which is unital (recall from Appendix \ref{app:Cstar} that this means that $\vi_{\alpha',\alpha}(1_\alpha)=1_{\alpha'}$, where,  for all  $\alpha\in\mJ$, we denote by $1_\alpha$ the identity of $\fA_\alpha$) and which satisfies the so-called cocycle condition, \ie, for all $\alpha, \alpha', \alpha''\in\mJ$ with $\alpha\preceq\alpha'$ and $\alpha'\preceq\alpha''$, 
\ba
\label{cocycle}
\vi_{\alpha'',\alpha'}\circ\vi_{\alpha',\alpha}
=\vi_{\alpha'',\alpha},
\ea
and let us call $\{\vi_{\alpha',\alpha}\}_{\alpha, \alpha'\in\mJ\hspace{-0.5mm}, \hspace{0.5mm}\alpha\preceq\alpha'}$ a family of isotonies (note that the right hand side of \eqref{cocycle} is well-defined due to the transitivity property of the partial ordering).
Next, consider the set
\ba
\label{defF}
\mF
:=\big\{f:\mJ\to\bigcup\nolimits_{\alpha\in\mJ}\fA_\alpha\,\big|\, \mbox{$f(\alpha)\in\fA_\alpha$ for all $\alpha\in\mJ$}\big\},
\ea
which, equipped with the addition $\mF\times\mF\to\mF$, the scalar multiplication $\C\times\mF\to\mF$, the multiplication $\mF\times\mF\to\mF$, and the involution $\mF\to\mF$ defined, for all $f,g\in\mF$, all $\lambda\in\C$, and all $\alpha\in\mJ$, by $(f+g)(\alpha):=f(\alpha)+g(\alpha)$, by $(\lambda f)(\alpha):=\lambda f(\alpha)$, by $(fg)(\alpha):=f(\alpha)g(\alpha)$, and by $(f^\ast)(\alpha):=(f(\alpha))^\ast$, respectively, becomes a unital \str algebra (below, we verify these and the following properties of the general construction in the concrete case at hand). Moreover, let the unital \str subalgebra $\mG\subseteq\mF$, defined by
\ba
\label{defG}
\mG
:=\{f\in\mF\,|\,&\mbox{there exists $\alpha\in\mJ$ such that}\nonumber\\
&\mbox{$f(\alpha')=\vi_{\alpha',\alpha}(f(\alpha))$ for all $\alpha'\in\mJ$ with $\alpha\preceq\alpha'$}\},
\ea
be equipped with the submultiplicative seminorm with \Cs property $|\cdot|:\mG\to\R$ defined, for all $f\in\mG$, by 
\ba
\label{defSN}
|f|
:=\lim_{\beta\in\mJ}{\|f(\beta)\|}_\beta,
\ea
where we used that, due to \eqref{piIso}, the limit $\lim_\beta{\|f(\beta)\|}_\beta$ of the net $\mJ\ni\beta\mapsto{\|f(\beta)\|}_\beta\in\R$ exists and equals ${\|f(\alpha)\|}_\alpha$ if $f\in\mG$. With the help of the 2-sided \str ideal $\mG_0\subseteq\mG$, defined by
\ba
\label{defG0}
\mG_0
:=\{f\in\mG\,|\, |f|=0\},
\ea
we define the vector space $\mG/\mG_0$ as in \eqref{V/W} (and use the notation from there). In addition, equipping $\mG/\mG_0$ with the multiplication $\mG/\mG_0\times\mG/\mG_0\to\mG/\mG_0$, the involution $\mG/\mG_0\to\mG/\mG_0$, and the submultiplicative norm with \Cs property $\|\cdot\|:\mG/\mG_0\to\R$ defined, for all $f,g\in\mG$, by $[f][g]:=[fg]$, by $[f]^\ast:=[f^\ast]$, and by $\|[f]\|:=|f|$, respectively, $\mG/\mG_0$ becomes a unital normed \str algebra. The \Cs completion (from Lemma \ref{lem:CsCmpl} of Appendix \ref{app:Cstar}) of the unital \str algebra $\mG/\mG_0$ equipped with the foregoing submultiplicative norm with \Cs property is denoted by $\fA$ and called the inductive limit of the net of  unital \Cs algebras $(\fA_\alpha)_{\alpha\in\mJ}$ with respect to the family of isotonies $\{\vi_{\alpha',\alpha}\}_{\alpha, \alpha'\in\mJ\hspace{-0.5mm}, \hspace{0.5mm}\alpha\preceq\alpha'}$ (sometimes, $\fA$ is denoted by $\lim_\alpha\{\fA_\alpha; \vi_{\alpha',\alpha}\,|\,\mbox{$\alpha, \alpha'\in\mJ$ with $\alpha\preceq\alpha'$}\}$, see, for example, \cite{Sakai}). Finally, there exists an increasing net of unital \Cs subalgebras $(\wt\fA_\alpha)_{\alpha\in\mJ}$  of $\fA$, \ie, $\wt \fA_\alpha\subseteq\wt\fA_{\alpha'}$ for all $\alpha, \alpha'\in\mJ$ with $\alpha\preceq\alpha'$, and a family of \str isomorphisms $\{\vi_\alpha\in\sIso(\fA_\alpha,\wt\fA_\alpha)\}_{\alpha\in\mJ}$ satisfying $\vi_{\alpha'}\circ\vi_{\alpha',\alpha}=\vi_\alpha$ for all  $\alpha, \alpha'\in\mJ$ with $\alpha\preceq\alpha'$ such that
\ba
\fA
=\clo_\fA\left(\bigcup\nolimits_{\alpha\in\mJ}\wt\fA_\alpha\right),
\ea
where $\clo_\fA$ stands for the closure with respect to the \Cs norm of $\fA$.\\

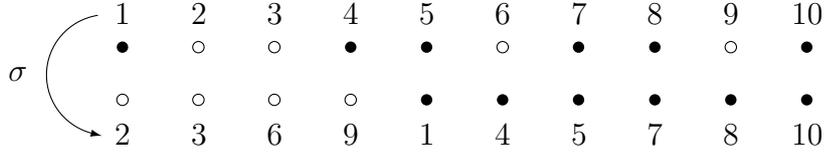
\begin{figure}
\vspace{10mm}
\setlength{\unitlength}{10mm}
\begin{picture}(0,0)(-3,0)
\put(1,0){\circle*{0.15}}
\put(0.9,0.3){$1$}
\put(2,0){\circle{0.15}}
\put(1.9,0.3){$2$}
\put(3,0){\circle{0.15}}
\put(2.9,0.3){$3$}
\put(4,0){\circle*{0.15}}
\put(3.9,0.3){$4$}
\put(5,0){\circle*{0.15}}
\put(4.9,0.3){$5$}
\put(6,0){\circle{0.15}}
\put(5.9,0.3){$6$}
\put(7,0){\circle*{0.15}}
\put(6.9,0.3){$7$}
\put(8,0){\circle*{0.15}}
\put(7.9,0.3){$8$}
\put(9,0){\circle{0.15}}
\put(8.9,0.3){$9$}
\put(10,0){\circle*{0.15}}
\put(9.8,0.3){$10$}
\put(1,-0.7){\circle{0.15}}
\put(0.9,-1.3){$2$}
\put(2,-0.7){\circle{0.15}}
\put(1.9,-1.3){$3$}
\put(3,-0.7){\circle{0.15}}
\put(2.9,-1.3){$6$}
\put(4,-0.7){\circle{0.15}}
\put(3.9,-1.3){$9$}
\put(5,-0.7){\circle*{0.15}}
\put(4.9,-1.3){$1$}
\put(6,-0.7){\circle*{0.15}}
\put(5.9,-1.3){$4$}
\put(7,-0.7){\circle*{0.15}}
\put(6.9,-1.3){$5$}
\put(8,-0.7){\circle*{0.15}}
\put(7.9,-1.3){$7$}
\put(9,-0.7){\circle*{0.15}}
\put(8.9,-1.3){$8$}
\put(10,-0.7){\circle*{0.15}}
\put(9.8,-1.3){$10$}
\put(0.8,-0.37){\arc[99,260]{0.8}}
\put(0.63,-1.155){\vector(2,-0.5){0.1}}
\put(-0.5,-0.45){$\sigma$}
\end{picture}
\vspace{15mm}
\caption{The permutation $\sigma\in\mS_{10}$ for $\Lm'=\{x_1, \ldots, x_{10}\}$ and $\Lm=\{x_2, x_3, x_6, x_9\}$.}
\label{fig:perm} 
\end{figure}

In the following, we specialize the foregoing general case to the case at hand, \ie, to the directed set of indices $\mJ:=\Fin(\Z)$ from \eqref{FinZ} with indices $\alpha:=\Lambda$ and to the net of unital \Cs algebras $(\fA_\Lambda)_{\Lambda\in\Fin(\Z)}$ from Lemma \ref{lem:ltp}. To this end, let $\Lm, \Lm'\in\Fin(\Z)$ with $\Lm\subseteq\Lm'$ and $\Lm\neq\Lm'$, set $n:=\card(\Lm)$ and $n':=\card(\Lm')$, and let $\Lm'=:\{x_1,\ldots,x_{n'}\}$ with $x_1<\ldots<x_{n'}$. Moreover, let $\sigma\in\mS_{n'}$ be the unique permutation satisfying $x_{\sigma(i)}\in\Lm$ for all $i\in\num{1}{n}$ with $\sigma(i)<\sigma(i+1)$ for all  $i\in\num{1}{n-1}$ if $n\ge 2$ and $x_{\sigma(i)}\in\Lm'\setminus\Lm$ for all $i\in\num{n+1}{n'}$ with $\sigma(i)<\sigma(i+1)$ for all $i\in\num{n+1}{n'-1}$ if $n'\ge n+2$, see Figure \ref{fig:perm}.
Then, due to Lemma \ref{lem:odot} \ref{odot:comm} and \ref{odot:assoc}, there exists a unique $\pi_{\Lm,\Lm'}\in\sIso(\fA_{\Lm'},\fA_{\Lm}\odot\fA_{\Lm'\setminus\Lm})$ such that, for all $A_i\in\fA_{x_i}$ with $i\in\num{1}{n'}$, 
\ba
\label{piLL}
\pi_{\Lm,\Lm'}(A_1\otimes\ldots\otimes A_{n'})
=(\otimes_{i\in\num{1}{n}} A_{\sigma(i)})\otimes(\otimes_{i\in\num{n+1}{n'}} A_{\sigma(i)}).
\ea
Finally, recall that, for all $\Lm\in\Fin(\Z)$, we denote by $1_\Lm$ the (multiplicative) identity of $\fA_\Lm$.

\bp[Infinite tensor product]
\label{prop:indlim}
For all $\Lm, \Lm'\in\Fin(\Z)$ with $\Lm\subseteq\Lm'$, let the map $\vi_{\Lm',\Lm}:\fA_\Lm\to\fA_{\Lm'}$ be defined, for all $A\in\fA_\Lm$, by
\ba
\label{viLL}
\vi_{\Lm',\Lm}(A)
:=
\begin{cases}
\hfill A, &  \Lm=\Lm',\\
\pi_{\Lm,\Lm'}^{-1}(A\otimes 1_{\Lm'\setminus\Lm}), & \Lm\neq\Lm'.
\end{cases}
\ea 
Then:
\bn[label=(\alph*), ref={\it (\alph*)}]
\setlength{\itemsep}{0mm}
\item 
\label{indlim:emb}
The family $\{\vi_{\Lm',\Lm}\}_{\Lm,  \Lm'\in\Fin(\Z), \hspace{0.3mm}\Lm\subseteq\Lm'}$ is a family of isotonies.

\item
\label{indlim:sub}
There exists a unital \Cs algebra $\fA$, an increasing net of unital \Cs subalgebras $(\wt\fA_\Lm)_{\Lm\in\Fin(\Z)}$ of $\fA$, and a family of \str isomorphisms $\{\vi_\Lm\in\sIso(\fA_\Lm, \wt\fA_\Lm)\}_{\Lm\in\Fin(\Z)}$ satisfying, for all $\Lm, \Lm'\in\Fin(\Z)$ with $\Lm\subseteq\Lm'$,
\ba
\label{indlim-2}
\vi_{\Lm'}\circ\vi_{\Lm',\Lm}
=\vi_\Lm,
\ea
such that 
\ba
\label{uhf}
\fA
=\clo_\fA\left(\bigcup\nolimits_{\Lm\in\Fin(\Z)}\wt\fA_\Lm\right).
\ea
The \Cs algebra $\fA$ is called the infinite tensor product of the net of local \Cs algebras $(\fA_\Lm)_{\Lm\in\Fin(\Z)}$. 
\en
\ep

\br
The \Cs algebra $\fA$ is also a so-called uniformly hyperfinite or Glimm algebra (see, for example, \cite{Emch, Sakai}).
\er

\br
The infinite tensor product $\fA$ is unique in the following sense (see, for example, \cite{Sakai} and also Proposition \ref{prop:AA1} below). Let $\fB$ be any unital \Cs algebra, $(\fB_\Lm)_{\Lm\in\Fin(\Z)}$ an increasing net of unital \Cs subalgebras  of $\fB$ such that $\fB=\clo_\fB(\bigcup_{\Lm\in\Fin(\Z)}\fB_\Lm)$, and let $\{\psi_\Lm\in\sIso(\fA_\Lm, \fB_\Lm)\}_{\Lm\in\Fin(\Z)}$ be a family of \str isomorphisms satisfying $\psi_{\Lm'}\circ\vi_{\Lm',\Lm}=\psi_\Lm$ for all $\Lm, \Lm'\in\Fin(\Z)$ with $\Lm\subseteq\Lm'$. Then, there exists $\Phi\in\sIso(\fA,\fB)$. 
\er

In the following, for all $\Lm\in\Fin(\Z)$, we denote by $0_{\Lm}$ the additive identity of $\fA_\Lm$. Moreover, recall from Lemma \ref{lem:ltp} \ref{ltp:iso} that $\xi_\Lm\in\sIso(\fA_\Lambda, \M{2^{n}})$ for all $\Lm\in\Fin(\Z)$ with $n=\card(\Lm)$.

\vspace{5mm}

\bprf
\ref{indlim:emb}\, 
Let $\Lm, \Lm'\in\Fin(\Z)$ with $\Lm\subseteq\Lm'$. Then, Lemma \ref{lem:odot} \ref{odot:ml} and Lemma \ref{lem:ltp} \ref{ltp:star} imply that $\vi_{\Lm',\Lm}\in\sHom(\fA_\Lm, \fA_{\Lm'})$. We next want to show that $\vi_{\Lm',\Lm}\in\sMon(\fA_\Lm, \fA_{\Lm'})$. If $\Lm=\Lm'$,  \eqref{viLL} yields the desired property. If $\Lm\neq\Lm'$ and if $A\in\fA_\Lm$ is such that $\vi_{\Lm',\Lm}(A)=0_{\Lm'}$, \eqref{viLL} yields $A\otimes 1_{\Lm'\setminus\Lm}=0\in\fA_{\Lm}\odot\fA_{\Lm'\setminus\Lm}$. Hence, Remark \ref{rem:tpvanish} implies that $f(A, 1_{\Lm'\setminus\Lm})=0$ for all vector spaces $\mU$ and all $f\in L_2(\fA_{\Lm},\fA_{\Lm'\setminus\Lm};\mU)$. In particular, if we pick $\mU:=\M{2^{n'}}$ and define $f(A,B):=\xi_\Lm(A)\oslash\xi_{\Lm'\setminus\Lm}(B)$ for all $A\in\fA_{\Lm}$ and all $B\in\fA_{\Lm'\setminus\Lm}$ (where $n:=\card(\Lm)$ and $n':=\card(\Lm')$), we get $\xi_\Lm(A)\oslash 1_{2^{n'-n}}=0_{2^{n'}}$ and Remark \ref{rem:inj} yields $\xi_\Lm(A)=0_{2^n}$, \ie, $A=0_\Lm$. Moreover, $\vi_{\Lm',\Lm}$ is also unital since, due to Remark \ref{rem:st} and Lemma \ref{lem:ltp} \ref{ltp:star}, we have $1_\Lm\otimes 1_{\Lm'\setminus\Lm}=1\in\fA_{\Lm}\odot\fA_{\Lm'\setminus\Lm}$. Finally, we directly check that,  due to \eqref{piLL}, the cocycle condition holds, too.

\ref{indlim:sub}\, 
We straightforwardly verify that \eqref{defF}, specialized to the case at hand, 
\ba
 \mF
 :=\big\{f:\Fin(\Z)\to\bigcup\nolimits_{\Lm\in\Fin(\Z)}\fA_\Lm \,\big|\, \mbox{$f_\Lm\in\fA_\Lm$ for all $\Lm\in\Fin(\Z)$}\big\},
 \ea
is a unital \str algebra with respect to the pointwise operations specified after \eqref{defF}, where we set $f_\Lm:=f(\Lm)$ for all $f\in\mF$ and all $\Lm\in\Fin(\Z)$. Next, we check that \eqref{defG}, given by
\ba
\label{Def-mG}
\mG
:=\{f\in\mF\,|\, &\mbox{there exists $\Lm\in\Fin(\Z)$ such that}\nonumber\\
&\mbox{$f_{\Lm'}=\vi_{\Lm',\Lm}(f_\Lm)$ for all $\Lm'\in\Fin(\Z)$ with $\Lm\subseteq\Lm'$}\},
\ea
is a unital \str subalgebra of $\mF$. To this end, let $f,g\in\mG$ and let $\Lm, \Gm\in\Fin(\Z)$ be such that $f_{\Lm'}=\vi_{\Lm',\Lm}(f_\Lm)$ for all $\Lm'\in\Fin(\Z)$ with $\Lm\subseteq\Lm'$ and $g_{\Lm'}=\vi_{\Lm',\Gm}(g_\Gm)$ for all $\Lm'\in\Fin(\Z)$ with $\Gm\subseteq\Lm'$. Since $\Fin(\Z)$ is equipped with a direction (see after \eqref{FinZ}), there exists $\Delta\in\Fin(\Z)$ such that $\Lm\subseteq\Delta$ and $\Gm\subseteq\Delta$. Hence, for all $\Lm'\in\Fin(\Z)$ with $\Delta\subseteq\Lm'$, the cocycle condition from \ref{indlim:emb} yields, for the addition on $\mF$, 
\ba
\label{GAdd}
(f+g)_{\Lm'}
&=f_{\Lm'}+g_{\Lm'}\nonumber\\
&=\vi_{\Lm',\Lm}(f_\Lm)+\vi_{\Lm',\Gm}(g_\Gm)\nonumber\\
&=(\vi_{\Lm',\Delta}\circ\vi_{\Delta,\Lm})(f_\Lm)+(\vi_{\Lm',\Delta}\circ\vi_{\Delta,\Gm})(g_\Gm)\nonumber\\
&=\vi_{\Lm',\Delta}(f_\Delta)+\vi_{\Lm',\Delta}(g_\Delta)\nonumber\\
&=\vi_{\Lm',\Delta}(f_\Delta+g_\Delta)\nonumber\\
&=\vi_{\Lm',\Delta}((f+g)_\Delta),
\ea
and, for the multiplication on $\mF$, we analogously get
$(fg)_{\Lm'}
=f_{\Lm'}g_{\Lm'}
=\vi_{\Lm',\Lm}(f_\Lm)\vi_{\Lm',\Gm}(g_\Gm)
=\vi_{\Lm',\Delta}(f_\Delta)\vi_{\Lm',\Delta}(g_\Delta)
=\vi_{\Lm',\Delta}((fg)_\Delta)$.
Moreover, $\mG$ is also invariant under the scalar multiplication on $\mF$ since
$(\lm f)_{\Lm'}
=\lm f_{\Lm'}
=\lm\vi_{\Lm',\Lm}(f_\Lm)
=\vi_{\Lm',\Lm}(\lm f_\Lm)
=\vi_{\Lm',\Lm}((\lm f)_\Lm)$ 
for all $\lm\in\C$ and all $\Lm'\in\Fin(\Z)$ with $\Lm'\subseteq\Lm$. Finally, as for the involution on $\mF$, we get 
$(f^\ast)_{\Lm'}
=(f_{\Lm'})^\ast
=(\vi_{\Lm',\Lm}(f_\Lm))^\ast
=\vi_{\Lm',\Lm}((f_\Lm)^\ast)
=\vi_{\Lm',\Lm}((f^\ast)_\Lm)$ 
for all $\Lm'\in\Fin(\Z)$ with $\Lm'\subseteq\Lm$.

We next check that \eqref{defSN} defines a submultiplicative seminorm having the \Cs property. Recall that, if $f\in\mG$ and $\Lm\in\Fin(\Z)$ are such that $f_{\Lm'}=\vi_{\Lm',\Lm}(f_\Lm)$ for all $\Lm'\in\Fin(\Z)$ with $\Lm\subseteq\Lm'$, then, for all $\Lm'\in\Fin(\Z)$ with $\Lm\subseteq\Lm'$, we have, on the one hand, ${\|f_{\Lm'}\|}_{\Lm'}={\|\vi_{\Lm',\Lm}(f_\Lm)\|}_{\Lm'}$ and, on the other hand, ${\|\vi_{\Lm',\Lm}(f_\Lm)\|}_{\Lm'}={\|f_\Lm\|}_{\Lm}$ due to \eqref{piIso} (since we know from {\it (a)} that $\vi_{\Lm',\Lm}\in\sMon(\fA_\Lm,\fA_{\Lm'})$). Hence, the net $\Fin(\Z)\ni\Gm\mapsto{\|f_{\Gm}\|}_{\Gm}\in\R$ converges to the (unique) limit $\lim_{\Gm\in\Fin(\Z)} {\|f_{\Gm}\|}_{\Gm}={\|f_\Lm\|}_{\Lm}$ and defines the map $|\cdot|:\mG\to\R$ from \eqref{defSN} given, for the case at hand and all $f\in\mG$, by
\ba
\label{limnet}
|f|
:=\lim_{\Gm\in\Fin(\Z)} {\|f_{\Gm}\|}_{\Gm}.
\ea
Note that \eqref{limnet} is well-defined since, if  $\wt\Lm\in\Fin(\Z)$ is such that $f_{\Lm'}=\vi_{\Lm',\wt\Lm}(f_{\wt\Lm})$ for all $\Lm'\in\Fin(\Z)$ with $\wt\Lm\subseteq\Lm'$, there exists $\Delta\in\Fin(\Z)$ such that $\Lm\subseteq\Delta$ and $\wt\Lm\subseteq\Delta$ and, hence, ${\|f_\Delta\|}_{\Delta}={\|f_\Lm\|}_{\Lm}$ and ${\|f_\Delta\|}_{\Delta}={\|f_{\wt\Lm}\|}_{\wt\Lm}$. Moreover, since ${\n{\cdot}}_{\Lm}$ is a \Cs norm on $\fA_\Lm$ for all $\Lm\in\Fin(\Z)$, \eqref{limnet} inherits the nonnegativity, the absolute homogeneity, the subadditivity, the submultiplicativity, and the \Cs property. However, \eqref{limnet} is not postive-definite since, if $f\in\mG$ and if $\Lm\in\Fin(\Z)$ is such that $f_\Lm=0$ and $f_{\Lm'}=\vi_{\Lm',\Lm}(f_{\Lm})$ for all $\Lm'\in\Fin(\Z)$ with $\Lm\subseteq\Lm'$, we have $|f|=0$ but $f_\Gm\in\fA_\Gm$ for $\Lm\nsubseteq\Gm$ is not necessarily equal to $0_\Gm$ (for example, set $\Lm:=\{0\}$ and define $f\in\mG$ by $f_\Gm:=0_\Gm$ if $0\in\Gm$ and $f_\Gm:=1_\Gm$ if $0\notin\Gm$).

With the help of \eqref{limnet}, we define $\mG_0:=\{f\in\mG\,|\, |f|=0\}$ as in \eqref{defG0} and we next check that $\mG_0$ is a 2-sided \str ideal of $\mG$. Since $\mG$ is a \str subalgebra of $\mF$, we have $f^\ast\in\mG$ for all $f\in\mG$ and, since \eqref{limnet} satisfies the \Cs property, we know that $|f^\ast|=|f|$ for all $f\in\mG$. Moreover, since \eqref{limnet} is absolutely homogeneous, subadditive, and submultiplicative, $\mG_0$ is a vector subspace of $\mG$ and satisfies the 2-sidedness property, too. As after \eqref{defG0}, we define the quotient space $\mG/\mG_0$ whose elements are the equivalence classes $[f]:=\{f'\in\mG\,|\, f'-f\in \mG_0\}$ for all $f\in\mG$. Since \eqref{limnet} is submultiplicative and has the \Cs property, the multiplication and involution as given after \eqref{defG0} are well-defined and $\mG/\mG_0$ becomes a unital \str algebra.

Next, we want to show that, for all  $f\in\mG$, the seminorm \eqref{limnet} satisfies $|f'|=|f|$ for all $f'\in [f]$. To this end, let $f\in\mG$ and $j\in\mG_0$ such that $f'=f+j$, and let $\Lm, \Gm\in\Fin(\Z)$ be such that $f_{\Lm'}=\vi_{\Lm',\Lm}(f_\Lm)$ for all $\Lm'\in\Fin(\Z)$ with $\Lm\subseteq\Lm'$ and $j_{\Lm'}=\vi_{\Lm',\Gm}(j_\Gm)$ for all $\Lm'\in\Fin(\Z)$ with $\Gm\subseteq\Lm'$. Hence, we have $|f|={\|f_\Lm\|}_\Lm={\|f_{\Lm'}\|}_{\Lm'}$ for all $\Lm'\in\Fin(\Z)$ with $\Lm\subseteq\Lm'$ and $|j|=\|j_\Gm\|_\Gm=\|j_{\Lm'}\|_{\Lm'}$ for all $\Lm'\in\Fin(\Z)$ with $\Gm\subseteq\Lm'$ and, since $j\in\mG_0$, the latter implies $j_{\Lm'}=0_{\Lm'}$ for all $\Lm'\in\Fin(\Z)$ with $\Gm\subseteq\Lm'$. Moreover, there exists $\Delta\in\Fin(\Z)$ such that $\Lm\subseteq\Delta$ and $\Gm\subseteq\Delta$ and \eqref{GAdd} yields $(f+j)_{\Lm'}=\vi_{\Lm',\Delta}((f+j)_\Delta)$ for all $\Lm'\in\Fin(\Z)$ with $\Delta\subseteq\Lm'$.  Hence, we arrive at $|f'|=|f+j|={\|(f+j)_\Delta\|}_{\Delta}={\|f_\Delta+j_\Delta\|}_{\Delta}={\|f_\Delta\|}_{\Delta}={\|f_\Lm\|}_\Lm=|f|$ as desired. Therefore, the map $\n{\cdot}:\mG/\mG_0\to\R$ defined, for all $f\in\mG$, by
\ba
\label{norm-FJ}
\|[f]\|
:=|f|,
\ea
is well-defined. Moreover, it is not only nonnegative, absolutely homogeneous, subadditive,  submultiplicative, and has the \Cs property but it is also positive definite by construction.

Since the \str algebra $\mG/\mG_0$ is equipped with a submultiplicative norm which has  the \Cs property, Lemma \ref{lem:CsCmpl} implies that its completion \eqref{V'}, denoted by
 \ba
 \label{fA}
 \fA
 :=(\mG/\mG_0)',
 \ea
 is a unital \Cs algebra (with respect to the operator norm on $(\mG/\mG_0)^{\ast\ast}$, see Appendix \ref{app:Cstar}). This concludes the construction of the inductive limit of the net of unital \Cs algebras $(\fA_\Lm)_{\Lm\in\Fin(\Z)}$ with respect to the family of isotonies $\{\vi_{\Lm',\Lm}\}_{\Lm,  \Lm'\in\Fin(\Z), \hspace{0.3mm}\Lm\subseteq\Lm'}$.

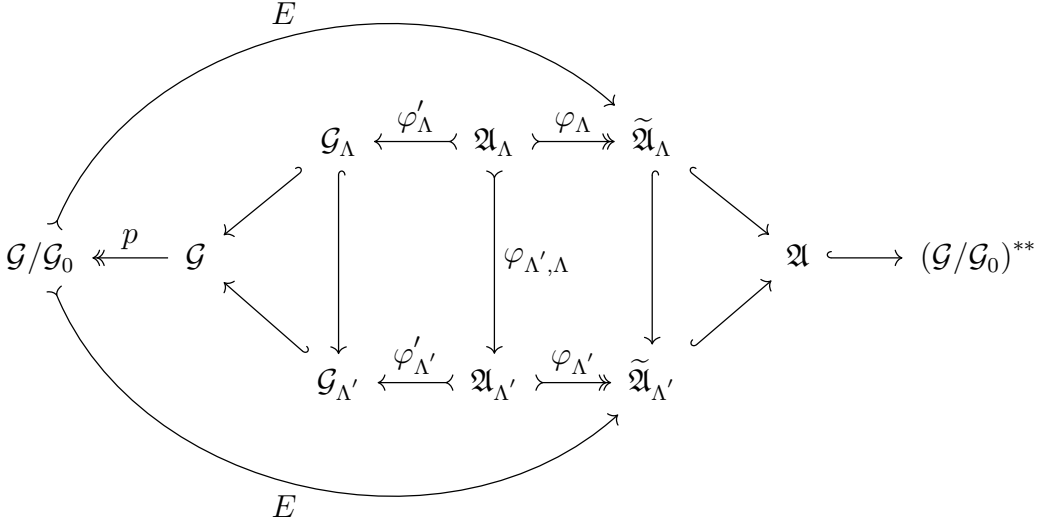
\begin{figure}
\begin{center}
\begin{tikzcd}
	&
	&\mG_\Lm
	\arrow[dd,  hookrightarrow]
	& \fA_\Lm
	\arrow[l, tail, "\textstyle\vi'_\Lm"']
	\arrow[r,   tail, two heads, "\textstyle\vi_\Lm"]
	\arrow[dd, tail, "\textstyle\vi_{\Lm',\Lm}"]
	&\tA_\Lm
	\arrow[rd,  hookrightarrow]
	\arrow[dd,  hookrightarrow]
	&
	&\\
\mG/\mG_0
\arrow[rrrru, tail, bend left=55, "\textstyle E"]
\arrow[rrrrd, tail, bend right=55, "\textstyle E"']
	&\mG
	\arrow[rd,  hookleftarrow]
	\arrow[ru,  hookleftarrow]
	\arrow[l,  two heads, "\textstyle p"']
	&
	&
	&
	&\fA
	\arrow[r,  hookrightarrow]
	&(\mG/\mG_0)^{\ast\ast}\\
	&
	&\mG_{\Lm'}
	& \fA_{\Lm'}
	\arrow[l, tail, "\textstyle\vi'_{\Lm'}"']
	\arrow[r,   tail, two heads, "\textstyle\vi_{\Lm'}"]
	&\tA_{\Lm'}
	\arrow[ru,  hookrightarrow]
	&
	&
\end{tikzcd}
\end{center}
\caption{The ingredients of (the proof of) Proposition \ref{prop:indlim}.} 
\label{fig:indlim}
\end{figure}

Finally, we have to construct the net of unital \Cs subalgebras $(\wt\fA_\Lm)_{\Lm\in\Fin(\Z)}$ of $\fA$ and the family of \str isomorphisms $\{\vi_\Lm\in\sIso(\fA_\Lm, \wt\fA_\Lm)\}_{\Lm\in\Fin(\Z)}$ which lead to \eqref{uhf}. In order to do so, we define, for all $\Lm\in\Fin(\Z)$, 
\ba
\label{Def-mGLm}
\mG_\Lm
:=\{f\in\mG\,|\, \mbox{$f_{\Lm'}=\vi_{\Lm',\Lm}(f_\Lm)$ for all $\Lm'\in\Fin(\Z)$ with $\Lm\subseteq\Lm'$}\},
\ea
and, as after \eqref{Def-mG},  we see that \eqref{Def-mGLm} is a unital \str subalgebra of $\mG$. Moreover,  due to the cocycle condition from \ref{indlim:emb}, we have $\mG_\Lm\subseteq\mG_{\Lm'}$ for all $\Lm, \Lm'\in\Fin(\Z)$ with $\Lm\subseteq\Lm'$. Next, for all $\Lm\in\Fin(\Z)$, we define the map $\vi_\Lm:\fA_\Lm\to\fA$, for all $A\in\fA_\Lm$, by
\ba
\label{DefviLm}
\vi_\Lm(A)
:=E([\vi'_\Lm(A)]),
\ea
where the isometry $E\in\mL(\mG/\mG_0, (\mG/\mG_0)^{\ast\ast})$ stems from \eqref{V'} and where, for all $\Lm\in\Fin(\Z)$, the map $\vi'_\Lm:\fA_\Lm\to\mG_\Lm$ is defined, for all $A\in\fA_\Lm$ and all $\Lm'\in\Fin(\Z)$, by
\ba
\label{DefviLm'}
(\vi'_\Lm(A))_{\Lm'}
:=\begin{cases}
\vi_{\Lm',\Lm}(A), & \Lm\subseteq\Lm',\\
\hfill 0_{\Lm'}, &  \Lm\nsubseteq\Lm'.
\end{cases}
\ea
Note that \eqref{DefviLm'} is well-defined because \eqref{viLL} yields $(\vi'_\Lm(A))_{\Lm'}=\vi_{\Lm',\Lm}(A)=\vi_{\Lm',\Lm}(\vi_{\Lm,\Lm}(A))=\vi_{\Lm',\Lm}((\vi'_\Lm(A))_{\Lm})$ for all  $\Lm,\Lm'\in\Fin(\Z)$ with $\Lm\subseteq\Lm'$ and all $A\in\fA_\Lm$. Moreover, $\vi_\Lm\in\sHom(\fA_\Lm,\fA)$ for all $\Lm\in\Fin(\Z)$ because $\vi_{\Lm',\Lm}\in\sHom(\fA_\Lm,\fA_{\Lm'})$ for all $\Lm, \Lm'\in\Fin(\Z)$ with $\Lm\subseteq\Lm'$ due to {\it (a)} and because $E\in\sHom(\mG/\mG_0,\fA)$ due to Lemma \ref{lem:CsCmpl}. Hence, setting, for all $\Lm\in\Fin(\Z)$, 
\ba
\label{Def-Atilde}
\wt\fA_\Lm
:=\ran(\vi_\Lm),
\ea
we know that \eqref{Def-Atilde} is a \Cs subalgebra of $\fA$ for all $\Lm\in\Fin(\Z)$. Since we actually have $\vi_{\Lm',\Lm}\in\sMon(\fA_\Lm,\fA_{\Lm'})$ for all $\Lm, \Lm'\in\Fin(\Z)$ with $\Lm\subseteq\Lm'$ due to {\it (a)} and $E\in\sMon(\mG/\mG_0,\fA)$ due to Lemma \ref{lem:CsCmpl}, we get, using \eqref{limnet}, that $\vi_\Lm\in\sIso(\fA_\Lm,\wt\fA_\Lm)$ for all $\Lm\in\Fin(\Z)$, see Figure \ref{fig:indlim}.

The family $\{\vi_\Lm\}_{\Lm\in\Fin(\Z)}$ also satisfies $\vi_{\Lm'}\circ\vi_{\Lm',\Lm}=\vi_\Lm$ for all $\Lm, \Lm'\in\Fin(\Z)$ with $\Lm\subseteq\Lm'$. Indeed, for all $\Lm, \Lm', \Lm''\in\Fin(\Z)$ with $\Lm\subseteq\Lm'\subseteq\Lm''$ and all $A\in\fA_\Lm$, we have, on the one hand, $(\vi'_{\Lm'}(\vi_{\Lm',\Lm}(A)))_{\Lm''}=\vi_{\Lm'',\Lm'}(\vi_{\Lm',\Lm}(A))=\vi_{\Lm'',\Lm}(A)$ and, on the other hand, $(\vi'_{\Lm}(A))_{\Lm''}=\vi_{\Lm'',\Lm}(A)$. Hence, since $\vi'_{\Lm'}(\vi_{\Lm',\Lm}(A))\in\mG_{\Lm'}\subseteq\mG_{\Lm''}$ and $\vi'_{\Lm}(A)\in\mG_{\Lm}\subseteq\mG_{\Lm''}$ for all $\Lm, \Lm',\Lm''\in\Fin(\Z)$ with $\Lm\subseteq\Lm'\subseteq\Lm''$, we have $|\vi'_{\Lm'}(\vi_{\Lm',\Lm}(A))-\vi'_{\Lm}(A)|=0$ for all $\Lm, \Lm'\in\Fin(\Z)$ with $\Lm\subseteq\Lm'$ and all $A\in\fA_\Lm$. Using \eqref{norm-FJ} and the fact that $E$ is an isometry,  we get $(\vi_{\Lm'}\circ\vi_{\Lm',\Lm})(A)-\vi_\Lm(A)=E([\vi'_{\Lm'}(\vi_{\Lm',\Lm}(A))-\vi'_{\Lm}(A)])=0$ for all $\Lm, \Lm'\in\Fin(\Z)$ with $\Lm\subseteq\Lm'$ and all $A\in\fA_\Lm$ as required. Furthermore, the property $\vi_{\Lm'}\circ\vi_{\Lm',\Lm}=\vi_\Lm$ for all $\Lm, \Lm'\in\Fin(\Z)$ with $\Lm\subseteq\Lm'$ implies that $(\wt\fA_\Lm)_{\Lm\in\Fin(\Z)}$ is an increasing net.

As for \eqref{uhf}, we first note that $\bigcup_{\Lm\in\Fin(\Z)}\wt\fA_\Lm$ is a \str subalgebra of $\fA$ because, if $A,B\in\bigcup_{\Lm\in\Fin(\Z)}\wt\fA_\Lm$, there exist $\Lm, \Lm'\in\Fin(\Z)$ such that $A\in\tA_\Lm$ and $B\in\tA_{\Lm'}$ and, hence, there exists $\Lm''\in\Fin(\Z)$ with $\Lm,\Lm'\subseteq\Lm''$ and $A, B\in\tA_{\Lm''}$ due to the fact that $(\wt\fA_\Lm)_{\Lm\in\Fin(\Z)}$ is an increasing net. Next, since 
$\clo_\fA(\ran(E))=\fA$, it is enough to show that $\ran(E)\subseteq\bigcup_{\Lm\in\Fin(\Z)}\wt\fA_\Lm$. To this end, let $[f]\in\mG/\mG_0$, where $f\in\mG$ and $\Lm\in\Fin(\Z)$ are such that $f_{\Lm'}=\vi_{\Lm',\Lm}(f_\Lm)$ for all  $\Lm'\in\Fin(\Z)$ with $\Lm\subseteq\Lm'$. Hence, since $(\vi'_\Lm(f_\Lm))_{\Lm'}=\vi_{\Lm',\Lm}(f_\Lm)=f_{\Lm'}$ for all $\Lm'\in\Fin(\Z)$ with $\Lm\subseteq\Lm'$, we again have $E([f])-\vi_\Lm(f_\Lm)=E([f-\vi'_\Lm(f_\Lm)])=0$, \ie, for all $A\in\mG/\mG_0$, there exists $\Lm\in\Fin(\Z)$ such that $E(A)\in\ran(\vi_\Lm)=\wt\fA_\Lm$.
\eprf

\section{Crossed product extension}
\label{sec:crossed}

In this section, we construct the so-called crossed product of a general \Cs algebra $\fA$ by the finite cyclic group $\Z_2$ (in Section \ref{sec:JW}, $\fA$ will play the role of the infinite  tensor product). To this end, we first briefly recall the data needed for the construction of a general crossed product (see, for example,  \cite{Williams, Lance}). Subsequently, we specialize to the concrete case at hand.

In the following, for any groups $G$ and $H$, we denote by $\Hom(G,H)$ the set of group homomorphisms between $G$ and $H$ (whereas $\sHom(\fA,\fB)$ stands for the set of \str homomorphisms between the \str algebras $\fA$ and $\fB$, see Appendix \ref{app:Cstar}). Moreover, if $G$ is any locally compact group and $\mV$ any topological vector space, we denote by $C_0(G,\mV)$ the complex vector space  (with respect to the usual pointwise addition and scalar multiplication) of $\mV$-valued continuous functions on $G$ with compact support. Finally, if  $\mH$ is any complex Hilbert space, $\mU(\mH)$ stands for the group of all unitary operators on $\mH$ (with respect to the composition of linear operators).

We start off with a so-called \Cs dynamical system, \ie, with a triple 
\ba
\label{CstarDyn}
(\fA, G, \alpha), 
\ea
where $\fA$ is a \Cs algebra, $G$ a locally compact group, and $\alpha\in\Hom(G,\sAut(\fA))$ a group homomorphism (between $G$ and the group $\sAut(\fA)$ of \str automorphisms of $\fA$, see Appendix \ref{app:Cstar}) which is strongly continuous, \ie, which has the property that, for all $A\in\fA$, the map $G\ni g\mapsto\alpha_g(A)\in\fA$ is continuous. Next, we make the complex vector space $C_0(G,\fA)$ into a \str algebra by equipping it with the multiplication and the involution defined, for all $f,g\in C_0(G,\fA)$ and all $s\in G$, by 
\ba
\label{C0-mult}
(fg)(s)
&:=\int_G\rd\mu_\fA(r) f(r)\alpha_r(g(r^{-1}s)),\\
\label{C0-invo}
f^\ast(s)
&:=\Delta(s^{-1})\alpha_s(f(s^{-1})^\ast),
\ea
where, in \eqref{C0-mult}, we use the fact that there exists a unique linear map $C_0(G,\fA)\to\fA$, written as $f\mapsto \int_G\rd\mu_\fA(s) f(s)$, such that $\eta(\int_G\rd\mu_\fA(s) f(s))=\int_G\rd\mu(s) \eta(f(s))$ for all $f\in C_0(G,\fA)$ and all $\eta\in\fA^\ast$, and $\mu:\mB_G\to\overline{\R}$  stands for the usual left-invariant Haar measure on $G$ (with $\mB_G$ the Borel $\sigma$-algebra generated by the topology on $G$ and $\overline{\R}$ the extended real line). In \eqref{C0-invo}, $\Delta\in\Hom(G,\R^+)$ denotes the usual modular function of $G$ (and $\R^+$ the multiplicative group of strictly positive real numbers).

Moreover, a covariant representation of the \Cs dynamical system $(\fA, G, \alpha)$ is defined to be a triple $(\mH, \pi, U)$, where $\mH$ is a complex Hilbert space, $\pi\in\sHom(\fA,\mL(\mH))$ a representation of $\fA$ on $\mH$, and $U\in\Hom(G,\mU(\mH))$ a strongly continuous unitary representation of $G$ on $\mH$ (\ie, as above, for all $\psi\in\mH$, the map $G\ni s\mapsto U_s\psi\in\mH$ is continuous) such that, for all $s\in G$ and all $A\in\fA$, 
\ba
\pi(\alpha_s(A))
=U_s\pi(A)U_s^\ast.
\ea
The collection of all covariant representations of the \Cs dynamical system $(\fA, G, \alpha)$ is denoted by $\Cov(\fA, G, \alpha)$. If $(\mH, \pi, U)\in\Cov(\fA, G, \alpha)$, the map $\pi\rtimes U\in\sHom(C_0(G,\fA),\mL(\mH))$ defined, for all $f\in C_0(G,\fA)$, by
\ba
\label{intf}
(\pi\rtimes U)(f)
:=\int_G\rd\mu_\mL(s)\, \pi(f(s))U_s,
\ea
is called the integrated form of $(\mH, \pi, U)$. Here, we used that there exists a unique linear map $C_0(G,\mL_s(\mH))\to\mL(\mH)$, denoted by $f\mapsto \int_G\rd\mu_\mL(s)\, f(s)$, such that $(\bar\rho( \int_G\rd\mu_\mL(s)\, f(s))\vi,\psi)=\int_G\rd\mu(s)\, (\bar\rho(f(s))\vi, \psi)$ for all nondegenerate  representations $(\mK, \rho)$ of $\mL^\infty(\mH)$ (where $\mK$ is a complex Hilbert space with scalar product $(\cdot,\cdot)$, $\rho\in\sHom(\mL^\infty(\mH), \mL(\mK))$, and $\mL^\infty(\mH)$ stands for the \Cs algebra of all compact operators on $\mH$), all $f\in C_0(G,\mL_s(\mH))$, and all $\vi,\psi\in\mK$ (and that the map $G\ni s\mapsto \pi(f(s))U_s\in\mL(\mH)$ in \eqref{intf} is an element of $C_0(G,\mL_s(\mH))$ due to the strong continuity of $U$). Moreover, we denote by $\bar\rho\in\sHom(\mL(\mH),\mL(\mK))$ the usual canonical extension to $\mL(\mH)$ of the representation $\rho$ of the two-sided closed ideal $\mL^\infty(\mH)$ of $\mL(\mH)$, and $\mL_s(\mH)$ stands for $\mL(\mH)$ equipped with the strict topology (\ie, the topology generated by the family of seminorms $\{\|\cdot K\|+\|K\cdot\|\,|\, K\in\mL^\infty(\mH)\}$ on $\mL(\mH)$, where $\|\cdot\|$ is the usual operator norm on $\mL(\mH)$, see Appendix \ref{app:Cstar}).

Finally, we equip the \str algebra $C_0(G,\fA)$ with the so-called universal norm defined, for all $f\in  C_0(G,\fA)$, by
\ba
\label{univ}
\|f\|
:=\hspace{-2mm}\sup_{\substack{(\mH,\pi, U)\in\\\Cov(\fA,G,\alpha)}} \hspace{-1mm}\|(\pi\rtimes U)(f)\|,
\ea
and we know that the supremum acts on a (bounded) subset of $\R$ due to the axiom of comprehension (or separation) of Zermelo-Fraenkel set theory (and that the supremum taken over the nondegenerate covariant representations leaves \eqref{univ} unchanged). Since we also know that \eqref{univ} is a submultiplicative norm which has the \Cs property, the \Cs completion of $C_0(G,\fA)$ with respect to \eqref{univ} (see Lemma \ref{lem:CsCmpl}) is a \Cs algebra which is called the crossed product of the \Cs dynamical system $(\fA, G, \alpha)$ and which we denote by $\fA\rtimes_\alpha G$ (as in \cite{Williams}).

We next apply the foregoing construction to the special case of a \Cs dynamical system $(\fA, G, \alpha)$, where $\fA$ is any \Cs algebra, 
\ba
\label{defZ2}
G
:=\Z_2,
\ea
 and $\alpha\in\Hom(\Z_2,\sAut(\fA))$ is any group homomorphism. Here, $\Z_2:=\{-1,1\}$ is the finite cyclic group of order $2$ whose multiplicative group law is the usual multiplication of real numbers and which is equipped with the discrete topology making it into a (locally) compact topological group. Of course, $\alpha_1(A)=A$ for all $A\in\fA$ but $\alpha_{-1}$ may be nontrivial. Moreover, $\alpha$ is automatically strongly continuous (as specified after \eqref{CstarDyn}) since $C(\Z_2,\fA)=\fA^{\Z_2}$, where $\fA^{\Z_2}$ is the $\Z_2$th power of $\fA$, \ie, the set of all functions from $\Z_2$ to $\fA$. Finally, we also have $C_0(\Z_2,\fA)=\fA^{\Z_2}$, of course. In order to make the complex vector space $\fA^{\Z_2}$ into a \str algebra by means of \eqref{C0-mult} and \eqref{C0-invo}, we use the fact that the Haar functional corresponding to the (normalized) Haar measure on $\Z_2$ is given, for all $\vi\in\C^{\Z_2}$ (again, the set of all functions from $\Z_2$ to $\C$), by
\ba
\label{HaarZ2}
\int_{\Z_2}\rd\mu(s) \vi(s)
=\frac12\sum_{s\in\Z_2} \vi(s).
\ea
Hence, since we know from above that there exists a unique linear map $\fA^{\Z_2}\to\fA$ satisfying the property introduced after \eqref{C0-invo}, \eqref{HaarZ2} implies that $\int_{\Z_2}\rd\mu_\fA(s) f(s)=\sum_{s\in\Z_2} f(s)/2$ for all $f\in\fA^{\Z_2}$ and \eqref{C0-mult}  therefore becomes, for all $f,g\in\fA^{\Z_2}$ and all $s\in\Z_2$, 
\ba
\label{Z2-mult}
(fg)(s)
&=\frac12\sum_{r\in\Z_2} f(r) \alpha_r(g(r^{-1}s))\nonumber\\
&=\frac12 \begin{cases}
f(1)g(1)+f(-1)\alpha_{-1}(g(-1)), & s=1,\\
f(1)g(-1)+f(-1)\alpha_{-1}(g(1)), & s=-1.
\end{cases}
\ea
Moreover, since $\Z_2$ is compact (or directly from \eqref{HaarZ2}), $\Z_2$ is unimodular, \ie, $\Delta(s)=1$ for all $s\in\Z_2$, and \eqref{C0-invo} becomes, for all $f\in\fA^{\Z_2}$ and all $s\in\Z_2$, 
\ba
\label{Z2-invo}
f^\ast(s)
&=\alpha_s(f(s^{-1})^\ast)\nonumber\\
&= \begin{cases}
\hfill f(1)^\ast, & s=1,\\
\alpha_{-1}(f(-1)^\ast), & s=-1.
\end{cases}
\ea
Hence, the complex vector space $\fA^{\Z_2}$ equipped with the multiplication \eqref{Z2-mult} and the involution \eqref{Z2-invo} becomes a \str algebra. Furthermore, since, for any Hilbert space $\mH$, we have $C_0(\Z_2,\mL_s(\mH))=\mL(\mH)^{\Z_2}$ and since we know from above that there exists a unique linear map $\mL(\mH)^{\Z_2}\to\mL(\mH)$ satisfying the property introduced after \eqref{intf}, \eqref{HaarZ2} implies that $\int_{\Z_2}\rd\mu_\mL(s)\, f(s)=\sum_{s\in\Z_2} f(s)/2$ for all $f\in\mL(\mH)^{\Z_2}$. Therefore, for all $(\mH,\pi,U)\in\Cov(\fA,\Z_2,\alpha)$ and all $f\in\fA^{\Z_2}$, \eqref{intf} becomes
\ba
\label{intf-Z2}
(\pi\rtimes U)(f)
=\frac12\sum_{s\in\Z_2} \pi(f(s)) U_s.
\ea

\vspace {2mm}
For the following, recall (see, for example, \cite{Lance}) that, for a given \Cs algebra $\fA$, a complex  vector space $\mV$ is called a scalar product $\fA$-module if, in addition, it is a right $\fA$-module with respect to the vector space addition $\mV\times\mV\to\mV$ and a so-called $\fA$-scalar right multiplication $\mV\times\fA\to\mV$ defined, for all $v,w\in\mV$ and all $A,B\in\fA$, by $v(A+B)=(vA)+(vB)$, $v(AB)=(vA)B$, and $(v+w)A=(vA)+(wA)$ (and $v1=v$ if $\fA$ is a unital \Cs algebra with identity $1$), if the scalar multiplication $\C\times\mV\to\mV$ is compatible with the $\fA$-scalar right multiplication in the sense that $\lambda (vA)=(\lambda v)A=v(\lambda A)$ for all $\lambda\in\C$, all $v\in\mV$, and all $A\in\fA$, and if it is equipped with a so-called $\fA$-scalar product $\l\cdot,\cdot\r:\mV\times\mV\to\fA$ defined, for all $u, v, w\in\mV$, all $\lambda,\mu\in\C$, and all $A\in\fA$, by $\l v,v\r\ge 0$ (\ie, $\l v,v\r\in\fA_+$, where $\fA_+$ denotes the convex cone of all positive elements of $\fA$), $\l v,v\r=0$ if and only if $v=0$, $\l v,\lambda w+\mu u\r=\lambda \l v,w\r+\mu\l v, u\r$ (the usual bilinearity), $\l v,w\r=\l w,v\r^\ast$, and
\ba
\l v,wA\r
=\l v,w\r A.
\ea
Finally, if, in addition, $\mV$ is equipped with the norm $\|\cdot\|:\mV\to\R$ defined, for all $v\in\mV$, by
\ba
\label{NormHM}
\|v\|
:=\sqrt{\|\l v,v\r\|},
\ea
where, on the right hand side of \eqref{NormHM}, $\|\cdot\|$ stands for the \Cs norm of $\fA$, and if $\mV$ is complete with respect to \eqref{NormHM}, then $\mV$ is called a Hilbert \Cs module over $\fA$.

For the following, let $\fA^2$ and $\fA^{2\times 2}$ stand for the set of $2$-vectors over $\fA$ and the set of $2\times 2$-matrices over $\fA$ whose elements are written as $v=[v_i]_{i\in\num{1}{2}}\in\fA^2$ and $X=[X_{ij}]_{i,j\in\num{1}{2}}\in\fA^{2\times 2}$, respectively  (using the same notation as for scalar entries from the beginning of Section \ref{sec:Local}). Moreover, for all $X\in\fA^{2\times 2}$, we define the map $t_X:\fA^2\to\fA^2$,  for all $v\in\fA^2$ and all $i\in\num{1}{2}$, by
\ba
\label{tX}
(t_Xv)_i
:=\sum_{j\in\num{1}{2}}X_{ij}v_j.
\ea

In order to construct the crossed product $\fA\rtimes_\alpha\Z_2$, we make use of the following.

\bl[Vectors and matrices over $\fA$]
\label{lem:Amat}
Let $\fA$ be any \Cs algebra. Then:
\bn[label=(\alph*), ref={\it (\alph*)}]
\setlength{\itemsep}{0mm}

\item
\label{Amod}
Equipped with the addition $\fA^2\times\fA^2\to\fA^2$, the scalar multiplication $\C\times\fA^2\to\fA^2$, the $\fA$-scalar right multiplication $\fA^2\times\fA\to\fA^2$, the $\fA$-scalar product $\l\cdot,\cdot\r:\fA^2\times\fA^2\to\fA$, and the norm $\n{\cdot}:\fA^2\to\R$ defined, for all $v,w\in\fA^2$, all $\lambda\in\C$, all $A\in\fA$, and all $i\in\num{1}{2}$, by $(v+w)_i:=v_i+w_i$, $(\lambda v)_i:=\lambda v_i$, 
\ba
\label{A2smult}
(vA)_i
&:=v_iA,\\
\label{A2sprod}
\l v,w\r
&:=\sum_{i\in\num{1}{2}} v_i^\ast w_i,\\
\label{A2norm}
\|v\|
&:=\sqrt{\|\l v,v\r\|},
\ea
respectively, $\fA^2$ becomes a Hilbert \Cs module over $\fA$.

\item 
\label{Amat-str}
Equipped with the addition $\fA^{2\times 2}\times\fA^{2\times 2}\to\fA^{2\times 2}$, the scalar multiplication $\C\times\fA^{2\times 2}\to\fA^{2\times 2}$, the multiplication $\fA^{2\times 2}\times\fA^{2\times 2}\to\fA^{2\times 2}$, and the involution $\fA^{2\times 2}\to\fA^{2\times 2}$ defined, for all $X, Y\in\fA^{2\times 2}$, all $\lambda\in\C$, and all $i,j\in\num{1}{2}$, by $(X+Y)_{ij}:=X_{ij}+Y_{ij}$, $(\lambda X)_{ij}:=\lambda X_{ij}$, $(XY)_{ij}:=\sum_{k\in\num{1}{2}}X_{ik}Y_{kj}$, and $(X^\ast)_{ij}:=X_{ji}^\ast$, respectively, $\fA^{2\times 2}$ becomes a \str algebra. Moreover, the norm $\n{\cdot}:\fA^{2\times 2}\to\R$ defined, for all $X\in\fA^{2\times 2}$, by 
\ba
\label{NormA22}
\|X\|
:=\sup_{\substack{v\in\fA^2 \\ \|v\|=1}} \|t_Xv\|,
\ea
is submultiplicative, has the \Cs property, and makes $\fA^{2\times 2}$ into a \Cs algebra.
\en
\el

\br
\label{rem:adjointable}
For a fixed $X\in\fA^{2\times 2}$, the map $t_X$ from \eqref{tX} is a so-called adjointable map from the $\fA$-module $\fA^2$ to itself (see, for example, \cite{Lance}), \ie, a map for which there exists a map $t_X^\ast:\fA^2\to\fA^2$ such that $\l t_Xv, w\r=\l v, t_X^\ast w\r$ for all $v, w\in\fA^2$ (here, $t_X^\ast=t_{X^\ast}$). We know that such a map is $\fA$-linear, \ie, $t_X$ is linear and, in addition, it satisfies $t_X(vA)=(t_X(v))A$ for all $v\in\fA^2$ and all $A\in\fA$. Moreover, $t_X$ is bounded (see Appendix \ref{app:Cstar}) which implies that \eqref{NormA22} is well-defined and that, for all $v\in\fA^2$, 
\ba
\label{tXBnd}
\n{t_Xv}
\le 
\n{X}\n{v}.
\ea
\er

\vspace{5mm}

\bprf
\ref{Amod}\, 
A direct check yields that $\fA^2$ is a complex vector space with respect to the addition and the scalar multiplication given before \eqref{A2smult}. We also directly verify that the $\fA$-scalar right multiplication \eqref{A2smult} is compatible with the scalar multiplication on $\fA^2$ and that it makes $\fA^2$ into a right $\fA$-module. Moreover, the $\fA$-scalar product \eqref{A2sprod} also has all the required properties. In particular, we have that $\l v,v\r=\sum_{i\in\num{1}{2}}v_i^\ast v_i\in\fA_+$ since $\fA_+$ is a convex cone and that $\l v,v\r=0$ implies $v=0$ since we also know that $\fA_+$ is pointed and salient, \ie, $\fA_+\cap(-\fA_+)=\{0\}$. Hence, $\fA^2$ becomes a scalar product $\fA$-module equipped with the norm \eqref{A2norm}. Finally, if $(v_n)_{n\in\N}$ is a Cauchy sequence in $\fA^2$, the sequences $(v_{n,i})_{n\in\N}$ with $i\in\num{1}{2}$ are Cauchy sequences in $\fA$ because, for all $m, n\in\N$ and all $i\in\num{1}{2}$, 
\ba
\label{vCauchy}
\n{v_n-v_m}^2
&=\big\|\sum\nolimits_{j\in\num{1}{2}}(v_{n,j}-v_{m,j})^\ast (v_{n,j}-v_{m,j})\big\|\nonumber\\
&\ge \|v_{n, i}-v_{m, i}\|^2,
\ea
where we used that $\|A\|\ge \|B\|$ for all $A,B\in\fA_+$ with $A\ge B$. Hence, since $\fA$ is complete with respect to its \Cs norm, there exists $v\in\fA^2$ such that $\|v_{n,i}-v_i\|\to 0$ for $n\to\infty$ and all $i\in\num{1}{2}$. Since $\n{v_n-v}^2\le \sum_{i\in\num{1}{2}}\|v_{n,i}-v_i\|^2$ for all $n\in\N$, we find that $\fA^2$ is complete with respect to \eqref{A2norm}.

\ref{Amat-str}\, 
A direct check yields that $\fA^{2\times 2}$ is a \str algebra with respect to the four specified operations. Next, we know from Remark \ref{rem:adjointable} that \eqref{NormA22} is well-defined, and we also directly verify that \eqref{NormA22} defines a norm which is submultiplicative. Moreover, \eqref{NormA22} has the \Cs property since,  on the one hand, we have the upper bound $\n{X^\ast X}\le \n{X^\ast}\n{X}$ for all $X\in\fA^{2\times 2}$. On the other hand, using that 
$\n{v}=\sup_{w\in\fA^2\hspace{-0.3mm},\hspace{0.3mm}\|w\|=1}\|\l v,w\r\|$ for all $v\in\fA^2$ (an identity which holds in general Hilbert \Cs modules over $\fA$), we can write, for all $X\in\fA^{2\times 2}$ and all $v\in\fA^2$ satisfying $\n{v}=1$, 
\ba
\label{tX*X}
\n{t_{X^\ast X}v}
&=\sup_{\substack{w\in\fA^2\\ \|w\|=1}}\|\l t_{X^\ast X}v,w\r\|\nonumber\\
&\ge \n{\l t_{X^\ast X}v,v\r}\nonumber\\
&=\n{t_Xv}^2,
\ea
which yields the lower bound $\n{X}^2\le\n{X^\ast X}$ for all $X\in\fA^{2\times 2}$. 

It remains to be shown that \eqref{NormA22} is a \Cs norm on $\fA^{2\times 2}$,  \ie, that $\fA^{2\times 2}$ is complete with respect to \eqref{NormA22} (see Appendix \ref{app:Cstar}). To this end, let $(X_n)_{n\in\N}$ be a Cauchy sequence in $\fA^{2\times 2}$. Again using that  $\|A\|\ge \|B\|$ for all $A,B\in\fA_+$ with $A\ge B$ and that $\|A\|=\sup_{B\in\fA, \|B\|=1}\n{A^\ast B}=\sup_{B\in\fA, \|B\|=1}\n{AB}$ for all $A\in\fA$ (the first equality being a special case of the identity mentioned before \eqref{tX*X} since $\fA$ is a Hilbert \Cs module over itself with respect to the $\fA$-scalar product $\fA\times\fA\ni(A,B)\mapsto \l A,B\r:=A^\ast B\in\fA$, see also \eqref{A2sprod}), we get, as in \eqref{vCauchy}, that, for all $m, n\in\N$ and all $i, j\in\num{1}{2}$, 
\ba
\label{EstEntries}
\n{X_n-X_m}
&= \sup_{\substack{v\in\fA^2\\\n{v}=1}}\sqrt{\big\|\sum\nolimits_{k\in\num{1}{2}}(t_{X_n-X_m}v)_k^\ast  (t_{X_n-X_m}v)_k\big\|}\nonumber\\
&\ge  \sup_{\substack{v\in\fA^2\\\n{v}=1}}\n{(t_{X_n-X_m}v)_i}\nonumber\\
&= \sup_{\substack{v\in\fA^2\\\n{v}=1}}\big\|\sum\nolimits_{l\in\num{1}{2}}(X_{n,il}-X_{m,il})v_l\big\|\nonumber\\
&\ge \n{X_{n, ij}-X_{m, ij}}.
\ea
Therefore, since $\fA$ is complete with respect to its \Cs norm, there exists $X\in\fA^{2\times 2}$ such that $\n{X_{n,ij}-X_{ij}}\to 0$ for $n\to\infty$ and all $i, j\in\num{1}{2}$. Moreover, since the subadditivity of the norm in $\fA$ and the Cauchy-Schwarz inequality for the Euclidean scalar product in $\R^2$ yields $\n{X_n-X}^2\le 2\sum_{i,j\in\num{1}{2}}\n{X_{n,ij}-X_{ij}}^2$ for all $n\in\N$, we find that $\fA^{2\times 2}$ is complete with respect to \eqref{NormA22}.
\eprf

For the following, recall from Appendix \ref{app:Cstar} that $\fA\cong\fB$ for two  \str algebras $\fA$ and $\fB$ means that there exists $\Phi\in\sIso(\fA,\fB)$. 

The following \Cs subalgebras of the \Cs algebra $\fA^{2\times 2}$ are used to establish the identification of the crossed product $\fA\rtimes_\alpha\Z_2$ which we want to use in the sequel.

\bl[Extending $\fA$]
\label{lem:WidehatA}
Let $(\fA,\Z_2,\alpha)$ be a \Cs dynamical system. Then:
\bn[label=(\alph*), ref={\it (\alph*)}]
\setlength{\itemsep}{0mm}
\item
\label{AHat}
The set 
\ba
\widehat\fA
:=\left\{\begin{bmatrix}A & B\\ \alpha_{-1}(B) & \alpha_{-1}(A)\end{bmatrix}\in\fA^{2\times 2}\,\bigg|\, A,B\in\fA\right\}
\ea
is a \Cs subalgebra of $\fA^{2\times 2}$.

\item
\label{AHat-A}
$\widehat\fA$ is an extension of $\fA$ in the sense that $\fA\cong\widehat\fA_0$, where the \Cs subalgebra $\widehat\fA_0$ of $\widehat\fA$ is defined by
\ba
\widehat\fA_0
:=\left\{\begin{bmatrix}A & 0\\ 0 & \alpha_{-1}(A)\end{bmatrix}\in\fA^{2\times 2}\,\bigg|\, A\in\fA\right\}.
\ea
\en
\el

\bprf
\ref{AHat}\, 
Using that $\alpha\in\Hom(\Z_2,\sAut(\fA)$, we straightforwardly check that $\widehat\fA$ is a \str subalgebra of $\fA^{2\times 2}$ with respect to the operations given in Lemma \ref{lem:Amat} \ref{Amat-str}. Moreover, due to \eqref{EstEntries} and since $\alpha_{-1}$ is bounded (see \eqref{cnt}), we also find that $\widehat\fA$ is closed with respect to \eqref{NormA22}.

\ref{AHat-A}\,
A direct check again yields that $\widehat\fA_0$ is a \str subalgebra of $\widehat\fA$ and that $\widehat\fA_0$ is also closed with respect to \eqref{NormA22}, \ie, $\widehat\fA_0$ is a  \Cs subalgebra of $\widehat\fA$. Moreover, we define the map $\psi: \fA\to\widehat\fA_0$,  for all $A\in\fA$, by
\ba
\label{IsoPsi}
\psi(A)
:=\begin{bmatrix}
A & 0\\ 
0 & \alpha_{-1}(A)
\end{bmatrix}. 
\ea
Since $\alpha_{-1}\in\sAut(\fA)$, we have $\psi\in\sHom(\fA,\widehat\fA_0)$ and since $\psi$ is clearly injective and surjective, we get $\psi\in\sIso(\fA,\widehat\fA_0)$. 
\eprf

We can now establish the following concrete identification of the crossed product $\fA\rtimes_\alpha\Z_2$ which, to the best of my knowledge, has not been used in the context of Araki's extension of the Jordan-Wigner transformation in the literature so far (see Section \ref{sec:JW}).

\bp[Crossed product]
\label{prop:cross}
Let $(\fA,\Z_2,\alpha)$ be a \Cs dynamical system. Then:
\bn[label=(\alph*), ref={\it (\alph*)}]
\setlength{\itemsep}{0mm}

\item 
\label{cross-gen}
If  $\fB$ is a \Cs algebra such that $\fB\cong\fA^{\Z_2}$, then $\fA\rtimes_\alpha\Z_2\cong \fB$.

\item 
\label{cross-spec}
The \Cs algebra $\widehat\fA$ from Lemma \ref{lem:WidehatA} \ref{AHat} yields
\ba
\fA\rtimes_\alpha\Z_2
\cong \widehat\fA. 
\ea
\en
\ep

\bprf
\ref{cross-gen}
For the sake of completeness, we want to somewhat elaborate the proof sketched in \cite{Williams}. We start off by defining the map $\eN{\cdot}:\fA^{\Z_2}\to\R$ by $\eN{f}:=\int_{\Z_2}\rd\mu(s)\, \|f(s)\|$ for all $f\in\fA^{\Z_2}$. With the help of \eqref{HaarZ2}, we have, for all $f\in\fA^{\Z_2}$, 
\ba
\label{L1norm}
\eN{f}
=\frac12\sum_{s\in\Z_2} \|f(s)\|,
\ea
and a  direct check yields that \eqref{L1norm} defines a norm on $\fA^{\Z_2}$, called $L^1$-norm,  which is submultiplicative with respect to the multiplication \eqref{Z2-mult}  and with respect to which the involution \eqref{Z2-invo} is an isometry. Moreover, if $(f_n)_{n\in\N}$ is a Cauchy sequence in $\fA^{\Z_2}$ with respect to \eqref{L1norm}, the sequences $(f_n(s))_{n\in\N}$ are Cauchy sequences in $\fA$ for all $s\in\Z_2$ since $\eN{f_n-f_m}
=\sum_{s\in\Z_2} \|f_n(s)-f_m(s)\|/2$ for all $n,m\in\N$. Hence, for all $s\in\Z_2$, there exists $f_s\in\fA$ such that $\|f_n(s)-f_s\|\to 0$ for $n\to\infty$. Defining $f\in\fA^{\Z_2}$ by $f(s):=f_s$ for all $s\in\Z_2$, we get $\eN{f_n-f}\to 0$ for $n\to\infty$ which yields that $\fA^{\Z_2}$ is complete with respect to $\eN{\cdot}$ and that $\fA^{\Z_2}$ is a Banach \str algebra with respect to $\eN{\cdot}$. Hence, \eqref{cnt} yields that, for all $f\in\fA^{\Z_2}$,
\ba
\label{L1nb}
\|\Phi(f)\|
\le\eN{f},
\ea
where $\Phi\in\sIso(\fA^{\Z_2},\fB)$ is the \str isomorphism underlying $\fA^{\Z_2}\cong \fB$ (and, as usual, $\n{\cdot}$ also denotes the \Cs norm on $\fB$). An element of $\sHom(\fA^{\Z_2},\fC)$, where  $\fC$ is any \Cs algebra, is called $L^1$-norm bounded if the corresponding inequality \eqref{L1nb} holds. Since there always exists $\gamma\in\sMon(\fB,\mL(\mH))$ for some Hilbert space $\mH$ (the \str monomorphism of the Gelfand-Naimark structure theorem, for example [see, for example, \cite{BR}]), we have $\gamma\circ\Phi\in\sMon(\fA^{\Z_2},\mL(\mH))$ and, due to \eqref{piIso}, $\|(\gamma\circ\Phi)(f)\|=\|\Phi(f)\|$ for all $f\in\fA^{\Z_2}$  because $\fB$ is a \Cs algebra. Moreover, since we also know that any $\varphi\in\sHom(\fA^{\Z_2},\mL(\mH))$ which is $L^1$-norm bounded is bounded with respect to the universal norm, too, \ie, satisfies $\|\varphi(f)\|\le \|f\|$ for all $f\in\fA^{\Z_2}$ (see \cite{Williams}), we get from \eqref{L1nb} that, for all $f\in\fA^{\Z_2}$,
\ba
\label{ineq-Bu}
\|\Phi(f)\|
&=\|(\gamma\circ\Phi)(f)\|\nonumber\\
&\le \|f\|.
\ea
We next want to reverse inequality \eqref{ineq-Bu}. To this end, let $(\mH, R)\in\Rep(\fA^{\Z_2})$, where, for all \str algebras $\fC$, we denote by $\Rep(\fC):=\{(\mH,R)\, |\, \mbox{$\mH$ is a Hilbert space and $R\in\sHom(\fC,\mL(\mH))$}\}$ the collection of all representations of $\fC$ on $\mH$. Then, setting $S:=R\circ\Phi^{-1}\in\sHom(\fB,\mL(\mH))$, \eqref{cnt} yields  that, for all $f\in\fA^{\Z_2}$,
\ba
\label{ineq-uB}
\|R(f)\|
&=\|S(\Phi(f))\|\nonumber\\
&\le \|\Phi(f)\|. 
\ea
Moreover, since, on the one hand, \eqref{intf-Z2} and \eqref{L1norm} straightforwardly yield that $(\mH, \pi\rtimes U)\in\Rep_1(\fA^{\Z_2})$ for all $(\mH,\pi, U)\in\Cov(\fA,\Z_2,\alpha)$, where the collection of all representations $(\mH,R)\in \Rep(\fA^{\Z_2})$ for which  $R\in\sHom(\fA^{\Z_2},\mL(\mH))$ is $L^1$-bounded is denoted by $\Rep_1(\fA^{\Z_2})$, and since, on the other hand, there exists a bijective correspondence between nondegenerate covariant representations of the dynamical system $(\fA,\Z_2,\alpha)$ and nondegenerate representations of its crossed product $\fA\rtimes_\alpha\Z_2$ through the map $\Cov(\fA,\Z_2,\alpha)\ni (\mH,\pi,U)\mapsto (\mH,(\pi\rtimes U)')\in\Rep(\fA\rtimes_\alpha\Z_2)$, where the prime stands for the extension from $\Rep_1(\fA^{\Z_2})$ to $\Rep(\fA\rtimes_\alpha \Z_2)$, we know (see \cite{Williams}) that \eqref{univ} reads, for all $f\in\fA^{\Z_2}$, 
\ba
\label{univ-2}
\|f\|
=\hspace{-2mm}\sup_{\substack{(\mH,R)\hspace{0.2mm}\in\\\Rep_1(\fA^{\Z_2})}}\hspace{-2mm}\|R(f)\|.
\ea
Therefore, \eqref{ineq-Bu}, \eqref{ineq-uB}, and \eqref{univ-2} yield that, for all $f\in\fA^{\Z_2}$,
\ba
\label{univ-3}
\|f\|
=\|\Phi(f)\|.
\ea
This implies that $\fA^{\Z_2}$ is complete with respect to the universal norm since, for any Cauchy sequence $(f_n)_{n\in\N}$ in $\fA^{\Z_2}$, the sequence $(\Phi(f_n))_{n\in\N}$ is a Cauchy sequence in the \Cs algebra $\fB$ due to \eqref{univ-3}. Hence, there exists $x\in\fB$ such that $\|f_n-\Phi^{-1}(x)\|=\|\Phi(f_n)-x\|\to 0$ for $n\to\infty$. 
Now, since $\fA^{\Z_2}$ (with respect to the universal norm) and $\fA\rtimes_\alpha\Z_2$ are two \Cs completions of $\fA^{\Z_2}$, we know that there exists an isometric $\Psi\in\sIso(\fA^{\Z_2},\fA\rtimes_\alpha\Z_2)$ (see after \eqref{V'}) and, hence, $\Phi\circ\Psi^{-1}\in\sIso(\fA\rtimes_\alpha\Z_2, \fB)$.

\ref{cross-spec}\,
Let the map $\Phi:\fA^{\Z_2}\to\widehat\fA$ be defined, for all $f\in \fA^{\Z_2}$, by
\ba
\label{IsoPhi}
\Phi(f)
&:=\begin{bmatrix}
f(1) & f(-1)\\
\alpha_{-1}(f(-1)) & \alpha_{-1}(f(1))
\end{bmatrix}.
\ea
Using that $\alpha_{-1}\in\sAut(\fA)$, a direct check yields that $\Phi\in\sHom(\fA^{\Z_2},\widehat\fA)$. Moreover, we see from \eqref{IsoPhi} that $\Phi$ is clearly bijective. Hence, since $\widehat\fA$ is a \Cs subalgebra of $\fA^{2\times 2}$ due to Lemma \ref{lem:WidehatA} \ref{AHat}, setting $\fB:=\widehat\fA$ and using part \ref{cross-gen} leads to the conclusion (see Figure \ref{fig:isos}).
\eprf

\begin{figure}
\begin{center}
\begin{tikzcd}
\fA\arrow[r, tail, two heads,"\textstyle\psi"]
	& \wh\fA_0\arrow[d,  hookrightarrow]
	&\\
\fA^{\Z_2}\arrow[r, tail, two heads, "\textstyle\Phi"]\arrow[rr, tail, two heads, bend right=55,"\textstyle\Psi"']
	& \wh\fA\arrow[r, tail, two heads]
	& \fA\rtimes_\alpha\Z_2
\end{tikzcd}
\end{center}
\caption{The identification of the crossed product $\fA\rtimes_\alpha\Z_2$ built out of the \Cs dynamical system $(\fA,\Z_2,\alpha)$.} 
\label{fig:isos}
\end{figure}

\section{Jordan-Wigner transformation}
\label{sec:JW}

In this section, $\fA$ stands for the infinite tensor product algebra from Proposition \ref{prop:indlim} \ref{indlim:sub}. Following \cite{Ara84}, we apply the general construction of Section \ref{sec:crossed} to the \Cs dynamical system $(\fA, \Z_2,\alpha)$, where the nontrivial part of $\alpha$ is chosen to be the rotation around the 3-axis by an angle of $\pi$ on all  nonpositive sites of $\Z$ (leaving the observables unchanged on the positive sites $x$, see Definition \ref{def:alpha} below). This construction yields the final ingredient for the definition of the observable algebra in the infinite system approach discussed in the Introduction.

For the following, also recall the notations from Sections \ref{sec:Local} and \ref{sec:infinite} and the definition of the Pauli matrices $\sigma_1, \sigma_2, \sigma_3\in\C^{2\times 2}$,
\ba
\sigma_1
:=\left[\begin{array}{cc}
0 & 1\\
1& 0
\end{array}\right],\quad
\sigma_2
:=\left[\begin{array}{cc}
0 &-\ii\\ 
\ii & 0
\end{array}\right],\quad
\sigma_3
:=\left[\begin{array}{cc}
1 & 0
\\ 0& -1
\end{array}\right],
\ea
which, together with $\sigma_0:=1_2\in\C^{2\times 2}$, constitute what we call the Pauli basis of $\C^{2\times 2}$.

\vspace{2mm}

\bd[Spin observables]
\label{def:spin}

\bn[label=(\alph*), ref={\it (\alph*)}]
\setlength{\itemsep}{0mm}
\item 
\label{def:spin-a}
Let $\Lambda=\{x_1,\ldots,x_n\}\in\Fin(\Z)$ for some $n\in\N$ be such that $x_1<\ldots<x_n$ if $n\ge 2$. For all $x\in\Lm$ and all $\kp\in\num{0}{3}$, we define $\sigma_\kp^{(x), \Lm}\in\fA_\Lm$ by
\ba
\label{def-sigmaL}
\sigma_\kp^{(x), \Lm}
:=\begin{cases}
\hfill \xi_x^{-1}(\sigma_\kp), & n=1,\\
1_{\{x_1\}}\otimes\ldots\otimes\xi_x^{-1}(\sigma_\kp)\otimes\ldots\otimes 1_{\{x_n\}}, &n\ge 2.
\end{cases}
\ea

\item
\label{def:spin-b}
For all $x\in\Z$, all $\kp\in\num{0}{3}$, and any set $\Lm\in\Fin(\Z)$ with $x\in\Lm$, we define $\sigma_\kp^{(x)}\in\wt\fA_\Lm$ by
\ba
\label{def-sigma}
\sigma_\kp^{(x)}
:=\vi_\Lm(\sigma_\kp^{(x), \Lm}).
\ea
\en
\ed

In the following, $\delta_{xy}$ with $x,y\in\Z$ stands for usual Kronecker symbol and $\veps_{\kp\lm\mu}$ with $\kp,\lm,\mu\in\num{1}{3}$ for the usual Levi-Civita symbol. Moreover, if $A,B$ are elements of any algebra, we define as usual the commutator and the anticommutator of $A$ and $B$ by $[A,B]:=AB-BA$ and $\{A,B\}:=AB+BA$, respectively.

We next collect the properties of the spin observables which are used in the sequel.

\bl[Spin properties]
\label{lem:spin}
\bn[label=(\alph*), ref={\it (\alph*)}]
\setlength{\itemsep}{0mm}
\item 
\label{lem:spin-a}
Let $\Lm, \Lm'\in\Fin(\Z)$ with $\Lm\subseteq\Lm'$, let $x,y\in\Lm$, and let $\kp, \lm\in\num{0}{3}$. Then, in $\fA_\Lm$ and $\fA_{\Lm'}$, 
\ba
\label{spin-a}
\sigma_0^{(x), \Lm}
&=1_\Lm,\\
\label{spin-b}
(\sigma_\kp^{(x), \Lm})^\ast
&=\sigma_\kp^{(x), \Lm},\\
\label{spin-c}
\sigma_\kp^{(x), \Lm}\sigma_\lm^{(x), \Lm}
&=\delta_{\kp\lm}1_\Lm+\ii\sum_{\mu\in\num{1}{3}}\veps_{\kp\lm\mu}\sigma_\mu^{(x), \Lm} 
\mbox{\hspace{1mm} if $\kp, \lm\in\num{1}{3}$},\\
\label{spin-d}
\{\sigma_\kp^{(x), \Lm},\sigma_\lm^{(x), \Lm}\}
&=2\delta_{\kp\lm} 1_\Lm 
\mbox{\hspace{1mm} if $\kp, \lm\in\num{1}{3}$},\\
\label{spin-e}
[\sigma_\kp^{(x), \Lm},\sigma_\lm^{(x), \Lm}]
&=2\ii \sum_{\mu\in\num{1}{3}}\veps_{\kp\lm\mu} \sigma_\mu^{(x), \Lm} 
\mbox{\hspace{1mm} if $\kp, \lm\in\num{1}{3}$},\\
\label{spin-f}
[\sigma_\kp^{(x), \Lm},\sigma_\lm^{(y), \Lm}]
&=0 
\mbox{\hspace{1mm} if $x\neq y$},\\
\label{spin-g}
\vi_{\Lm',\Lm}(\sigma_\kp^{(x), \Lm})
&=\sigma_\kp^{(x), \Lm'}.
\ea

\item
\label{lem:spin-b}
Let $\Lm, \Lm'\in\Fin(\Z)$ with $\Lm\subseteq\Lm'$, let $x\in\Lm$, and let $\kp\in\num{0}{3}$. Then, in $\wt\fA_\Lm\subseteq\wt\fA_{\Lm'}$, 
\ba
\label{spin-h}
\vi_\Lm(\sigma_\kp^{(x), \Lm})=\vi_{\Lm'}(\sigma_\kp^{(x), \Lm'}).
\ea

\item
\label{lem:spin-c}
Let $x, y\in\Z$ and let $\kp, \lm\in\num{0}{3}$. Then, in $\fA$,
\ba
\label{spin-i}
\sigma_0^{(x)}
&=1,\\
\label{spin-j}
(\sigma_\kp^{(x)})^\ast
&=\sigma_\kp^{(x)},\\
\label{spin-k}
\sigma_\kp^{(x)}\sigma_\lm^{(x)}
=&\delta_{\kp\lm} 1
+\ii\sum_{\mu\in\num{1}{3}} \veps_{\kp\lm\mu}\sigma_\mu^{(x)}
\mbox{\hspace{1mm} if $\kp, \lm\in\num{1}{3}$},\\
\label{spin-l}
\{\sigma_\kp^{(x)},\sigma_\lm^{(x)}\}
&=2\delta_{\kp,\lm} 1
\mbox{\hspace{1mm} if $\kp, \lm\in\num{1}{3}$},\\
\label{spin-m}
[\sigma_\kp^{(x)},\sigma_\lm^{(x)}]
&=2\ii \sum_{\mu\in\num{1}{3}}\veps_{\kp\lm\mu} \sigma_\mu^{(x)} 
\mbox{\hspace{1mm} if $\kp, \lm\in\num{1}{3}$},\\
\label{spin-n}
[\sigma_\kp^{(x)},\sigma_\lm^{(y)}]
&=0 
\mbox{\hspace{1mm} if $x\neq y$}.
\ea
\en
\el

\br
\label{rem:spin}
Let $x\in\Z$ and let $\Lm_1, \Lm_2\in\Fin(\Z)$ be such that $x\in\Lm_1\cap\Lm_2$. Setting $\Gm:=\Lm_1\cup\Lm_2$ and using \eqref{spin-h}, we get $\vi_\Gm(\sigma_\kp^{(x), \Gm})=\vi_{\Lm_1}(\sigma_\kp^{(x), \Lm_1})$ and also  $\vi_\Gm(\sigma_\kp^{(x), \Gm})=\vi_{\Lm_2}(\sigma_\kp^{(x), \Lm_2})$. Hence, the notation $\sigma_\kp^{(x)}$ from Definition \ref{def:spin} \ref{def:spin-b} is reasonable. In particular, we can write $\sigma_\kp^{(x)}=\vi_{\{x\}}(\sigma_\kp^{(x), \{x\}})\in\wt\fA_{\{x\}}\subseteq\fA$ for all $\kp\in\num{0}{3}$ and all $x\in\Z$.
\er

\br
\label{rem:formalT}
Let $\Lambda=\{x_1,\ldots,x_n\}, \Gm=\{x_1,\ldots,x_n, x_{n+1},\ldots,x_{m}\}\in\Fin(\Z)$ for some $n, m\in\N$ with $2\le n<m$ and $x_1<\ldots<x_n<x_{n+1}<\ldots <x_{m}$. Then, using \eqref{def-sigma}, \eqref{KTiso}, \eqref{piIso}, and Lemma \ref{lem:kron} \ref{kron:cross}, we find
\ba
\|\sigma_3^{(x_1)}\ldots\sigma_3^{(x_{m})}-\sigma_3^{(x_1)}\ldots\sigma_3^{(x_{n})}\|
&=\|\vi_{\Gm}(\xi_{\Gm}^{-1}(\sigma_3\oslash\ldots\oslash\sigma_3\oslash(\sigma_3\oslash\ldots\oslash\sigma_3-1_2\oslash\ldots\oslash 1_2)))\|\nonumber\\
&=\|\sigma_3\|^n \|\sigma_3\oslash\ldots\oslash\sigma_3-1_2\oslash\ldots\oslash 1_2\|\nonumber\\
&=2.
\ea
In the last equality, we used the fact that, for all $k\in\N$ and all $A\in\C^{k\times k}$, the spectral norm \eqref{specnorm} equals $\|A\|=\max_{\lm\in\spec(A^\ast A)}\sqrt{\lm}$ ($\spec$ stands for the spectrum of the matrix in question) and, hence, $\|\sigma_3\|=1$ and $\|\sigma_3\oslash\ldots\oslash\sigma_3-1_2\oslash\ldots\oslash 1_2\|=2$ since \eqref{KP} yields $\spec((\sigma_3\oslash\ldots\oslash\sigma_3-1_2\oslash\ldots\oslash 1_2)^\ast (\sigma_3\oslash\ldots\oslash\sigma_3-1_2\oslash\ldots\oslash 1_2))=\{0,4\}$.

\er

\vspace{2mm}

\bprf
\ref{lem:spin-a}\,
As for \eqref{spin-a}, for all $x\in\Lm$, we have $\sigma_0^{(x), \Lm}=\xi_x^{-1}(\sigma_0)\in\fA_{\{x\}}$ if $n=1$ and $\sigma_0^{(x), \Lm}=1_{\{x_1\}}\otimes\ldots\otimes\xi_x^{-1}(\sigma_0)\otimes\ldots\otimes 1_{\{x_n\}}\in\fA_\Lm$ if $n\ge 2$ and if $\Lambda=\{x_1,\ldots,x_n\}$ is such that $x_1<\ldots<x_n$. Since, from \eqref{copy}, $\xi_x\in\sIso(\fA_{x},\C^{2\times 2})$ for all $x\in\Z$, we get $\xi_x^{-1}(\sigma_0)=1_{\{x\}}\in\fA_{x}$ for all $x\in\Z$. If $n\ge 2$, \eqref{KTiso} yields $1_{\{x_1\}}\otimes\ldots\otimes 1_{\{x_n\}}=\xi_\Lm^{-1}(\xi_{x_1}(1_{\{x_1\}})\oslash\ldots\oslash\xi_{x_n}(1_{\{x_n\}}))=\xi_\Lm^{-1}(1_{2^n})=1_\Lm$. 
Next, since, in $\M{2}$, we have $(\sigma_\kp)^\ast=\sigma_\kp$ for all $\kp\in\num{0}{3}$, \eqref{copy} and \eqref{ALmInvo} yield \eqref{spin-b}.
Since, in $\M{2}$, we also have $\sigma_\kp\sigma_\lm=\delta_{\kp\lm} 1_2+\ii\sum_{\mu\in\num{1}{3}}\veps_{\kp\lm\mu} \sigma_\mu$ \hspace{0.5mm} for all $\kp, \lm\in\num{1}{3}$, \eqref{copy} and \eqref{ALmProd} (and Lemma \ref{lem:odot} \ref{odot:ml}) yield \eqref{spin-c}.
Moreover, \eqref{spin-c} yields \eqref{spin-d} and \eqref{spin-e}, and \eqref{def-sigmaL} and \eqref{ALmProd} yield  \eqref{spin-f}.
Finally, as for \eqref{spin-g}, if $\Lm'=\{x\}$ for any $x\in\Z$, we have $\Lm=\Lm'$ and \eqref{viLL} implies $\vi_{\Lm,\Lm}(\sigma_\kp^{(x), \Lm})=\sigma_\kp^{(x), \Lm}$. If $n'\ge 2$ and $\Lm'=\{x_1,\ldots, x_{n'}\}$ with $x_1<\ldots<x_{n'}$ and if $\Lm=\{x_{i_1},\ldots, x_{i_n}\}$ with $x_{i_1}<\ldots<x_{i_n}$ for some $n<n'$,  \eqref{viLL} yields, for all $x\in\Lm$ and all $\kp\in\num{0}{3}$, 
\ba
\vi_{\Lm',\Lm}(\sigma_\kp^{(x), \Lm})
&=\vi_{\Lm',\Lm}(1_{\{x_{i_1}\}}\otimes\ldots\otimes\xi_x^{-1}(\sigma_\kp)\otimes\ldots\otimes1_{\{x_{i_n}\}})\nonumber\\
&=\pi_{\Lm,\Lm'}^{-1}((1_{\{x_{i_1}\}}\otimes\ldots\otimes\xi_x^{-1}(\sigma_\kp)\otimes\ldots\otimes1_{\{x_{i_n}\}})\otimes 1_{\Lm'\setminus\Lm})\nonumber\\
&=1_{\{x_1\}}\otimes\ldots\otimes\xi_x^{-1}(\sigma_\kp)\otimes\ldots\otimes1_{\{x_{n'}\}}\nonumber\\
&=\sigma_\kp^{(x), \Lm'}.
\ea

\ref{lem:spin-b}\,
Using \eqref{indlim-2} and \eqref{spin-g}, we get $\vi_\Lm(\sigma_\kp^{(x), \Lm})=\vi_{\Lm'}(\vi_{\Lm',\Lm}(\sigma_\kp^{(x), \Lm}))=\vi_{\Lm'}(\sigma_\kp^{(x), \Lm'})$ for all $x\in\Lm$ and all $\kp\in\num{0}{3}$.

\ref{lem:spin-c}\,
As for \eqref{spin-i}, let $A\in\fA$. Then, due to \eqref{uhf}, there exists a sequence $(\Lm_n)_{n\in\N}$ in $\Fin(\Z)$ and a sequence $(A_n)_{n\in\N}$ in $\fA$ with $A_n\in\wt\fA_{\Lm_n}$ for all $n\in\N$ such that $\|A_n-A\|\to0$ for $n\to\infty$. Setting $\Gamma_n:=\Lm_n\cup\{x\}$ for all $n\in\N$, Remark \ref{rem:spin} yields $\sigma_0^{(x)}=\vi_{\Gm_n}(\sigma_0^{(x), \Gm_n})$ for all $n\in\N$. Moreover, since $\Lm_n\subseteq\Gm_n$ for all $n\in\N$, \eqref{spin-a} yields 
$\sigma_0^{(x)}A_n
=\vi_{\Gm_n}(1_{\Gm_n})A_n=A_n$ for all $n\in\N$ because $\vi_{\Gm_n}\in\sIso(\fA_{\Gm_n},\wt\fA_{\Gm_n})$ for all $n\in\N$ (see Lemma \ref{prop:indlim} \ref{indlim:sub}). Hence, $\|\sigma_0^{(x)}A-A\|\le \|\sigma_0^{(x)}\|\|A-A_n\|+\|\sigma_0^{(x)}A_n-A\|$ for all $n\in\N$ and together with the analogous argument for $A\sigma_0^{(x)}-A$, we get \eqref{spin-i}.
Finally,  setting $\Lm:=\{x,y\}$, applying $\vi_\Lm$ to \eqref{spin-b}-\eqref{spin-f}, and using again that $\vi_{\Lm}\in\sHom(\fA_\Lm,\wt\fA_\Lm)$ and \eqref{spin-i}, we arrive at \eqref{spin-j}-\eqref{spin-n}.
\eprf

The first ingredient used for the construction of $TS_x\sigma_-^{(x)}$ from \eqref{AJWFermion} is the so-called lowering operator. 

\bd[Raising and lowering operators]
For all $x\in\Z$, we define $\sigma_\pm^{(x)}\in\fA$ by
\ba
\label{def-raislow}
\sigma_\pm^{(x)}
:=\frac12(\sigma_1^{(x)}\pm \ii \sigma_2^{(x)}),
\ea
and call them raising and lowering operator (at site $x$), respectively.
\ed

\br
\label{rem:FermiBose}
Using Lemma \ref{lem:spin} \ref{lem:spin-c}, we get, in $\fA$, for all $x,y\in\Z$ with $x\neq y$,
\ba
\label{FB-1}
\{\sigma_+^{(x)}, \sigma_+^{(x)}\}
&=0,\\
\label{FB-2}
\{\sigma_+^{(x)}, \sigma_-^{(x)}\}
&=1,\\
\label{FB-3}
[\sigma_+^{(x)}, \sigma_+^{(y)}]
&=0,\\
\label{FB-4}
[\sigma_+^{(x)}, \sigma_-^{(y)}]
&=0,
\ea
\ie, the raising and lower operators are of fermionic nature at the same sites and of bosonic nature at different sites. This is the reason for the introduction of the next ingredient.
\er

The second ingredient of \eqref{AJWFermion} is the following extended string.

\bd[Nonlocal multiplicator]
For all $x\in\Z$, we define $S_x\in\fA$ by
\ba
\label{def-Sx}
S_x
:=\begin{cases}
\prod_{y\in\num{1}{x-1}} \sigma_3^{(y)}, & x\ge 2,\\
\hfill 1, 
& x=1,\\
\hfill \prod_{y\in\num{x}{0}} \sigma_3^{(y)}, & x\le 0,
\end{cases}
\ea
and call it the nonlocal multiplicator (in $\fA$).
\ed

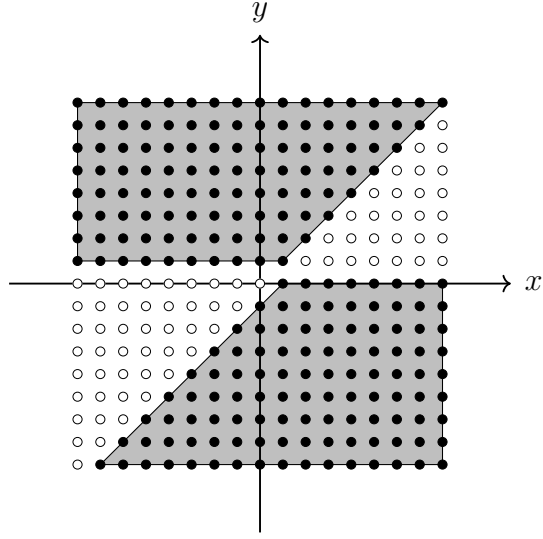
\begin{figure}[h!]
\begin{center}
\begin{tikzpicture}
\filldraw[fill=lightgray](-2.4,0.3) -- (0.3,0.3) -- (2.4,2.4) -- (-2.4,2.4) -- cycle;
\filldraw[fill=lightgray](-2.1,-2.4) -- (2.4,-2.4) -- (2.4,0) -- (0.3,0) -- cycle;
\draw [->, line width=0.25mm] (-3.3,0) -- (3.3,0);
\draw [->, line width=0.25mm](0,-3.3) -- (0,3.3);
\foreach \n in {0,...,8} {
	 	\filldraw[white][line width=0.1mm] (-\n*0.3,0) circle (0.6mm);}
\foreach \n in {1,...,8} {
	\foreach \m in {1,...,8} {
		\ifthenelse{\n>\m;}{\draw[line width=0.1mm] (\n*0.3,\m*0.3) circle (0.6mm);}{;}}}
\foreach \n in {0,...,8} {
	\foreach \m in {0,...,8} {
	 	\ifthenelse{\n>\m \OR \n=\m;}{\draw[line width=0.1mm] (-\n*0.3,-\m*0.3) circle (0.6mm);}{;}}}
\foreach \n in {0,...,8} {
	\foreach \m in {1,...,8} {
	 	\filldraw[line width=0.1mm] (-\n*0.3,\m*0.3) circle (0.6mm);}}
\foreach \n in {1,...,8} {
	\foreach \m in {0,...,8} {
	 	\filldraw[line width=0.1mm] (\n*0.3,-\m*0.3) circle (0.6mm);}}
\foreach \n in {0,...,8} {
	\foreach \m in {0,...,8} {
	 	\ifthenelse{\n<\m;}{\filldraw[line width=0.1mm] (-\n*0.3,-\m*0.3) circle (0.6mm);}{;}}}
\foreach \n in {1,...,8} {
	\foreach \m in {1,...,8} {
	 	\ifthenelse{\n<\m \OR \n=\m;}{\filldraw[line width=0.1mm] (\n*0.3,\m*0.3) circle (0.6mm);}{;}}}
\node at (3.6,0) {$x$};
\node at (0,3.6) {$y$};
\end{tikzpicture}
\caption{The prefactor function $\Z^2\ni(x,y)\mapsto\veps_{xy}\in\{-1,1\}$ from \eqref{prefac}. It equals $1$ at the filled circles (grey area) and $-1$ at the open circles (white area).}
\label{fig:epsxy}
\end{center}
\end{figure}

For the following, for all $x,y\in\Z$, we set $\sign(x):=1$ if $x\ge0$ and $\sign(x):=-1$ if $x<0$, and the prefactor function $\veps_{xy}\in\{-1,1\}$ is defined by (see Figure \ref{fig:epsxy})
\ba
\label{prefac}
\veps_{xy}
:=-\sign(y-x)\sign(-y).
\ea

The nonlocal multiplicator has the following properties. 

\bl[Nonlocal multiplicator]
Let $x, y\in\Z$ and let $\kp\in\num{1}{2}$. Then, in $\fA$,
\ba
\label{S-1}
S_x^\ast
&=S_x,\\
\label{S-2}
S_x^2
&=1,\\
\label{S-3}
[S_x, S_y]
&=0,\\
\label{S-4}
[S_x, \sigma_3^{(y)}]
&=0,\\
\label{S-5}
S_x\sigma_\kp^{(y)}
&=\veps_{xy}\sigma_\kp^{(y)} S_x.
\ea
\el

\bprf
Using \eqref{spin-i}-\eqref{spin-k} and \eqref{spin-n}, a direct computation leads to the assertion.
\eprf

In order to be able to introduce the ingredients allowing for the definition of the crossed product extension for the concrete case at hand, we need the following.

\bl[Extension of local \str automorphisms]
\label{lem:ExtLoc}
Let $\{\theta_\Lm\in\sAut(\fA_\Lm)\}_{\Lm\in\Fin(\Z)}$ be a family of \str automorphisms having the property that, for all $\Lm, \Lm'\in\Fin(\Z)$ with $\Lm\subseteq\Lm'$, 
\ba
\label{cond-theta}
\vi_{\Lm',\Lm}\circ\theta_\Lm
=\theta_{\Lm'}\circ\vi_{\Lm',\Lm}.
\ea
Then, the map $\theta:\fA\to\fA$ defined, for all $A\in\fA$, by
\ba
\label{lim-theta}
\theta(A)
:=\lim_{n\to\infty} \wt\theta_{\Lm_n}(A_n),
\ea
satisfies $\theta\in\sAut(\fA)$, where $\wt\theta_\Lm:=\vi_\Lm\circ\theta_\Lm\circ\vi_\Lm^{-1}\in\sAut(\wt\fA_\Lm)$ for all $\Lm\in\Fin(\Z)$ and where $(\Lm_n)_{n\in\N}$ is a sequence in $\Fin(\Z)$ and $(A_n)_{n\in\N}$ a sequence in $\fA$ with $A_n\in\wt\fA_{\Lm_n}$ for all $n\in\N$ such that $\|A_n-A\|\to0$ for $n\to\infty$. 
We call $\theta\in\sAut(\fA)$ the extension to $\fA$ of the family $\{\theta_\Lm\in\sAut(\fA_\Lm)\}_{\Lm\in\Fin(\Z)}$.
\el

\bprf
We start off by verifying that \eqref{lim-theta} is well-defined. First, we show that the limit exists, \ie, that the sequence $(\wt\theta_{\Lm_n}(A_n))_{n\in\N}$ is a Cauchy sequence in $\fA$. To this end, let $\Lm, \Lm'\in\Fin(\Z)$ with $\Lm\subseteq\Lm'$ and plug $\vi_{\Lm',\Lm}=\vi_{\Lm'}^{-1}\circ\vi_\Lm$ from \eqref{indlim-2} into \eqref{cond-theta}. Then, we get $\vi_\Lm\circ\theta_\Lm\circ\vi_\Lm^{-1}=\vi_{\Lm'}\circ\theta_{\Lm'}\circ\vi_{\Lm'}^{-1}$ on $\wt\fA_\Lm$, \ie, for all $A\in\wt\fA_\Lm$,
\ba
\label{TildeTheta}
\wt\theta_\Lm(A)
=\wt\theta_{\Lm'}(A).
\ea
Setting $\Lm'_{nm}:=\Lm_n\cup\Lm_m$ for all $n,m\in\N$ and using \eqref{TildeTheta}, we thus have, for all $n,m\in\N$, 
\ba
\label{ThetaAnCauchy}
\|\wt\theta_{\Lm_n}(A_n)-\wt\theta_{\Lm_m}(A_m)\|
&=\|\wt\theta_{\Lm'_{nm}}(A_n-A_m)\|\nonumber\\
&=\|A_n-A_m\|,
\ea
where we used \eqref{piIso}. Moreover, the limit in \eqref{lim-theta} is independent of the choice of the sequence $(A_n)_{n\in\N}$. In order to show this, let $(\Gm_n)_{n\in\N}$ be a sequence in $\Fin(\Z)$ and $(B_n)_{n\in\N}$ a sequence in $\fA$ with $B_n\in\wt\fA_{\Gm_n}$ for all $n\in\N$ such that $\|B_n-A\|\to0$ for $n\to\infty$ and set $\theta'(A):=\lim_{n\to\infty} \wt\theta_{\Gm_n}(B_n)$.
Then, since 
$\|\theta(A)-\theta'(A)\|
\le \|\theta(A)-\wt\theta_{\Lm_n}(A_n)\|
+\|\wt\theta_{\Lm_n}(A_n)-\wt\theta_{\Gm_n}(B_n)\|
+\|\wt\theta_{\Gm_n}(B_n)-\theta'(A)\|$ for all $n\in\N$, setting $\Lm'_n:=\Lm_n\cup\Gm_n$ for all $n\in\N$ and writing $\|\wt\theta_{\Lm_n}(A_n)-\wt\theta_{\Gm_n}(B_n)\|=\|\wt\theta_{\Lm'_n}(A_n-B_n)\|=\|A_n-B_n\|$ for all $n\in\N$ (with the help of \eqref{TildeTheta}) yields the conclusion.

We next prove that $\theta\in\sHom(\fA)$. Since, due to  Proposition \ref{prop:indlim} \ref{indlim:sub}, $\wt\fA_{\Lm_n}$is a \Cs subalgebra of $\fA$ for all $n\in\N$,  we have $\lm A_n, A_n^\ast\in\wt\fA_{\Lm_n}$ for all $\lm\in\C$ and all $n\in\N$. Hence, since $\wt\theta_{\Lm_n}\in\sAut(\wt\fA_{\Lm_n})$ for all $n\in\N$ and since the scalar multiplication and the involution of $\fA$ are continuous (with respect to the \Cs norm of $\fA$), \eqref{lim-theta} implies that $\theta$ preserves  the scalar multiplication and the involution of $\fA$. 
As for the addition and the multiplication, let $B\in\fA$ and let $(\Gm_n)_{n\in\N}$ be a sequence in $\Fin(\Z)$ and $(B_n)_{n\in\N}$ a sequence in $\fA$ with $B_n\in\wt\fA_{\Gm_n}$ for all $n\in\N$ such that $\|B_n-B\|\to0$ for $n\to\infty$. Setting $\Lm'_n:=\Lm_n\cup\Gm_n$ for all $n\in\N$, the sequence $(A_n+B_n)_{n\in\N}$ in $\fA$ satisfies $A_n+B_n\in\wt\fA_{\Lm'_n}$ for all $n\in\N$. Moreover,  the fact that $\wt\theta_{\Lm'_n}\in\sAut(\wt\fA_{\Lm'_n})$ for all $n\in\N$ and \eqref{TildeTheta} yield that,  for all $n\in\N$,
\ba
\label{ThetaAdd}
\wt\theta_{\Lm'_n}(A_n+B_n)
&=\wt\theta_{\Lm'_n}(A_n)+\wt\theta_{\Lm'_n}(B_n)\nonumber\\
&=\wt\theta_{\Lm_n}(A_n)+\wt\theta_{\Gm_n}(B_n).
\ea
Due to the continuity of the addition of $\fA$, \eqref{ThetaAdd} and \eqref{lim-theta} imply that $\theta$ also preserves  the addition of $\fA$. Analogously, since $\wt\theta_{\Lm'_n}(A_nB_n)=\wt\theta_{\Lm_n}(A_n)\wt\theta_{\Gm_n}(B_n)$ for all $n\in\N$ and since the multiplication of $\fA$ is continuous, $\theta$ also preserves  the multiplication of $\fA$. 

Finally, we show that $\theta$ is bijective. To this end, let $A\in\fA$, let $(\Lm_n)_{n\in\N}$ be a sequence in $\Fin(\Z)$ and $(A_n)_{n\in\N}$ a sequence in $\fA$ with $A_n\in\wt\fA_{\Lm_n}$ for all $n\in\N$ such that $\|A_n-A\|\to0$ for $n\to\infty$. First, suppose that $\theta(A)=0$. Then, since $\wt\theta_{\Lm_n}$ is an isometry for all $n\in\N$ (see \eqref{piIso}), we get $0=\|\theta(A)\|=\lim_{n\to\infty}\|\wt\theta_{\Lm_n}(A_n)\|=\lim_{n\to\infty}\|A_n\|=\|A\|$, \ie, $\theta$ is injective. As for its surjectivity, set $B_n:=\wt\theta_{\Lm_n}^{-1}(A_n)\in\wt\fA_{\Lm_n}$ for all $n\in\N$. Since, for all $\Lm, \Lm'\in\Fin(\Z)$ with $\Lm\subseteq\Lm'$, we have $\wt\theta_{\Lm}^{-1}\in\sAut(\wt\fA_\Lm)$ and $\wt\theta_\Lm^{-1}=\wt\theta_{\Lm'}^{-1}$ on $\wt\fA_\Lm$ as in \eqref{TildeTheta}, $(B_n)_{n\in\N}$ is a Cauchy sequence in $\fA$ as in \eqref{ThetaAnCauchy}. Writing $B:=\lim_{n\to\infty} B_n$, \eqref{lim-theta} yields $\theta(B)=\lim_{n\to\infty}\wt\theta_{\Lm_n}(B_n)=A$. 
\eprf

For the following, we define $\veps_\kp\in\{-1,1\}$ for all $\kp\in\num{0}{3}$ by
\ba
\veps_\kp
:=\begin{cases}
\hfill1,  & \kp\in \{0,3\},\\
-1, & \kp\in \{1,2\}.
\end{cases}
\ea

The \str automorphisms of the next proposition are at the heart of the extension from \cite{Ara84} of the Jordan-Wigner transformation. They correspond to a rotation around the 3-axis by an angle of $\pi$ on all sites of $\Z$ and on all nonpositive sites of $\Z$, respectively.

\bp[Spin rotations]
\label{prop:rot}
For all $\Lm\in\Fin(\Z)$, let $\Theta_\Lm, \Theta_\Lm'\in\sHom(\fA_\Lm)$ be the unique \str homomorphisms defined, for all $x\in\Lm$ and all $\kp\in\num{0}{3}$, by
\ba
\label{ThetaLm}
\Theta_\Lm(\sigma_\kp^{(x), \Lm})
&:=\veps_\kp \sigma_\kp^{(x), \Lm},\\
\label{Theta'Lm}
\Theta'_\Lm(\sigma_\kp^{(x), \Lm})
&:=\begin{cases}
\hfill\sigma_\kp^{(x), \Lm}, & \mbox{$\Lm\cap\N\neq\emptyset$ and $x\ge 1$},\\
\veps_\kp \sigma_\kp^{(x), \Lm}, & \mbox{otherwise}.
\end{cases}
\ea
Then:
\bn[label=(\alph*), ref={\it (\alph*)}]
\setlength{\itemsep}{0mm}
\item 
\label{rot:lim}
The families $\{\Theta_\Lm\in\sAut(\fA_\Lm)\}_{\Lm\in\Fin(\Z)}$ and $\{\Theta'_\Lm\in\sAut(\fA_\Lm)\}_{\Lm\in\Fin(\Z)}$ satisfy \eqref{cond-theta}. 

\item
\label{rot:spin}
For all $x\in\Z$ and all $\kp\in\num{0}{3}$, the respective extensions $\Theta, \Theta'\in\sAut(\fA)$ satisfy
\ba
\label{Theta-sigma}
\Theta(\sigma_\kp^{(x)})
&=\veps_\kp \sigma_\kp^{(x)},\\
\label{Theta'-sigma}
\Theta'(\sigma_\kp^{(x)})
&= \begin{cases}
\hfill\sigma_\kp^{(x)}, & x\ge 1,\\
\veps_\kp \sigma_\kp^{(x)}, & x\le 0.
\end{cases}
\ea
Moreover, we have 
\ba
\label{Tinv}
\Theta^2
&=1,\\
\label{T'inv}
\Theta'^2
&=1,\\
\label{CommTT'}
\Theta\Theta'
&=\Theta'\Theta.
\ea
\en
\ep

\bprf
We first note that, if, for a fixed $\Lambda=\{x_1,\ldots,x_n\}$ for some $n\in\N$ which is such that $x_1<\ldots<x_n$ if $n\ge 2$, we assume that $\theta_\Lm\in\sHom(\fA_\Lm)$ and that $\theta_\Lm(\sigma_\kp^{(x), \Lm})$ is given for all $x\in\Lm$ and all $\kp\in\num{0}{3}$, then $\theta_\Lm$ is unique because $\{\xi_{x_1}^{-1}(\sigma_{\kp_1})\otimes\ldots\otimes \xi_{x_n}^{-1}(\sigma_{\kp_n})\}_{\kp_1,\ldots, \kp_n\in\num{0}{3}}$ is a basis of $\fA_\Lm$ due to Lemma \ref{lem:ltp} \ref{ltp:iso} implying that any $A\in\fA_\Lm$ can be written, for some 
$a_K\in\C$ with $K=(\kp_1,\ldots,\kp_n)\in\num{0}{3}^n$, as
\ba
\label{AExp}
A
=\sum_{K\in\num{0}{3}^n} a_K\prod_{i\in\num{1}{n}}\sigma_{\kp_i}^{(x_i), \Lm},
\ea
from which $\theta_\Lm(A)=\sum_{K\in\num{0}{3}^n} a_K\prod_{i\in\num{1}{n}}\theta_\Lm(\sigma_{\kp_i}^{(x_i), \Lm})$. Hence, \eqref{ThetaLm} and \eqref{Theta'Lm} uniquely determine $\Theta_\Lm, \Theta_\Lm'\in\sHom(\fA_\Lm)$, respectively. 

\ref{rot:lim}\, 
In order to make use of Lemma \ref{lem:ExtLoc}, we have to check that $\Theta_\Lm, \Theta'_\Lm\in\sAut(\fA_\Lm)$ for all $\Lm\in\Fin(\Z)$ and that \eqref{cond-theta} holds for both families.  But, for all $\Lm\in\Fin(\Z)$,  we have $\Theta_\Lm\circ\Theta_\Lm=1$ and $\Theta'_\Lm\circ\Theta'_\Lm=1$ (here, $1$ stands for the identity element of the group $\sAut(\fA_\Lm)$) since $\veps_\kp^2=1$ for all $\kp\in\num{0}{3}$. Moreover, let $\Lm, \Lm'\in\Fin(\Z)$ with $\Lm\subseteq\Lm'$. Since $\vi_{\Lm',\Lm}\circ\Theta_\Lm, \Theta_{\Lm'}\circ\vi_{\Lm',\Lm}, \vi_{\Lm',\Lm}\circ\Theta'_\Lm, \Theta'_{\Lm'}\circ\vi_{\Lm',\Lm}\in\sHom(\fA_\Lm, \fA_{\Lm'})$, it is enough to check \eqref{cond-theta} on $\sigma_\kp^{(x), \Lm}$ for all $x\in\Lm$ and all $\kp\in\num{0}{3}$. Using  \eqref{ThetaLm} and \eqref{spin-g}, we get $\vi_{\Lm',\Lm}(\Theta_\Lm(\sigma_\kp^{(x), \Lm}))=\veps_\kp\vi_{\Lm',\Lm}(\sigma_\kp^{(x), \Lm})=\veps_\kp\sigma_\kp^{(x), \Lm'}=\Theta_{\Lm'}(\sigma_\kp^{(x), \Lm'})=\Theta_{\Lm'}(\vi_{\Lm',\Lm}(\sigma_\kp^{(x), \Lm}))$ for all $x\in\Lm$ and all $\kp\in\num{0}{3}$ and, analogously, \eqref{Theta'Lm} yields $\vi_{\Lm',\Lm}(\Theta'_\Lm(\sigma_\kp^{(x), \Lm}))=\Theta'_{\Lm'}(\vi_{\Lm',\Lm}(\sigma_\kp^{(x), \Lm}))$ for all $x\in\Lm$ and all $\kp\in\num{0}{3}$, \ie, both families satisfy \eqref{cond-theta}. Hence, Lemma \ref{lem:ExtLoc} yields the respective extensions $\Theta, \Theta'\in\sAut(\fA)$. 

\ref{rot:spin}\,  
With the help of Definition \ref{def:spin} \ref{def:spin-b}, \eqref{lim-theta}, and \eqref{ThetaLm}, we get $\Theta(\sigma_\kp^{(x)})=\Theta(\vi_\Lm(\sigma_\kp^{(x),\Lm}))=\wt\Theta_\Lm(\vi_\Lm(\sigma_\kp^{(x),\Lm}))=\vi_\Lm(\Theta_\Lm(\sigma_\kp^{(x),\Lm}))=\veps_\kp\sigma_\kp^{(x)}$ for all $\kp\in\num{0}{3}$, all $x\in\Z$, and all $\Lm\in\Fin(\Z)$ with $x\in\Lm$, and \eqref{Theta'-sigma} follows analogously from \eqref{Theta'Lm}. Moreover, since again $\veps_\kp^2=1$ for all $\kp\in\num{0}{3}$, we have $\Theta\circ\Theta=\Theta'\circ\Theta'=1$ (here, $1$ stands for the identity element of the group $\sAut(\fA)$). Finally, since 
$\Theta_\Lm\circ \Theta'_\Lm=\Theta'_\Lm\circ\Theta_\Lm$ for all $\Lm\in\Fin(\Z)$ due to \eqref{ThetaLm}, \eqref{Theta'Lm}, and \eqref{AExp}, we get, for all $A\in\fA$, 
\ba
\Theta(\Theta'(A))
-\Theta'(\Theta(A))
&=\lim_{n\to\infty}
(\wt\Theta_{\Lm_n}(\wt\Theta'_{\Lm_n}(A_n))
-\wt\Theta'_{\Lm_n}(\wt\Theta_{\Lm_n}(A_n)))\nonumber\\
&=\lim_{n\to\infty}
(\vi_{\Lm_n}(\Theta_{\Lm_n}(\Theta'_{\Lm_n}(\vi_{\Lm_n}^{-1}(A_n))))
-\vi_{\Lm_n}(\Theta'_{\Lm_n}(\Theta_{\Lm_n}(\vi_{\Lm_n}^{-1}(A_n)))))\nonumber\\
&=0,
\ea
where $(\Lm_n)_{n\in\N}$ is a sequence in $\Fin(\Z)$ and $(A_n)_{n\in\N}$ a sequence in $\fA$ with $A_n\in\wt\fA_{\Lm_n}$ for all $n\in\N$ such that $\|A_n-A\|\to0$ for $n\to\infty$.
\eprf

For the following, recall from Proposition \ref{prop:indlim} \ref{indlim:sub} that the infinite tensor product $\fA$ is a unital \Cs algebra. 

We next introduce the third ingredient of \eqref{AJWFermion}. To this end, we make use of Lemma \ref{lem:WidehatA} and  Proposition \ref{prop:cross} for the following special choice (recall Proposition \ref{prop:rot} \ref{rot:spin}).

\bd[Anchor element]
\label{def:alpha}
Let $\alpha\in\Hom(\Z_2,\sAut(\fA))$ be defined by 
\ba
\label{choice}
\alpha_{-1}
:=\Theta'.
\ea
\bn[label=(\alph*), ref={\it (\alph*)}]
\setlength{\itemsep}{0mm}
\item 
As in Lemma \ref{lem:WidehatA} \ref{AHat} and \ref{AHat-A}, we set 
\ba
\label{choice-1}
\widehat\fA
&:=\left\{\begin{bmatrix}A & B\\ \Theta'(B) & \Theta'(A)\end{bmatrix}\in\fA^{2\times 2}\,\bigg|\, A,B\in\fA\right\},\\
\label{choice-2}
\widehat\fA_0
&:=\left\{\begin{bmatrix}A & 0\\ 0 & \Theta'(A)\end{bmatrix}\in\fA^{2\times 2}\,\bigg|\, A\in\fA\right\},
\ea
and we note that \eqref{choice-1} and \eqref{choice-2} are unital \Cs algebras with identity 
\ba
\begin{bmatrix} 1 & 0\\0 & 1\end{bmatrix}\in\wh\fA_0\subseteq\wh\fA\subseteq\fA^{2\times 2}.
\ea 
Moreover, the \str isomorphism $\psi\in\sIso(\fA,\widehat\fA_0)$ is defined as in \eqref{IsoPsi}.  

\item
We call  anchor element (of $\widehat\fA$) the element $T\in\widehat\fA\setminus\widehat\fA_0$ defined by
\ba
\label{def-T}
T
:=\begin{bmatrix} 0 & 1\\ 1 & 0\end{bmatrix}.
\ea
\en
\ed

For the following, if $\mV$ is any vector space and $\mW_1$ and $\mW_2$ are vector subspaces of $\mV$, we denote by $\mW_1+\mW_2$ and $\mW_1\oplus\mW_2$ their usual sum and (internal) direct sum, respectively.

The anchor element has the following properties.

\bl[Anchor element]
\label{lem:T}
\bn[label=(\alph*), ref={\it (\alph*)}]
\setlength{\itemsep}{0mm}
\item 
\label{lem:T-a}
For all $A\in\fA$, we have
\ba
\label{T-1}
T^2
&=1,\\
\label{T-2}
T^\ast
&=T,\\
\label{T-3}
T\psi(A)
&=\psi(\Theta'(A))T.
\ea

\item
\label{lem:T-b}
Setting $\wh\fA_0 T:=\{AT\,|\, A\in\wh\fA_0\}$, we have the decomposition
\ba
\label{dcomp-1}
\widehat\fA
=\widehat\fA_0\oplus\widehat\fA_0 T.
\ea
\en
\el

\bprf
\ref{lem:T-a}\, 
Using  Lemma \ref{lem:Amat} \ref{Amat-str},  \eqref{IsoPsi}, and Proposition \ref{prop:rot} \ref{rot:spin}, we get \eqref{T-1}-\eqref{T-3}.

\ref{lem:T-b}\, 
Since $\begin{bmatrix}B & 0\\ 0 & \Theta'(B)\end{bmatrix}\begin{bmatrix} 0 & 1\\ 1 & 0\end{bmatrix}=\begin{bmatrix} 0 & B\\ \Theta'(B) & 0\end{bmatrix}\in\widehat\fA_0 T$ for all $B\in\fA$, we have $\widehat\fA_0\cap\widehat\fA_0 T=\{0\}$ and \eqref{choice-1} and \eqref{choice-2} yield the assertion.
\eprf

We now arrive at Araki's extension \eqref{AJWFermion} of the Jordan-Wigner transformation discussed in the Introduction, \ie, at the definition of what we call the Araki-Jordan-Wigner fermion. This generalized annihilation operator lies outside of $\hA_0$ and constitutes the building block for the construction of the CAR algebra over the configuration space $\Z$.

\bd[Araki-Jordan-Wigner fermion]
\label{def:nihil}
For all $x\in\Z$, we define $a_x\in\hA\setminus\hA_0$ by
\ba
\label{def-nihil}
a_x
:=T\psi(S_x\sigma_-^{(x)}),
\ea
and we call it the annihilation operator (at site $x$).  Moreover, the element $a_x^\ast:=(a_x)^\ast\in\widehat\fA$ for all $x\in\Z$ is called creation operator (at site $x$).
\ed

Unlike the raising and lowering operators in Remark \ref{rem:FermiBose}, the annihilation and creation operators indeed satisfy the CARs.

\bp[CAR]
For all $x,y\in\Z$, we have, in $\widehat\fA$, that  $a_x^\ast=T\psi(S_x\sigma_+^{(x)})$ and that
\ba
\label{CAR-1}
\{a_x, a_y\}
&=0,\\
\label{CAR-2}
\{a_x, a_y^\ast\}
&=\delta_{xy} 1.
\ea
\ep

\bprf
Using \eqref{T-3},  \eqref{T-1}, \eqref{Theta'-sigma} written, for all $\kp\in\num{1}{2}$ and all $x\in\Z$, as (note that $\veps_{xx}=\sign(x-1)$ for all $x\in\Z$, see Figure \ref{fig:epsxy})
\ba
\label{Theta'sigma}
\Theta'(\sigma_\kp^{(x)})
=\veps_{xx} \sigma_\kp^{(x)},
\ea
\eqref{S-5}, \eqref{S-3}, and the involution of \eqref{FB-1} and \eqref{FB-3} written as $\sigma_-^{(x)}\sigma_-^{(y)}=(1-2\delta_{xy})\sigma_-^{(y)}\sigma_-^{(x)}$ for all  $x,y\in\Z$, we get, for all $x,y\in\Z$, 
\ba
a_xa_y
&=T\psi(S_x\sigma_-^{(x)})T\psi(S_y\sigma_-^{(y)})\nonumber\\
&=\psi(\Theta'(S_x)\Theta'(\sigma_-^{(x)})S_y\sigma_-^{(y)})\nonumber\\
&=\veps_{xx}\psi(S_x\sigma_-^{(x)}S_y\sigma_-^{(y)})\nonumber\\
&=\veps_{xx}\veps_{yx}\psi(S_yS_x\sigma_-^{(x)}\sigma_-^{(y)})\nonumber\\
&=(1-2\delta_{xy})\veps_{xx}\veps_{xy}\veps_{yx}\psi(S_y\sigma_-^{(y)}S_x\sigma_-^{(x)})\nonumber\\
&=(1-2\delta_{xy})\veps_{xx}\veps_{xy}\veps_{yx}T\psi(\Theta'(S_y)\Theta'(\sigma_-^{(y)}))T\psi(S_x\sigma_-^{(x)})\nonumber\\
&=(1-2\delta_{xy})\veps_{xx}\veps_{yy}\veps_{xy}\veps_{yx}a_y a_x\nonumber\\
&=-a_ya_x,
\ea
where we used that $\veps_{xx}\veps_{yy}\veps_{xy}\veps_{yx}=2\delta_{xy}-1$ for all $x,y\in\Z$ (see  Figure \ref{fig:epsxy}), \ie, we get \eqref{CAR-1}.
As for \eqref{CAR-2}, since, similarly,
$a_x^\ast
=T\psi(\Theta'(\sigma_+^{(x)} S_x))
=T\psi(\Theta'(\veps_{xx}S_x\sigma_+^{(x)}))
=T\psi(\veps_{xx}^2 S_x\sigma_+^{(x)})
=T\psi(S_x\sigma_+^{(x)})$ for all $x\in\Z$
and since $\sigma_-^{(x)}\sigma_+^{(y)}=(1-2\delta_{xy})\sigma_+^{(y)}\sigma_-^{(x)}+\delta_{xy}1$ for all  $x,y\in\Z$ due to \eqref{FB-2} and \eqref{FB-4}, we analogously get $a_xa_y^\ast=(1-2\delta_{xy})\veps_{xx}\veps_{yy}\veps_{xy}\veps_{yx}a_y^\ast a_x+\veps_{xx}^2\delta_{xy}\psi(S_x^2)=-a_y^\ast a_x+\delta_{xy}1$ for all $x,y\in\Z$.
\eprf

\br
\label{rem:axForm}
With  the help of \eqref{def-T}, \eqref{IsoPsi}, \eqref{choice}, and \eqref{Theta'-sigma} (written as in \eqref{Theta'sigma} if $\kp\in\num{1}{2}$), the annihilation and creation operators read, for all $x\in\Z$, 
\ba
\label{AJWa}
a_x
&=
\begin{bmatrix} 
0& \sign(x-1) S_x\sigma_-^{(x)} \\ 
S_x\sigma_-^{(x)} & 0
\end{bmatrix},\\
\label{AJWas}
a_x^\ast
&=
\begin{bmatrix} 
0& \sign(x-1) S_x\sigma_+^{(x)} \\ 
S_x\sigma_+^{(x)} & 0
\end{bmatrix},
\ea
\ie, $a_x, a_x^\ast\in\widehat\fA\setminus\widehat\fA_0$ for all $x\in\Z$.
\er

We next define the following subset of $\widehat\fA$.

\bd[CAR subalgebra]
\label{def:CAR}
Let $\fB:=\{a_x\,|\,x\in\Z\}\subseteq\widehat\fA$.
\bn[label=(\alph*), ref={\it (\alph*)}]
\setlength{\itemsep}{0mm}
\item 
\label{def:CAR-1}
The \Cs subalgebra of $\widehat\fA$ generated by $\fB$ is defined by
\ba
\label{def-A1}
\widehat\fA_1
&:=
\hspace{-5mm}\bigcap_{\substack{\fB':\,\textup{\Cs subalgebra of }\widehat\fA\\\fB\subseteq\fB'}}\hspace{-10mm}\fB'.
\ea

\item
\label{def:CAR-2}
A polynomial in $\widehat\fA$ generated by $\fB$ is an element of $\widehat\fA$ of the form
\ba
\label{Pol}
\alpha 1+\sum_{l\in\num{1}{N}}\sum_{\substack{X_l\in\Z^l\\ \sharp_l\in\{\pm\}^l}}\alpha_{X_l, \sharp_l} \prod_{i\in\num{1}{l}} a_{x_{l, i}}^{\sharp_{l, i}},
\ea
where $a_x^{-}:=a_x$ and  $a_x^{+}:=a_x^\ast$ for all $x\in\Z$ and where $\alpha\in\C$, $N\in\N$, and, for all $l\in\num{1}{N}$, $\alpha_{X_l, \sharp_l} \in\C$ for all $X_l=(x_{l,1},\ldots, x_{l,l})\in\Z^l$ and all $\sharp_l=(\sharp_{l,1},\ldots,\sharp_{l,l})\in\{\pm\}^l$ and $\card(\{X_l\in\Z^l\,|\, \alpha_{X_l, \sharp_l}\neq 0\})\in\N_0$ for all $\sharp_l\in\{\pm\}^l$. 
The set of all the polynomials in $\widehat\fA$ generated by $\fB$ is denoted by $\Pol(\fB)$.
\en
\ed

\bp[CAR subalgebra]
\label{prop:cloPol}
$\widehat\fA_1$ is a unital \Cs subalgebra of $\widehat\fA$ and 
\ba
\label{cloPol}
\widehat\fA_1
=\clo_{\widehat\fA}(\Pol(\fB)).
\ea
\ep

\bprf
The set \eqref{def-A1} is a \str subalgebra of $\widehat\fA$ which is closed with respect to the norm of $\widehat\fA$. Moreover, $\widehat\fA_1$ is unital due to \eqref{CAR-2} and the fact that the anticommutator satisfies $\{a_x, a_x^\ast\}\in\fB'$ for all $x\in\Z$ and all \Cs subalgebras $\fB'$ of $\wh\fA$ with $\fB\subseteq\fB'$. Next, since $\Pol(\fB)$ is a \str subalgebra of $\widehat\fA$ and since $\Pol(\fB)\subseteq\fB'$ for all \Cs subalgebras $\fB'$ of $\widehat\fA$ satisfying $\fB\subseteq\fB'$, it follows that $\Pol(\fB)$ is a \str subalgebra of $\widehat\fA_1$, too. Hence, $\clo_{\widehat\fA}(\Pol(\fB))\subseteq\widehat\fA_1$ since $\widehat\fA_1$ is closed with respect to the \Cs norm of $\wh\fA$. In order to deduce the converse inclusion, it is enough to show that $\clo_{\widehat\fA}(\Pol(\fB))$ is a \Cs subalgebra of $\widehat\fA$ which contains $\fB$. But $\clo_{\widehat\fA}(\Pol(\fB))$ is clearly a \str  subalgebra of $\widehat\fA$ which not only contains $\fB$ but which is closed, too. 
\eprf

For the following, recall that $\Theta\in\sAut(\fA)$ is the spin rotation from Proposition \ref{prop:rot} \ref{rot:spin}. Moreover, if $\fA$ is any \Cs algebra, $\fB$ a subset of $\fA$, and $\pi\in\sAut(\fA)$, we set $\pi(\fB):=\{\pi(B)\,|\, B\in\fB\}$.

The even and odd parts (with respect to the involutive $\Theta$) of the \Cs subalgebras $\wh\fA_0$ and $\wh\fA_1$ of $\wh\fA$ are related as follows. They play an important role in many applications (see Remark \ref{rem:appl}).

\bp[Decomposition]
\label{prop:dec}
\bn[label=(\alph*), ref={\it (\alph*)}]
\setlength{\itemsep}{0mm}
\item 
\label{dec-a}
There exists a unique $\widehat\Theta\in\sAut(\widehat\fA)$ satisfying
\ba
\label{dec-a1}
\widehat\Theta\circ\psi
&=\psi\circ\Theta,\\
\label{dec-a2}
\widehat\Theta(T)
&=T.
\ea

\item
\label{dec-b}
We have $\widehat\Theta(\widehat\fA_0)\subseteq\widehat\fA_0$ and $\widehat\Theta(\widehat\fA_1)\subseteq\widehat\fA_1$ and the so-called even and odd parts (with respect to $\wh\Theta$) of $\wh\fA$, $\wh\fA_0$, and $\wh\fA_1$ are defined by
\ba
\widehat\fA_\pm
&:=\{A\in\widehat\fA\,|\,\widehat\Theta(A)=\pm A\},\\
\widehat\fA_{0,\pm}
&:=\{A\in\widehat\fA_0\,|\,\widehat\Theta(A)=\pm A\},\\
\widehat\fA_{1,\pm}
&:=\{A\in\widehat\fA_1\,|\,\widehat\Theta(A)=\pm A\}.
\ea
The even parts $\widehat\fA_+$, $\widehat\fA_{0,+}$, and $\widehat\fA_{1,+}$ are unital \Cs subalgebras of $\widehat\fA$. The odd parts $\widehat\fA_-$, $\widehat\fA_{0,-}$, and $\widehat\fA_{1,-}$ are invariant under the addition, the scalar multiplication, and the involution of $\widehat\fA$ and they are closed with respect to the \Cs norm of $\widehat\fA$.

\item
\label{dec-c}
We have $\widehat\fA=\widehat\fA_+ \oplus\widehat\fA_-$, $\widehat\fA_0=\widehat\fA_{0,+}\oplus\widehat\fA_{0,-}$, and $\widehat\fA_1=\widehat\fA_{1,+}\oplus\widehat\fA_{1,-}$, and the even and odd parts of $\widehat\fA_0$ and $\widehat\fA_1$ are related by
\ba
\label{A10}
\widehat\fA_{1,+}
&=\widehat\fA_{0,+},\\
\label{A10T}
\widehat\fA_{1,-}
&=\widehat\fA_{0,-}T.
\ea
Moreover, $\wh\fA$ also has the decomposition
\ba
\label{decAA1}
\widehat\fA
=\widehat\fA_1\oplus\widehat\fA_1T.
\ea
\en
\ep

\bprf
\ref{dec-a}\, 
Let $\widehat\Theta_0:=\psi\circ\Theta\circ\psi^{-1}\in\sAut(\widehat\fA_0)$ and let $A\in\widehat\fA$. Then, Lemma \ref{lem:T} \ref{lem:T-b} implies that there exist $B,C\in\widehat\fA_0$ such that $A=B+CT$ and that this decomposition is unique because $\widehat\fA_0\cap\widehat\fA_0T=\{0\}$. Hence, for all $A=B+CT\in\widehat\fA$ with $B,C\in\widehat\fA_0$, we define $\widehat\Theta:\widehat\fA\to\widehat\fA$ by
\ba
\label{def-ThetaHat}
\widehat\Theta(A)
&:=\widehat\Theta_0(B)+\widehat\Theta_0(C)T, 
\ea
and, since $\widehat\Theta_0\in\sAut(\widehat\fA_0)$ and due to Proposition \ref{prop:rot} \ref{rot:spin} and Lemma \ref{lem:T} \ref{lem:T-a}, we get $\widehat\Theta\in\sHom(\widehat\fA)$.  For instance, since, for all $B,C\in\widehat\fA_0$, we can write 
$\widehat\Theta((B+CT)^\ast)
=\widehat\Theta(B^\ast+\psi(\Theta'(\psi^{-1}(C^\ast)))T)
=\widehat\Theta_0(B^\ast)+\widehat\Theta_0(\psi(\Theta'(\psi^{-1}(C^\ast))))T
=\widehat\Theta_0(B^\ast)+\psi(\Theta(\Theta'(\psi^{-1}(C^\ast))))T$ 
and 
$\widehat\Theta(B+CT)^\ast
=\widehat\Theta_0(B^\ast)+\psi(\Theta'(\Theta(\psi^{-1}(C^\ast))))T$, \eqref{CommTT'} implies that $\wh\Theta$ preserves the involution. Moreover, Lemma \ref{lem:T} \ref{lem:T-b} also implies that $\widehat\Theta$ is injective and, since $B+CT=\widehat\Theta(\widehat\Theta_0^{-1}(B)+\widehat\Theta_0^{-1}(C)T)$ for all $B,C\in\widehat\fA_0$, we get $\widehat\Theta\in\sAut(\widehat\fA)$ (see Figure \ref{fig:ThetaHat}). 
\begin{figure}
\begin{center}
\begin{tikzcd}
\fA\arrow[r, tail, two heads, "\textstyle\psi"] \arrow[d, tail, two heads, "\textstyle\Theta"'] 
	& \widehat\fA_0 \arrow[d, tail, two heads, "\textstyle\widehat\Theta_0"]\arrow[r, hookrightarrow] 		& \widehat\fA=\widehat\fA_0\oplus\widehat\fA_0 T\arrow[d, tail, two heads, "\textstyle\widehat\Theta"]  \\
\fA\arrow[r, tail, two heads, "\textstyle\psi"]  
	& \widehat\fA_0\arrow[r, hookrightarrow] 
	& \widehat\fA=\widehat\fA_0\oplus\widehat\fA_0 T 
\end{tikzcd}
\caption{The construction of $\widehat\Theta\in\sAut(\widehat\fA)$ from (the proof of) Proposition \ref{prop:dec} \ref{dec-a}.}
\label{fig:ThetaHat}
\end{center}
\end{figure}
As for \eqref{dec-a1} and \eqref{dec-a2}, we have $\widehat\Theta(\psi(A))=\widehat\Theta_0(\psi(A))=\psi(\Theta(A))$ for all $A\in\fA$ and $\widehat\Theta(T)=\widehat\Theta_0(1)T=T$, respectively. 
Finally, Lemma \ref{lem:T} \ref{lem:T-b},  \eqref{dec-a1}, and \eqref{dec-a2} yield that $\widehat\Theta$ is unique.

\ref{dec-b}\, 
Due to \eqref{dec-a1}, we have $\widehat\Theta(\widehat\fA_0)\subseteq\widehat\fA_0$. Moreover, using \eqref{def-nihil}, \eqref{T-3}, \eqref{def-ThetaHat}, \eqref{CommTT'}, \eqref{Theta-sigma}, and \eqref{T'inv}, we get, for all $x\in\Z$,
\ba
\label{Theta-ax}
\widehat\Theta(a_x)
&=\widehat\Theta(T\psi(S_x\sigma_-^{(x)}))\nonumber\\
&=\widehat\Theta(\psi(\Theta'(S_x\sigma_-^{(x)}))T)\nonumber\\
&=\widehat\Theta_0(\psi(\Theta'(S_x\sigma_-^{(x)})))T\nonumber\\
&=\psi(\Theta(\Theta'(S_x\sigma_-^{(x)})))T\nonumber\\
&=\psi(\Theta'(\Theta(S_x\sigma_-^{(x)})))T\nonumber\\
&=-\psi(\Theta'(S_x\sigma_-^{(x)}))T\nonumber\\
&=-T\psi(S_x\sigma_-^{(x)})\nonumber\\
&=-a_x.
\ea
Hence, $\widehat\Theta(a_x^\ast)=-a_x^\ast$ for all $x\in\Z$ and $\widehat\Theta(\Pol(\fB))\subseteq\Pol(\fB)$ due to \eqref{Pol}. Since $\widehat\Theta$ is continuous (see \eqref{cnt}), we get $\widehat\Theta(\widehat\fA_1)\subseteq\widehat\fA_1$. 
Moreover, $\widehat\fA_+$ is clearly a unital \str subalgebra of $\widehat\fA$ which is also closed due to the continuity of $\widehat\Theta$. Analogously, since $\widehat\fA_0$ and $\widehat\fA_1$ are unital \Cs subalgebras of $\widehat\fA$,  it follows that $\widehat\fA_{0,+}$ and $\widehat\fA_{1,+}$ are unital \Cs subalgebras of $\widehat\fA$, too.
Similarly, $\widehat\fA_-$, $\widehat\fA_{0,-}$, and $\widehat\fA_{1,-}$ are invariant under the addition, the scalar multiplication, and the involution of $\widehat\fA$ but they are, in general, not invariant under the multiplication on $\widehat\fA$. However, they are also closed with respect to the norm of $\widehat\fA$ due again to the continuity of $\widehat\Theta$. 

\ref{dec-c}\,
 For all $A\in\widehat\fA$, we have $A=A_++A_-$, where we set
 \ba
 \label{Apm}
 A_\pm
 :=\frac{A\pm\widehat\Theta(A)}{2}
 \in\widehat\fA_\pm.
 \ea
Since $\widehat\fA_\pm$ are both vector subspaces of $\widehat\fA$ due to \ref{dec-b}, we have $\widehat\fA=\widehat\fA_+ +\widehat\fA_-$. Moreover, since $\widehat\fA_+\cap\widehat\fA_-=\{0\}$, we get $\widehat\fA=\widehat\fA_+\oplus\widehat\fA_-$. The arguments for $\widehat\fA_0$ and $\widehat\fA_1$ are analogous.

We next show that $\widehat\fA_{1,+}\subseteq\widehat\fA_{0,+}$ and $\widehat\fA_{1,-}\subseteq\widehat\fA_{0,-}T$. To this end, we first note that, since $\widehat\Theta(\Pol(\fB))\subseteq\Pol(\fB)$ due to \eqref{Theta-ax}, we can define $\Pol(\fB)_\pm:=\{A\in\Pol(\fB)\,|\,\widehat\Theta(A)=\pm A\}$. Let $P\in\Pol(\fB)$. Hence, using \eqref{Pol} (see there for the notation) and \eqref{Apm}, we have
\ba
\label{P+}
 P_+
 &=\alpha 1+\sum_{\substack{l\in\num{1}{N}\\\textup{$l$ even}}}\sum_{\substack{X_l\in\Z^l\\ \sharp_l\in\{\pm\}^l}}\alpha_{X_l, \sharp_l} \prod_{i\in\num{1}{l}} a_{x_{l, i}}^{\sharp_{l, i}},\\
 \label{P-}
 P_-
 &=\sum_{\substack{l\in\num{1}{N}\\\textup{$l$ odd}}}\sum_{\substack{X_l\in\Z^l\\ \sharp_l\in\{\pm\}^l}}\alpha_{X_l, \sharp_l} \prod_{i\in\num{1}{l}} a_{x_{l, i}}^{\sharp_{l, i}}.
\ea
Since, for all $x\in\Z$, we have  $a_x^\pm=T\psi(S_x\sigma_\pm^{(x)})=\psi(\Theta'(S_x\sigma_\pm^{(x)}))T$ due to \eqref{T-3}, $\psi(S_x\sigma_\pm^{(x)})\in\widehat\fA_{0,-}$ due to \eqref{dec-a1}, and $\psi(\Theta'(S_x\sigma_\pm^{(x)}))\in\widehat\fA_{0,-}$ due to \eqref{CommTT'}, \eqref{T-1} implies that $\prod_{i\in\num{1}{l}} a_{x_{l, i}}^{\sharp_{l, i}}\in\widehat\fA_{0,+}$ if $l$ is even and that $\prod_{i\in\num{1}{l}} a_{x_{l, i}}^{\sharp_{l, i}}\in\widehat\fA_{0,-}T$ if $l$ is odd. Hence, since we know from \ref{dec-b} that $\widehat\fA_{0,\pm}$ are vector subspaces of $\widehat\fA$, \eqref{P+} and \eqref{P-} yield $P_+\in\widehat\fA_{0,+}$ and $P_-\in\widehat\fA_{0,-}T$, respectively, \ie, 
\ba
\label{Pol+}
\Pol(\fB)_+
&\subseteq \widehat\fA_{0,+},\\
\label{Pol-}
\Pol(\fB)_-
&\subseteq\widehat\fA_{0,-}T.
\ea
Next, let $A\in\widehat\fA_1$. Due to \eqref{cloPol}, there exists a sequence $(P_n)_{n\in\N}$ in $\Pol(\fB)$ such that $\|A-P_n\|\to0$ for $n\to\infty$. Since,  for all $n\in\N$, we have $P_{n,+}\in\widehat\fA_{0,+}$ due to \eqref{Pol+} and $P_{n,-}T\in\widehat\fA_{0,-}$ due to \eqref{Pol-}, we get, for all $n\in\N$, that $\|A_+-P_{n,+}\|\le \|A-P_n\|/2+\|\widehat\Theta(A)-\widehat\Theta(P_n)\|/2=\|A-P_n\|$ (using \eqref{piIso}) and $\|A_-T-P_{n,-}T\|\le\|A-P_n\|\|T\|$. Hence, since $\widehat\fA_{0,\pm}$ are Banach spaces with respect to the norm of $\widehat\fA$ (due to \ref{dec-b}), we find $A_+\in\widehat\fA_{0,+}$ and $A_-T\in\widehat\fA_{0,-}$, \ie, $\widehat\fA_{1,+}\subseteq\widehat\fA_{0,+}$ and $\widehat\fA_{1,-}\subseteq\widehat\fA_{0,-}T$. Conversely, we now show that $\widehat\fA_{0,+}\subseteq\widehat\fA_{1,+}$ and $\widehat\fA_{0,-}\subseteq\widehat\fA_{1,-}T$. To this end, let $\Lambda=\{x_1,\ldots,x_n\}\in\Fin(\Z)$ for some $n\in\N$ be such that $x_1<\ldots<x_n$ if $n\ge 2$. Since $\{\xi_{x_1}^{-1}(\sigma_{\kp_1})\otimes\ldots\otimes \xi_{x_n}^{-1}(\sigma_{\kp_n})\}_{\kp_1,\ldots, \kp_n\in\num{0}{3}}$ is a basis of $\fA_\Lm$ (due to \eqref{KTiso}), $\{\sigma_{\kp_1}^{(x_1)}\hspace{-1mm}\ldots\sigma_{\kp_n}^{(x_n)}\}_{\kp_1,\ldots, \kp_n\in\num{0}{3}}$ is a basis of $\wt\fA_\Lm$. Hence, since any $A\in\wt\fA_\Lm\subseteq\fA$ has an expansion 
$A=\sum_{K\in\num{0}{3}^n}a_K\prod_{i\in\num{1}{n}}\sigma_{\kp_i}^{(x_i)}$ for some $a_K\in\C$ with $K=(\kp_1,\ldots,\kp_n)\in\num{0}{3}^n$, any $B\in\psi(\wt\fA_\Lm)\subseteq\widehat\fA_0$ reads, for some $a_K\in\C$ with $K=(\kp_1,\ldots,\kp_n)\in\num{0}{3}^n$,
\ba
\label{BExp}
B
=\sum_{K\in\num{0}{3}^n} \hspace{-1mm}a_K\hspace{0.5mm} \prod\nolimits_{i\in\num{1}{n}}\psi(\sigma_{\kp_i}^{(x_i)}).
\ea
Moreover, since \eqref{dec-a1} and \eqref{Theta-sigma} imply that, for all $\kp_1,\ldots, \kp_n\in\num{0}{3}$,
\ba
\label{Theta-psi}
\widehat\Theta\big(\prod\nolimits_{i\in\num{1}{n}}\psi(\sigma_{\kp_i}^{(x_i)})\big) 
=\big(\prod\nolimits_{i\in\num{1}{n}}\veps_{\kp_i}\big) \prod\nolimits_{i\in\num{1}{n}}\psi(\sigma_{\kp_i}^{(x_i)}),
\ea
we get $\widehat\Theta(\psi(\wt\fA_\Lm))\subseteq\psi(\wt\fA_\Lm)$ which allows us to define $\psi(\wt\fA_\Lm)_\pm:=\{B\in\psi(\wt\fA_\Lm)\,|\,\widehat\Theta(B)=\pm B\}\subseteq\widehat\fA_{0,\pm}$. 
With the help of \eqref{BExp}, \eqref{Theta-psi}, and \eqref{Apm}, we then have
\ba
\label{B+}
B_+
&=\sum_{\substack{K\in\num{0}{3}^n\\ C_K\,\textup{even}}}\hspace{-1mm}a_K\hspace{0.5mm}
\prod\nolimits_{i\in\num{1}{n}}\psi(\sigma_{\kp_i}^{(x_i)}),\\
\label{B-}
B_-
&=\sum_{\substack{K\in\num{0}{3}^n\\ C_K\,\textup{odd}}}\hspace{-1mm}a_K\hspace{0.5mm}\prod\nolimits_{i\in\num{1}{n}}\psi(\sigma_{\kp_i}^{(x_i)}),
\ea
where we set $C_K:=\card(\{i\in\num{1}{n}\,|\,\kp_i\in\num{1}{2}\})$. We next express $\psi(\sigma_{\kp}^{(x)})$ for all $\kp\in\num{0}{3}$ and all $x\in\Z$ through \eqref{def-nihil}. Using \eqref{spin-i} for the 0-direction, \eqref{def-nihil}, \eqref{def-raislow}, \eqref{T-3}, and \eqref{Theta'-sigma} for the 1- and 2-direction (starting from $a_x\pm a_x^\ast$), and \eqref{def-nihil}, \eqref{T-3}, \eqref{Theta'-sigma}, \eqref{S-5}, and $\sigma_+^{(x)}\sigma_-^{(x)}=(\sigma_3^{(x)}+1)/2$ for all $x\in\Z$ for the 3-direction (starting from $a_x^\ast a_x$), we get, for all $x\in\Z$, 
\ba
\label{sig0}
\psi(\sigma_0^{(x)})
&=1,\\
\label{sig1}
\psi(\sigma_1^{(x)})
&=T\psi(S_x) (a_x+a_x^\ast),\\
\label{sig2}
\psi(\sigma_2^{(x)})
&=T\psi(S_x) (a_x-a_x^\ast),\\
\label{sig3}
\psi(\sigma_3^{(x)})
&=2a_x^\ast a_x-1,
\ea
and we note that, due to \eqref{sig3} and \eqref{def-Sx}, $\psi(S_x)\in\Pol(\fB)_+$ for all $x\in\Z$. Hence, with the help of \eqref{T-3}, \eqref{T-1}, \eqref{Theta'-sigma},  $T a_x^\pm T=\veps_{xx} a_x^\pm$ for all $x\in\Z$ due to \eqref{Theta'sigma}, and \eqref{Theta-ax}, it follows that $\prod_{i\in\num{1}{n}} \psi(\sigma_{\kp_i}^{(x_i)})\in\widehat\fA_{1,+}$ if $C_K$ is even and that $\prod_{i\in\num{1}{n}} \psi(\sigma_{\kp_i}^{(x_i)})\in\widehat\fA_{1,-}T$ if $C_K$ is odd. Hence, since we know from \ref{dec-b} that $\widehat\fA_{1,\pm}$ are vector subspaces of $\widehat\fA$, \eqref{B+} and \eqref{B-} yield $B_+\in\widehat\fA_{1,+}$ and $B_-\in\widehat\fA_{1,-}T$, respectively, \ie, we get,  for all $\Lm\in\Fin(\Z)$,
\ba
\label{psi+}
\psi(\wt\fA_\Lm)_+
&\subseteq\widehat\fA_{1,+},\\
\label{psi-}
\psi(\wt\fA_\Lm)_-
&\subseteq\widehat\fA_{1,-}T.
\ea
Now, let $B\in\widehat\fA_0$. Then, there exists $A\in\fA$ such that $B=\psi(A)$ and, due to \eqref{uhf}, there exists a sequence $(\Lm_n)_{n\in\N}$ in $\Fin(\Z)$ and a sequence $(A_n)_{n\in\N}$ in $\fA$ with $A_n\in\wt\fA_{\Lm_n}$ for all $n\in\N$ such that $\|A-A_n\|\to0$ for $n\to\infty$. Hence, using \eqref{piIso}, we have $\|B-\psi(A_n)\|\to0$ for $n\to\infty$. 
Since,  for all $n\in\N$, we have $\psi(A_n)_+\in\widehat\fA_{1,+}$ due to \eqref{psi+} and $\psi(A_n)_-T\in\widehat\fA_{1,-}$ due to \eqref{psi-}, 
we get, for all $n\in\N$, that $\|B_+-\psi(A_n)_+\|\le \|B-\psi(A_n)\|$ and $\|B_-T-\psi(A_n)_-T\|\le\|B-\psi(A_n)_+\|\|T\|$.  Since $\widehat\fA_{1,\pm}$ are Banach spaces with respect to the norm of $\widehat\fA$ (due to \ref{dec-b}), we find $B_+\in\widehat\fA_{1,+}$ and $B_-T\in\widehat\fA_{1,-}$, \ie, $\widehat\fA_{0,+}\subseteq\widehat\fA_{1,+}$ and $\widehat\fA_{0,-}\subseteq\widehat\fA_{1,-}T$. Hence, we arrive at \eqref{A10} and \eqref{A10T}.

Finally, since Lemma \ref{lem:T} \ref{lem:T-b}, the foregoing decomposition $\widehat\fA_0=\widehat\fA_{0,+}\oplus\widehat\fA_{0,-}$, and \eqref{T-1} yield
$\widehat\fA
=\widehat\fA_0\oplus\widehat\fA_0T
=(\widehat\fA_{0,+}\oplus \widehat\fA_{0,-})\oplus ((\widehat\fA_{0,+}\oplus \widehat\fA_{0,-})T)
=(\widehat\fA_{0,+}\oplus \widehat\fA_{0,-})\oplus (\widehat\fA_{0,+}T\oplus \widehat\fA_{0,-}T)
=\widehat\fA_{0,+}\oplus \widehat\fA_{0,-}\oplus \widehat\fA_{0,+}T\oplus \widehat\fA_{0,-}T$, we get, using  \eqref{A10} and \eqref{A10T}, 
\ba
\widehat\fA
=\widehat\fA_{1,+}\oplus \widehat\fA_{1,-}T\oplus \widehat\fA_{1,+}T\oplus \widehat\fA_{1,-},
\ea
\ie, we arrive at \eqref{decAA1}.
\eprf

In order to construct the Jordan-Wigner transformation, we  make use of the following local version of Definition \ref{def:CAR} (recall the notations from there).

\bd[Local CAR subalgebras]
\label{def:CARLm}
Let $\Lm\in\Fin(\Z)$ and set $\fB_\Lm:=\{a_x\,|\,x\in\Lm\}\subseteq\hA$.
\bn[label=(\alph*), ref={\it (\alph*)}]
\setlength{\itemsep}{0mm}
\item 
The \Cs subalgebra of $\widehat\fA$ generated by $\fB_\Lm$ is defined by
\ba
\label{def-A1Lm}
\widehat\fA_{1,\Lm}
&:=\hspace{-5mm}\bigcap_{\substack{\fB':\,\textup{\Cs subalgebra of }\wh\fA\\\fB_\Lm\subseteq\fB'}}
\hspace{-2mm}\fB'.
\ea

\item
A polynomial in $\widehat\fA$ generated by $\fB_\Lm$ is an element of $\widehat\fA$ of the form
\ba
\label{PolLm}
\alpha 1+\sum_{l\in\num{1}{N}}\sum_{\substack{X_l\in\Lm^l\\ \sharp_l\in\{\pm\}^l}}\alpha_{X_l, \sharp_l} \prod_{i\in\num{1}{l}} a_{x_{l, i}}^{\sharp_{l, i}},
\ea
where $\alpha\in\C$, $N\in\N$, and, for all $l\in\num{1}{N}$, $\alpha_{X_l, \sharp_l} \in\C$ for all $X_l=(x_{l,1},\ldots, x_{l,l})\in\Lm^l$ and all $\sharp_l=(\sharp_{l,1},\ldots,\sharp_{l,l})\in\{\pm\}^l$. The set of all the polynomials in $\widehat\fA$ generated by $\fB_\Lm$ is denoted by $\Pol(\fB_\Lm)$.
\en
\ed

For the following, recall that, if $\fA$ is a unital \Cs algebra and $N\in\N$, the elements of the family $\{e_{\alpha\beta}\}_{\alpha,\beta\in\num{1}{N}}\subseteq\fA$ are called $N\times N$ matrix units in $\fA$  (see, for example, \cite{Glimm60}) if, for all $\alpha,\beta,\gamma,\delta\in\num{1}{N}$, 
\ba
\label{MUnt-1}
e_{\alpha\beta} e_{\gamma\delta}
&=\delta_{\beta\gamma} e_{\alpha\delta},\\
\label{MUnt-2}
e_{\alpha\beta}^\ast
&=e_{\beta\alpha},\\
\label{MUnt-3}
\sum_{\alpha\in\num{1}{N}}e_{\alpha\alpha}
&=1.
\ea

The CAR algebra $\wh\fA_1$ has the following local structure.

\bp[Local CAR structure]
\label{prop:LocCARs}
The net $(\wh\fA_{1,\Lm})_{\Lm\in\Fin(\Z)}$ is an increasing net of unital \Cs subalgebras of $\wh\fA_1$ satisfying $\widehat\fA_{1,\Lm}=\Pol(\fB_\Lm)$ for all $\Lm\in\Fin(\Z)$ and
\ba
\label{cloCAR}
\wh\fA_1
=\clo_{\wh\fA}\left(\bigcup\nolimits_{\Lm\in\Fin(\Z)}\wh\fA_{1,\Lm}\right).
\ea
\ep

\bprf
Let $\Lm\in\Fin(\Z)$ be fixed. Then, $\wh\fA_{1,\Lm}$ as given in \eqref{def-A1Lm} defines a \str subalgebra of $\widehat\fA$ which is closed with respect to the norm of $\widehat\fA$ (as in the proof of Proposition \ref{prop:cloPol}). Moreover, the \Cs algebra $\widehat\fA_{1,\Lm}$ is unital due to \eqref{CAR-2} and the fact that the anticommutator satisfies $\{a_x, a_x^\ast\}\in\fB'$ for all $x\in\Lm$ and all \Cs subalgebras $\fB'$ of $\wh\fA$ with $\fB_\Lm\subseteq\fB'$. Since $\fB_\Lm\subseteq\fB$ (from Definition \ref{def:CAR}), we also have $\wh\fA_{1,\Lm}\subseteq\wh\fA_1$, \ie, $\wh\fA_{1,\Lm}$ is a unital \Cs subalgebra of $\wh\fA_1$. 
Furthermore, for all $\Lm,\Lm'\in\Fin(\Z)$ with $\Lm\subseteq\Lm'$, we have $\fB_\Lm\subseteq\fB_{\Lm'}$ and, thus, $\wh\fA_{1,\Lm}\subseteq\wh\fA_{1,\Lm'}$, \ie, the net $(\wh\fA_{1,\Lm})_{\Lm\in\Fin(\Z)}$ is increasing.

Next,  let $\Lm\in\Fin(\Z)$ again be fixed. Since $\Pol(\fB_\Lm)$ is a \str subalgebra of $\widehat\fA$ and since $\Pol(\fB_\Lm)\subseteq\fB'$ for all \Cs subalgebras $\fB'$ of $\widehat\fA$ satisfying $\fB_\Lm\subseteq\fB'$, it follows that $\Pol(\fB_\Lm)$ is a \str subalgebra of the \Cs algebra $\widehat\fA_{1,\Lm}$, too. Moreover, $\clo_{\widehat\fA}(\Pol(\fB_\Lm))\subseteq\widehat\fA_{1,\Lm}$ because $\widehat\fA_{1,\Lm}$ is closed and, since the \Cs subalgebra $\clo_{\widehat\fA}(\Pol(\fB_\Lm))$ of $\widehat\fA$ contains $\fB_\Lm$, we also get the inverse inclusion. We next show that $\clo_{\widehat\fA}(\Pol(\fB_\Lm))=\Pol(\fB_\Lm)$. To this end, let $P\in\Pol(\fB_\Lm)$ be written as in \eqref{PolLm}. Hence, $P$ is a (finite) linear combination of monomials of the form $\prod_{i\in\num{1}{l}} a_{y_i}^{\sharp_{i}}$, where $l\in\N$, $(y_1,\ldots,y_l)\in\Lm^l$, and $(\sharp_1,\ldots,\sharp_l)\in\{\pm\}^l$ (we use the notation from Definition \ref{def:CAR}). Using that, due to \eqref{CAR-1}-\eqref{CAR-2}, we have $\{a_x^\sharp, a_y^\flat\}=0$ for all $x,y\in\Lm$ with $x\neq y$ and all $\sharp,\flat\in\{\pm\}$, we first group together (at an arbitrary position) all annihilation and creation operators having the same site index, say  $x$, by anticommuting them across all annihilation and creation operators having a site index different from $x$. Afterwards, we may order all these groupings increasingly with respect to the total order on $\Z$ (these two operations only change the global sign of the initial monomial). Now consider a fixed grouping with site index $x$. Such a grouping has the form of a monomial $\prod_{i\in\num{1}{m}} a_x^{\flat_i}$, where $m\in\N$ and $\flat_i\in\{\pm\}$ for all $i\in\num{1}{m}$, and, using \eqref{CAR-1}-\eqref{CAR-2}, we get
\ba
\prod_{i\in\num{1}{m}} a_x^{\flat_i}
\in\big(\{0\}\cup\{f^{(x)}_{\alpha\beta}\}_{\alpha, \beta\in\num{1}{2}}\big),
\ea
where, for all $x\in\Z$ and all $\alpha,\beta\in\num{1}{2}$, the elements $f^{(x)}_{\alpha\beta}\in\wh\fA_1$ are defined by
\ba
\label{fab}
f^{(x)}_{\alpha\beta}
:=\begin{cases}
\hfill a_x^\ast a_x, & (\alpha, \beta)=(1,1),\\
\hfill a_x^\ast, & (\alpha, \beta)=(1,2),\\
\hfill a_x, & (\alpha, \beta)=(2,1),\\
\hfill a_x a_x^\ast, & (\alpha, \beta)=(2,2).
\end{cases}
\ea
A direct check yields that, for any fixed $x\in\Z$, the family $\{f^{(x)}_{\alpha\beta}\}_{\alpha, \beta\in\num{1}{2}}$ is a family of $2\times 2$ matrix units in $\wh\fA_1$. Hence, if $\Lambda=\{x_1,\ldots,x_n\}\in\Fin(\Z)$ for some $n\in\N$ is such that $x_1<\ldots<x_n$ if $n\ge 2$, any $P\in\Pol(\fB_\Lm)$ can be written as
\ba
P
=\sum_{\Gamma\in\num{1}{2}^{2n}}\alpha_\Gamma \prod_{i\in\num{1}{n}} f^{(x_i)}_{\alpha_i\beta_i},
\ea
where $\alpha_\Gamma\in\C$ for all $\Gamma=(\alpha_1,\beta_1,\ldots,\alpha_n,\beta_n)\in\num{1}{2}^{2n}$ (and where we used \eqref{CAR-2}, \ie, $f^{(x)}_{11}+f^{(x)}_{22}=1$ for all $x\in\Z$ [see \eqref{MUnt-3}], to compensate for a missing site index in the initial monomial), \ie, 
\ba
\label{Polspan}
\Pol(\fB_\Lm)
=\spa\Big(\Big\{\prod\nolimits_{i\in\num{1}{n}} f^{(x_i)}_{\alpha_i\beta_i}\Big\}
_{\Gamma\in\num{1}{2}^{2n}}\Big).
\ea
Since the argument of the linear span on the right hand side of \eqref{Polspan} contains only a finite number of elements of $\wh\fA_1$, the \str algebra $\Pol(\fB_\Lm)$ is a finite dimensional vector subspace of the Banach space $\wh\fA_1$, \ie, $\Pol(\fB_\Lm)$ is closed (with respect to the \Cs norm of $\wh\fA$ and, hence, of $\wh\fA_1$).

Finally, since $\Pol(\fB)=\bigcup_{\Lm\in\Fin(\Z)}\Pol(\fB_\Lm)$, the foregoing fact that $\Pol(\fB_\Lm)=\wh\fA_{1,\Lm}$ for all $\Lm\in\Fin(\Z)$ and \eqref{cloPol} lead to \eqref{cloCAR}.
\eprf

In the following, for all $\Lm, \Lm'\in\Fin(\Z)$ with $\Lm\subseteq\Lm'$, we denote by $i_\Lm$ and $i_{\Lm',\Lm}$ the canonical inclusion maps $\tA_\Lm\hookrightarrow\fA$ and $\tA_\Lm\hookrightarrow\tA_{\Lm'}$, respectively (see Figure \ref{fig:iso}).

We next show that $\fA$ and $\hA_1$ are isomorphic.

\bp[Spin-CAR isomorphism]
\label{prop:AA1}
\bn[label=(\alph*), ref={\it (\alph*)}]
\setlength{\itemsep}{0mm}
\item
\label{AA1-a}
For all $\Lm\in\Fin(\Z)$, there exists $\vartheta_\Lm\in$ \linebreak $\sIso(\fA_\Lm,\hA_{1,\Lm})$ such that, for all $\Lm'\in\Fin(\Z)$ with $\Lm\subseteq\Lm'$, 
\ba
\label{VthetaVi}
\vartheta_{\Lm'}\circ\vi_{\Lm',\Lm}
=\vartheta_\Lm.
\ea

\item 
\label{AA1-b}
There exists $\phi\in\sIso(\fA, \hA_1)$ such that, for all $\Lm\in\Fin(\Z)$,
\ba
\label{AA1-1}
\phi(\tA_\Lm)
&=\hA_{1,\Lm},\\
\label{AA1-2}
\phi \circ i_\Lm
&=\vartheta_\Lm\circ\vi_\Lm^{-1}.
\ea
\en
\ep

\bprf
\ref{AA1-a}\, 
Let $\Lambda=\{x_1,\ldots,x_n\}\in\Fin(\Z)$ for some $n\in\N$ with $x_1<\ldots<x_n$ if $n\ge 2$. For all $\alpha,\beta\in\num{1}{2}$, let $E_{\alpha\beta}\in\C^{2\times 2}$ be defined by
\ba
E_{11}:=\begin{bmatrix} 1&0\\0&0\end{bmatrix}, \hspace{2mm}
E_{12}:=\begin{bmatrix} 0&1\\0&0\end{bmatrix}, \hspace{2mm}
E_{21}:=\begin{bmatrix} 0&0\\1&0\end{bmatrix}, \hspace{2mm}
E_{22}:=\begin{bmatrix} 0&0\\0&1\end{bmatrix},
\ea
and note that $\{E_{\alpha\beta}\}_{\alpha,\beta\in\num{1}{2}}$ is a family of $2\times 2$ matrix units in $\C^{2\times 2}$. Hence, due to \eqref{KTiso}, $\{\xi_{x_1}^{-1}(E_{\alpha_1\beta_1})\otimes\ldots\otimes\xi_{x_n}^{-1}(E_{\alpha_n\beta_n})\}_{\Gamma\in\num{1}{2}^{2n}}$, where  $\Gamma=(\alpha_1,\beta_1,\ldots,\alpha_n,\beta_n)$, is a basis of $\fA_\Lm$. Since we want $\vartheta_\Lm$ to be a \str homomorphism, \ie, in particular, since we want $\vartheta_\Lm$ to preserve the multiplication, we cannot define $\vartheta_\Lm(\xi_{x_1}^{-1}(E_{\alpha_1\beta_1})\otimes\ldots\otimes\xi_{x_n}^{-1}(E_{\alpha_n\beta_n}))$ by $\prod_{i\in\num{1}{n}} f^{(x_i)}_{\alpha_i\beta_i}$ (the set $\{\prod_{i\in\num{1}{n}} f^{(x_i)}_{\alpha_i\beta_i}\}_{\Gamma\in\num{1}{2}^{2n}}$ is a basis of $\hA_{1,\Lm}$, see below before \eqref{A1Lmspan} and after \eqref{eGstar}) since the family of $2\times 2$ matrix units \eqref{fab} is not commuting at different sites, \ie, for all $x,y\in\Z$ with $x\neq y$ and all $\alpha,\beta,\gamma,\delta\in\num{1}{2}$, 
\ba
f^{(x)}_{\alpha\beta}f^{(y)}_{\gamma\delta}
=(-1)^{(\alpha+\beta)(\gamma+\delta)}f^{(y)}_{\gamma\delta}f^{(x)}_{\alpha\beta}.
\ea
Therefore, for all $i\in\num{1}{n}$ and all $\alpha,\beta\in\num{1}{2}$, let $e_{\alpha\beta}^{(i)}\in\wh\fA$ be defined by
\ba
\label{jw-1}
e_{\alpha\beta}^{(i)}
:=\begin{cases}
\hfill a_{x_i}^\ast a_{x_i}, & (\alpha, \beta)=(1,1),\\
\hfill\psi(R_i) a_{x_i}^\ast, & (\alpha, \beta)=(1,2),\\
\hfill\psi(R_i)a_{x_i}, & (\alpha, \beta)=(2,1),\\
\hfill a_{x_i} a_{x_i}^\ast, & (\alpha, \beta)=(2,2),
\end{cases}
\ea
where, for all $i\in\num{1}{n}$, we define $R_i\in\fA$ by (compare to \eqref{def-Sx}) 
\ba
\label{jw-2}
R_i
:=\begin{cases}
\hfill 1, & i=1,\\
\prod_{j\in\num{1}{i-1}}\sigma_3^{(x_j)}, & i\in\num{2}{n}.
\end{cases}
\ea
Using that $a_{x_i}^\sharp\psi(R_j)=\veps_{x_jx_i} \psi(R_j)a_{x_i}^\sharp$ for all $i,j\in\num{1}{n}$ and all $\sharp\in\{\pm\}$ and that, for all  $i,j\in\num{1}{n}$, we have $[\psi(R_i), \psi(R_j)]=0$ and $(\psi(R_i))^2=1$, a direct computation yields that $\{e_{\alpha\beta}^{(i)}\}_{\alpha,\beta\in\num{1}{2}}$ is a family of $2\times 2$ matrix units which commutes at different sites, \ie, for all $i,j\in\num{1}{n}$ with $i\neq j$ and all $\alpha,\beta,\gamma,\delta\in\num{1}{2}$, 
\ba
\label{eiCom}
e_{\alpha\beta}^{(i)}e_{\gamma\delta}^{(j)}
=e_{\gamma\delta}^{(j)}e_{\alpha\beta}^{(i)}.
\ea
Hence, setting, for all $\Gamma=(\alpha_1,\beta_1,\ldots,\alpha_n,\beta_n)\in\num{1}{2}^{2n}$, 
\ba
e_\Gamma:=\prod_{i\in\num{1}{n}}e_{\alpha_i\beta_i}^{(i)},
\ea
we define the map $\vartheta_\Lm:\fA_\Lm\to\wh\fA$, for all $\Gamma\in\num{1}{2}^{2n}$, by
\ba
\label{vtheta}
\vartheta_\Lm(\xi_{x_1}^{-1}(E_{\alpha_1\beta_1})\otimes\ldots\otimes\xi_{x_n}^{-1}(E_{\alpha_n\beta_n}))
:=e_\Gamma,
\ea
and by linear extension to the whole of $\fA_\Lm$. 
We next verify that $\vartheta_\Lm$ has the required properties. First, since $\wh\fA_{1,\Lm}=\Pol(\fB_\Lm)$ due to Proposition \ref{prop:LocCARs}, since \eqref{Polspan} holds, 
and since  (if $n\ge2$) $\prod_{i\in\num{1}{n}}f_{\alpha_i\beta_i}^{(x_i)}=e^{(1)}_{\alpha_1\beta_1}(e^{(1)}_{11}-e^{(1)}_{22})^{N_1}e^{(2)}_{\alpha_2\beta_2}(e^{(2)}_{11}-e^{(2)}_{22})^{N_2}\ldots e^{(n-1)}_{\alpha_{n-1}\beta_{n-1}}(e^{(n-1)}_{11}-e^{(n-1)}_{22})^{N_{n-1}}e^{(n)}_{\alpha_n\beta_n}$ for all $\Gamma\in\num{1}{2}^{2n}$ with $N_i:=\sum_{j\in\num{i+1}{n}}(\alpha_j+\beta_j)$ for all $i\in\num{1}{n-1}$, where we used $f^{(x_i)}_{\alpha\beta}=\psi(R_i)^{\alpha+\beta}e^{(i)}_{\alpha\beta}$ for all $i\in\num{1}{n}$ and all $\alpha,\beta\in\num{1}{2}$, the fact that $\psi(\sigma_3^{(x_i)})=2a_{x_i}^\ast a_{x_i}-1=e^{(i)}_{11}-e^{(i)}_{22}$ for all $i\in\num{1}{n}$, and \eqref{eiCom}, \eqref{MUnt-1} and $(e^{(i)}_{11}-e^{(i)}_{22})^{N}=1$ for all $i\in\num{1}{n}$ if $N\in\N$ is even lead to $\prod_{i\in\num{1}{n}}f_{\alpha_i\beta_i}^{(x_i)}=\pm\prod_{i\in\num{1}{n}}e_{\alpha_i\beta_i}^{(i)}$ for all $\Gamma\in\num{1}{2}^{2n}$. Hence, we get 
\ba
\label{A1Lmspan}
\wh\fA_{1,\Lm}
=\spa\Big(\{e_\Gamma\}_{\Gamma\in\num{1}{2}^{2n}}\Big),
\ea 
\ie, $\ran(\vartheta_\Lm)\subseteq\wh\fA_{1,\Lm}$. Moreover, using that,  for all $\Gamma=(\alpha_1,\beta_1,\ldots,\alpha_n,\beta_n)\in\num{1}{2}^{2n}$ and all $\Gamma'=(\alpha_1',\beta_1',\ldots,\alpha_n',\beta_n')\in\num{1}{2}^{2n}$, 
\ba
\label{eGeG'}
e_\Gamma e_{\Gamma'}
&=\Big(\prod\nolimits_{i\in\num{1}{n}}\delta_{\alpha_i' \beta_i}\Big) e_{(\alpha_1,\beta_1',\ldots,\alpha_n,\beta_n')},\\
\label{eGstar}
e_\Gamma^\ast
&=e_{(\beta_1,\alpha_1,\ldots,\beta_n,\alpha_n)},
\ea
a direct check yields that $\vartheta_\Lm\in\sHom(\fA_\Lm, \wh\fA_{1,\Lm})$.  Next, we want to show that $\{e_\Gamma\}_{\Gamma\in\num{1}{2}^{2n}}$ is a basis of $\wh\fA_{1,\Lm}$. Due to \eqref{A1Lmspan}, it is enough to show that $\{e_\Gamma\}_{\Gamma\in\num{1}{2}^{2n}}$ is linearly independent. To this end, let $\sum_{\Gm\in\num{1}{2}^{2n}}\alpha_\Gm e_\Gm=0$, where $\alpha_\Gm\in\C$ for all $\Gm\in\num{1}{2}^{2n}$. Then, for all $\Gamma'=(\alpha_1',\beta_1',\ldots,\alpha_n',\beta_n')\in\num{1}{2}^{2n}$ and all $\Gamma''=(\alpha_1'',\beta_1'',\ldots,\alpha_n'',\beta_n'')\in\num{1}{2}^{2n}$, multiplying this equation by $e_{\Gm'}$ from the left and by $e_{\Gm''}$ from the right and using \eqref{eGeG'}, we get $\alpha_{(\beta_1',\alpha_1'',\ldots,\beta_n',\alpha_n'')}e_{(\alpha_1',\beta_1'',\ldots,\alpha_n',\beta_n'')}=0$, \ie, $\alpha_\Gm=0$ for all $\Gamma\in \num{1}{2}^{2n}$ (if there exists $\Gm_0\in\num{1}{2}^{2n}$ such that $e_{\Gm_0}=0$, then $e_\Gm=0$ for all $\Gamma\in \num{1}{2}^{2n}$ due to \eqref{eGeG'} and, hence, $\wh\fA=\{0\}$ due to \eqref{A1Lmspan}, \eqref{cloCAR}, and \eqref{decAA1}). Therefore, since, due to \eqref{vtheta}, $\vartheta_\Lm$ maps a basis of $\fA_\Lm$ to a basis of $\wh\fA_{1,\Lm}$, it follows that $\vartheta_\Lm$ is bijective and we arrive at $\vartheta_\Lm\in\sIso(\fA_\Lm, \wh\fA_{1,\Lm})$.

As for \eqref{VthetaVi}, let now $\Lm,\Lm'\in\Fin(\Z)$ with $\Lm\subseteq\Lm'$ be such that $\Lm'=\{x_1,\ldots,x_{n'}\}$ for some $n'\in\N$ (\ie, $n'\ge 2$) with $x_1<\ldots<x_{n'}$ and let $n:=\card(\Lm)$. Moreover, let $\sigma\in\mS_{n'}$ be the unique permutation satisfying $x_{\sigma(i)}\in\Lm$ for all $i\in\num{1}{n}$ with $\sigma(i)<\sigma(i+1)$ for all  $i\in\num{1}{n-1}$ if $n\ge 2$ and $x_{\sigma(i)}\in\Lm'\setminus\Lm$ for all $i\in\num{n+1}{n'}$ with $\sigma(i)<\sigma(i+1)$ for all $i\in\num{n+1}{n'-1}$ if $n'\ge n+2$, see Figure \ref{fig:perm}. Then, on the one hand, we have $\vartheta_\Lm(\xi_{x_{\sigma(1)}}^{-1}(E_{\alpha_{\sigma(1)}\beta_{\sigma(1)}})\otimes\ldots\otimes\xi_{x_{\sigma(n)}}^{-1}(E_{\alpha_{\sigma(n)}\beta_{\sigma(n)}}))=e_{(\alpha_{\sigma(1)},\beta_{\sigma(1)},\ldots, \alpha_{\sigma(n)},\beta_{\sigma(n)})}$ due to \eqref{vtheta}. On the other hand, using \eqref{piLL}-\eqref{viLL}, we can write $\vi_{\Lm',\Lm}(\xi_{x_{\sigma(1)}}^{-1}(E_{\alpha_{\sigma(1)}\beta_{\sigma(1)}})\otimes\ldots\otimes\xi_{x_{\sigma(n)}}^{-1}(E_{\alpha_{\sigma(n)}\beta_{\sigma(n)}}))=A_1\otimes\ldots\otimes A_{n'}$, where $A_i\in\{\xi_{x_i}^{-1}(E_{\alpha_i\beta_i}), 1_{\{x_i\}}\}$ for all $i\in\num{1}{n'}$. Hence, since, for all $i\in\num{1}{n'}$, we have $1_{\{x_i\}}=\xi_{x_i}^{-1}(E_{11}+E_{22})$ and $e_{11}^{(i)}+e_{22}^{(i)}=1$ (see \eqref{MUnt-3}),  \eqref{vtheta} leads to $\vartheta_{\Lm'}(\vi_{\Lm',\Lm}(\xi_{x_{\sigma(1)}}^{-1}(E_{\alpha_{\sigma(1)}\beta_{\sigma(1)}})\otimes\ldots\otimes\xi_{x_{\sigma(n)}}^{-1}(E_{\alpha_{\sigma(n)}\beta_{\sigma(n)}})))=e_{(\alpha_{\sigma(1)},\beta_{\sigma(1)},\ldots, \alpha_{\sigma(n)},\beta_{\sigma(n)})}$, \ie, we find \eqref{VthetaVi}.

\ref{AA1-b}\, 
Since $\wh\fA_1$ is a unital \Cs algebra due to Proposition \ref{prop:cloPol} and $(\wh\fA_{1,\Lm})_{\Lm\in\Fin(\Z)}$ an increasing net of unital \Cs subalgebras of $\wh\fA_1$ satisfying \eqref{cloCAR}, we know from \cite{Sakai} that part \ref{AA1-a}\, leads to the desired assertion. For the sake of completeness, we briefly discuss the construction from \cite{Sakai}. We start off by defining $\phi_\Lm:=\vartheta_\Lm\circ\vi_\Lm^{-1}\in\sIso(\wt\fA_\Lm,\wt\fA_{1,\Lm})$ for all $\Lm\in\Fin(\Z)$ (see Figure \ref{fig:iso}). Next, let $\Lm, \Lm'\in\Fin(\Z)$ with $\Lm\subseteq\Lm'$. Then, using \eqref{indlim-2} and \eqref{VthetaVi}, we get, for all $A\in\wt\fA_\Lm$,
\ba
\label{phiLmLm'}
\phi_{\Lm'}(A)
&=\vartheta_{\Lm'}(\vi_{\Lm'}^{-1}(A))\nonumber\\
&=\vartheta_{\Lm'}(\vi_{\Lm'}^{-1}(\vi_{\Lm'}(\vi_{\Lm',\Lm}(\vi_{\Lm}^{-1}(A)))))\nonumber\\
&=\vartheta_{\Lm'}(\vi_{\Lm',\Lm}(\vi_{\Lm}^{-1}(A)))\nonumber\\
&=\phi_\Lm(A),
\ea
 \ie, we have $\phi_\Lm=\phi_{\Lm'}\circ i_{\Lm',\Lm}$ for all $\Lm, \Lm'\in\Fin(\Z)$ with $\Lm\subseteq\Lm'$. 
 
 Next, we define the map $\phi':\bigcup_{\Lm\in\Fin(\Z)}\tA_\Lm\to\bigcup_{\Lm\in\Fin(\Z)}\hA_{1,\Lm}$ by $\phi'(A):=\phi_\Lm(A)$ for all $\Lm\in\Fin(\Z)$ and all $A\in\tA_\Lm$. If $A\in\tA_\Lm$ for some $\Lm\in\Fin(\Z)$ and $A\in\tA_{\Lm'}$ for some $\Lm'\in\Fin(\Z)$, there exists $\Lm''\in\Fin(\Z)$ such that $\Lm,\Lm'\subseteq\Lm''$ and, due to \eqref{phiLmLm'}, we get $\phi_\Lm(A)=\phi_{\Lm''}(A)$ and $\phi_{\Lm'}(A)=\phi_{\Lm''}(A)$, \ie, $\phi'$ is well-defined.  Moreover, since $\phi_\Lm\in\sIso(\tA_\Lm,\hA_{1,\Lm})$ for all $\Lm\in\Fin(\Z)$, the definition of $\phi'$ directly implies that $\phi'$ preserves the scalar multiplication and the involution. As for the addition and the multiplication, if $A,B\in\bigcup_{\Lm\in\Fin(\Z)}\wt\fA_\Lm$, there exist $\Lm, \Lm'\in\Fin(\Z)$ such that $A\in\tA_\Lm$ and $B\in\tA_{\Lm'}$ and, hence, there exists $\Lm''\in\Fin(\Z)$ with $\Lm,\Lm'\subseteq\Lm''$ and $A, B\in\tA_{\Lm''}$ since $(\wt\fA_\Lm)_{\Lm\in\Fin(\Z)}$ is an increasing net. Hence, due to \eqref{phiLmLm'}, $\phi'$ preserves the addition and the multiplication, too, \ie, $\phi'$ is a \str homomorphism.  In addition, if $A\in\tA_\Lm$ for some $\Lm\in\Fin(\Z)$, we have $\|\phi'(A)\|=\|\phi_\Lm(A)\|=\|A\|$ due to $\phi_\Lm\in\sIso(\tA_\Lm,\hA_{1,\Lm})$ and \eqref{piIso}, \ie, $\phi'$ is isometric and, therefore, injective. On the other hand, if $B\in\hA_{1,\Lm}$ for some $\Lm\in\Fin(\Z)$, we can write $B=\phi_\Lm(\phi_\Lm^{-1}(B))=\phi'(\phi_\Lm^{-1}(B))$, \ie,  $\phi'$ is also surjective and we arrive at $\phi'\in\sIso(\bigcup_{\Lm\in\Fin(\Z)}\tA_\Lm, \bigcup_{\Lm\in\Fin(\Z)}\hA_{1,\Lm})$. 

Now, we define the map $\phi: \fA\to\hA_1$, for all $A\in\fA$, by
\ba
\label{phiAJW}
\phi(A)
:=\lim_{n\to\infty}\phi'(A_n),
\ea
where $(\Lm_n)_{n\in\N}$ is a sequence in $\Fin(\Z)$ and $(A_n)_{n\in\N}$ a sequence in $\fA$ with $A_n\in\tA_{\Lm_n}$ for all $n\in\N$ such that $\|A_n-A\|\to 0$ for $n\to\infty$ (see \eqref{uhf}). Note that the limit in \eqref{phiAJW} exists in the \Cs algebra $\hA_1$ since $\phi'$ is isometric. Moreover, the limit in \eqref{phiAJW} is independent of the choice of the sequence $(A_n)_{n\in\N}$ because $\|\phi(A)-\phi'(B_n)\|\le\|\phi(A)-\phi'(A_n)\|+\|A_n-B_n\|$ for all $n\in\N$, where $(\Lm_n')_{n\in\N}$ is a sequence in $\Fin(\Z)$ and $(B_n)_{n\in\N}$ a sequence in $\fA$ with $B_n\in\tA_{\Lm'_n}$ for all $n\in\N$ such that $\|B_n-A\|\to 0$ for $n\to\infty$.
 Next, the fact that $\phi'$ is a \str homomorphism immediately yields that $\phi\in\sHom(\fA,\hA_1)$ and we want to show that $\phi$ is bijective, too. First,  $\phi$ is isometric since $\|\phi(A)\|=\lim_{n\to\infty}\|\phi'(A_n)\|=\lim_{n\to\infty}\|A_n\|=\|A\|$ for all $A\in\fA$ and all sequences $(\Lm_n)_{n\in\N}$ in $\Fin(\Z)$ and $(A_n)_{n\in\N}$ in $\fA$ with $A_n\in\tA_{\Lm_n}$ for all $n\in\N$ such that $\|A_n-A\|\to 0$ for $n\to\infty$. As for the surjectivity, if $B\in\hA_1$, there exist sequences $(\Lm_n)_{n\in\N}$ in $\Fin(\Z)$ and $(B_n)_{n\in\N}$ in $\hA_1$ with $B_n\in\hA_{1,\Lm_n}$ for all $n\in\N$ such that $\|B_n-B\|\to 0$ for $n\to\infty$. Since $\phi_{\Lm_n}^{-1}(B_n)\in\tA_{\Lm_n}$ for all $n\in\N$ and since the sequence $(\phi_{\Lm_n}^{-1}(B_n))_{n\in\N}$ in the \Cs algebra $\fA$ satisfies 
$\|\phi_{\Lm_n}^{-1}(B_n)-\phi_{\Lm_m}^{-1}(B_m)\|=\|\phi'(\phi_{\Lm_n}^{-1}(B_n)-\phi_{\Lm_m}^{-1}(B_m))\|=\|B_n-B_m\|$ for all $n,m\in\N$, there exists $A\in\fA$   such that $\|A-\phi_{\Lm_n}^{-1}(B_n)\|\to 0$ for $n\to\infty$. Therefore,
$\|B-\phi(A)\|
\le \|B-\phi'(\phi_{\Lm_n}^{-1}(B_n))\|+\|\phi(\phi_{\Lm_n}^{-1}(B_n)-A)\|
=\|B-B_n\|+\|\phi_{\Lm_n}^{-1}(B_n)-A\|$ for all  $n\in\N$ leads to $\phi\in\sIso(\fA,\hA_1)$.

Finally, since $\phi(A)=\phi'(A)=\phi_\Lm(A)\in\hA_{1,\Lm}$ for all $\Lm\in\Fin(\Z)$ and all $A\in\tA_\Lm$ and since $B=\phi_\Lm(\phi_\Lm^{-1}(B))$ for all $\Lm\in\Fin(\Z)$ and all $B\in\hA_{1,\Lm}$, we get \eqref{AA1-1}. Since, by definition, $\phi_\Lm=\vartheta_\Lm\circ\vi_\Lm^{-1}$ for all $\Lm\in\Fin(\Z)$, we also get \eqref{AA1-2}.
\eprf

\begin{figure}
\begin{center}
\begin{tikzcd}
	&\hA_{1,\Lm}
	\arrow[dd,  hookrightarrow]
	&\fA_\Lm
	\arrow[l, tail, two heads, "\textstyle\vartheta_\Lm"]
	\arrow[r, tail, two heads, "\textstyle\vi_\Lm"']
	\arrow[dd, tail, "\textstyle\vi_{\Lm',\Lm}"]	
	&\tA_\Lm
	\arrow[rd,  hookrightarrow, "\textstyle i_\Lm"]
	\arrow[dd,  hookrightarrow, "\textstyle i_{\Lm',\Lm}"]
	\arrow[ll, tail, two heads, bend right=37, "\textstyle\phi_\Lm"]
	&
	&
	\\
\hA_1
\arrow[ru,  hookleftarrow]
\arrow[rd,  hookleftarrow]
	&
	&
	&
	&\fA
	\arrow[r, tail, two heads, "\textstyle\psi"]
	\arrow[llll, tail, two heads, bend right=90, "\textstyle\phi"']
	&\hA_0
	\arrow[lllll, tail, two heads, bend left=85, "\textstyle\Phi"]
	\\
	&\hA_{1,\Lm'}
	&\fA_{\Lm'}
	\arrow[l, tail, two heads, "\textstyle\vartheta_{\Lm'}"']
	\arrow[r, tail, two heads, "\textstyle\vi_{\Lm'}"]
	&\tA_{\Lm'}
	\arrow[ru,  hookrightarrow, , "\textstyle i_{\Lm'}"']
	\arrow[ll, tail, two heads, bend left=40, "\textstyle\phi_{\Lm'}"']
	&
	&
\end{tikzcd}
\caption{The ingredients of (the proof of) Proposition \ref{prop:AA1}.}
\label{fig:iso}
\end{center}
\end{figure}

\br
The definition of the local \str isomorphism \eqref{vtheta} is the analog of the classical Jordan-Wigner transformation (see \cite{LSM1961}).
\er

The \Cs algebras $\fA$ and $\hA_1$ are isomorphic due to Proposition \ref{prop:AA1} \ref{AA1-b}. But there is no \str homomorphism which respects the spin structure in the following sense. 

\bp[Nonpreservation]
There exists no $\Pi\in\sHom(\fA,\hA_1)$ such that $\Pi(\sigma_-^{(x)})=a_x$ for at least two different $x\in\Z$. 
\ep

\bprf
Let $x,y\in\Z$ with $x\neq y$ and assume that there exists $\Pi\in\sHom(\fA,\hA_1)$ such that $\Pi(\sigma_-^{(x)})=a_x$ and $\Pi(\sigma_-^{(y)})=a_y$. Then, on the one hand, we have $[a_x,a_y]=[\Pi(\sigma_-^{(x)}), \Pi(\sigma_-^{(y)})]=\Pi([\sigma_-^{(x)},\sigma_-^{(y)}])=0$ due to \eqref{spin-n}. On the other hand, using \eqref{CAR-1}, we also have $[a_x,a_y]=2a_xa_y$, and we thus get $a_xa_y=0$. Multiplying this equation by $a_y^\ast$ once from the left and once from the right, adding the resulting two equations, and using \eqref{CAR-2}, we get $a_x=0$, \ie, $\hA=\{0\}$ again due to \eqref{CAR-2}.
\eprf

\br
Defining the maps $\C^{3\times 3}\times\fA^3\to\fA^3$, $\l\cdot,\cdot\r:\fA^3\times\fA^3\to\fA$, $\wedge:\fA^3\times\fA^3\to\fA^3$, and $\C^3\times\fA^3\to\fA$ for all $M=[m_{ij}]_{i,j\in\num{1}{3}}\in\C^{3\times 3}$, all $m=[m_i]_{i\in\num{1}{3}}\in\C^3$, and all $A=[A_i]_{i\in\num{1}{3}}, B=[B_i]_{i\in\num{1}{3}}\in\fA^3$ by $MA:=[\sum_{j\in\num{1}{3}}m_{ij}A_j]_{i\in\num{1}{3}}$, $\l A, B\r:=\sum_{i\in\num{1}{3}} A_i^\ast B_i$ (see Lemma \ref{lem:Amat} \ref{Amod}), $A\wedge B:=[\sum_{j,k\in\num{1}{3}}\veps_{ijk}A_jB_k]_{i\in\num{1}{3}}$, and $mA:=\sum_{i\in\num{1}{3}}m_i A_i$, respectively, and decomposing any $J=[J_{ij}]_{i,j\in\num{1}{3}}\in\R^{3\times 3}$ as $J=J_d+J_s+J_a$, where $J_d, J_s, J_a\in\R^{3\times 3}$ read $J_d:=\diag[J_{11}, J_{22}, J_{33}]$, $J_s:=(J+J^T)/2-J_d$, and $J_a:=(J-J^T)/2$, we get, for all $A=[A_i]_{i\in\num{1}{3}}, B=[B_i]_{i\in\num{1}{3}}\in\fA^3$, 
\ba
\l A, J B\r
=\l A, J_d B\r+\l A, J_s B\r+\l A, J_a B\r,
\ea
 where the three contributions are given by
\ba
\label{AdB}
\l A, J_d B\r
&= \sum_{i\in\num{1}{3}} J_{ii} A_i^\ast B_i,\\
\label{AsB}
\l A, J_s B\r
&=\sum_{\substack{i, j\in\num{1}{3}\\ i<j}} \frac{J_{ij}+J_{ji}}{2}\hspace{0.5mm} (A_i^\ast B_j+A_j^\ast B_i),\\
\label{AaB}
\l A, J_a B\r
&=\sum_{\substack{i, j\in\num{1}{3}\\ i<j}}
\frac{J_{ij}-J_{ji}}{2}\hspace{0.5mm} (A_i^\ast B_j-A_j^\ast B_i).
\ea

In particular, setting $\sigma^{(x)}:=[\sigma^{(x)}_1,\sigma^{(x)}_2,0]^T\in\fA^3$ for all $x\in\Z$ (using the same notation as for scalar entries) and plugging $A=\sigma^{(x)}$ and $B=\sigma^{(x+1)}$ for any $x\in\Z$ into \eqref{AdB}-\eqref{AaB},  the so-called direct, symmetric, and antisymmetric (Dzyaloshinskii-Moriya) parts of the nearest neighbor magnetic exchange interaction have respectively the form
\ba
\label{SdS}
\l \sigma^{(x)}, J_d \sigma^{(x+1)}\r
&=\frac{J_{11}+J_{22}}{2}\hspace{0.5mm} \big(\sigma^{(x)}_1\sigma^{(x+1)}_1+\sigma^{(x)}_2\sigma^{(x+1)}_2 \big) \nonumber\\
&+\frac{J_{11}-J_{22}}{2}\hspace{0.5mm} \big(\sigma^{(x)}_1\sigma^{(x+1)}_1-\sigma^{(x)}_2\sigma^{(x+1)}_2 \big),\\
\label{SsS}
\l \sigma^{(x)}, J_s \sigma^{(x+1)}\r
&=\frac{J_{12}+J_{21}}{2}\hspace{0.5mm} \big(\sigma^{(x)}_1\sigma^{(x+1)}_2+\sigma^{(x)}_2\sigma^{(x+1)}_1 \big),\\
\label{SaS}
\l \sigma^{(x)}, J_a \sigma^{(x+1)}\r
&=\frac{J_{12}-J_{21}}{2}\hspace{0.5mm} \big(\sigma^{(x)}_1\sigma^{(x+1)}_2-\sigma^{(x)}_2\sigma^{(x+1)}_1 \big),
\ea
and note that \eqref{SaS} can also be written as $\alpha(\sigma^{(x)}\wedge\sigma^{(x+1)})$, where $\alpha\in\R^3$ is given by $\alpha=[0,0,(J_{12}-J_{21})/2]^T$. Now, \eqref{SdS}-\eqref{SaS} become quadratic forms in the Araki-Jordan-Wigner creation and annihilation operators \eqref{AJWa}-\eqref{AJWas}, since, for all $x\in\Z$ and all $n\in\N$, we have
\ba
\label{SecQ-0}
\lefteqn{\psi^{-1}(a_x^\ast a_{x+n}-a_{x+n}^\ast a_x)}\nonumber\\
&=\begin{cases}
\hfill \frac\ii 2 \big(\sigma_1^{(x)}\sigma_2^{(x+1)}- \sigma_2^{(x)}\sigma_1^{(x+1)}\big), & n=1,\\
\frac\ii 2 \big(\sigma_1^{(x)}\big(\prod_{i\in\num{1}{n-1}}\sigma_3^{(x+i)}\big) \sigma_2^{(x+n)}- \sigma_2^{(x)}\big(\prod_{i\in\num{1}{n-1}}\sigma_3^{(x+i)}\big)\sigma_1^{(x+n)}\big) , & n\ge 2,
\end{cases}
\ea

\ba
\label{SecQ-1}
 \lefteqn{\psi^{-1}(a_x^\ast a_{x+n}^\ast-a_{x+n} a_x)}\nonumber\\
 &=\begin{cases}
\hfill -\frac\ii 2 \big(\sigma_1^{(x)}\sigma_2^{(x+1)}+ \sigma_2^{(x)}\sigma_1^{(x+1)}\big), & n=1,\\
-\frac\ii 2 \big(\sigma_1^{(x)}\big(\prod_{i\in\num{1}{n-1}}\sigma_3^{(x+i)}\big) \sigma_2^{(x+n)}+ \sigma_2^{(x)}\big(\prod_{i\in\num{1}{n-1}}\sigma_3^{(x+i)}\big)\sigma_1^{(x+n)}\big) , & n\ge 2,
\end{cases}\\
\label{SecQ-2}
\lefteqn{\psi^{-1}(a_x^\ast a_{x+n}^\ast+a_{x+n} a_x)}\nonumber\\
&=\begin{cases}
\hfill -\frac12 \big(\sigma_1^{(x)}\sigma_1^{(x+1)}- \sigma_2^{(x)}\sigma_2^{(x+1)}\big), & n=1,\\
-\frac12 \big(\sigma_1^{(x)}\big(\prod_{i\in\num{1}{n-1}}\sigma_3^{(x+i)}\big) \sigma_1^{(x+n)}- \sigma_2^{(x)}\big(\prod_{i\in\num{1}{n-1}}\sigma_3^{(x+i)}\big)\sigma_2^{(x+n)}\big) , & n\ge 2,
\end{cases}\\
\label{SecQ-3}
\lefteqn{\psi^{-1}(a_x^\ast a_{x+n}+a_{x+n}^\ast a_x)}\nonumber\\
&=\begin{cases}
\hfill -\frac12 \big(\sigma_1^{(x)}\sigma_1^{(x+1)}+ \sigma_2^{(x)}\sigma_2^{(x+1)}\big), & n=1,\\
-\frac12 \big(\sigma_1^{(x)}\big(\prod_{i\in\num{1}{n-1}}\sigma_3^{(x+i)}\big) \sigma_1^{(x+n)}+ \sigma_2^{(x)}\big(\prod_{i\in\num{1}{n-1}}\sigma_3^{(x+i)}\big)\sigma_2^{(x+n)}\big) , & n\ge 2,
\end{cases}
\ea
and recall from \eqref{sig3} that $\psi^{-1}(2a_x^\ast a_x-1)=\sigma_3^{(x)}$ for all $x\in\Z$. The so-called quasifree fermionic systems whose Hamiltonian densities have the form \eqref{SecQ-2}-\eqref{SecQ-3} are called (generalized [if $n\ge 2$]) XY models or Suzuki models (see also \cite{A21}).
\er

\br
\label{rem:appl}
Let the physical system under consideration be specified by the triple $(\fA, \omega, \tau)$, where $\omega$ is a given state (\ie, a normalized positive linear functional) on the algebra of observables $\fA$ and $\tau\in\Hom(\R,\sAut(\fA))$ a group of time evolution automorphisms (see \eqref{CstarDyn}, $\R$ being considered as the additive group of real numbers). If the system is even, \ie, if $\omega\circ\Theta=\omega$ and $\tau_t\circ\Theta=\Theta\circ\tau_t$ for all $t\in\R$, it is possible, for many quantities of physical interest, to restrict one's study to $\hA_{0,+}$. But then, due to 
\eqref{A10}, one is left with a purely fermionic system $(\hA_{1,+}, \omega_+, \tau_+)$, where $\omega_+$ and $\tau_+$ denote the restrictions to  $\hA_{0,+}=\hA_{1,+}$. Moreover, if, in addition, this system is quasifree, one has powerful tools at one's disposal (see, for example, \cite{A21} and references therein).
\er

\vspace{10mm}

\noindent{\it Acknowledgments}\,
I would like to thank the anonymous referee for his careful reading of the manuscript.

\begin{appendix}
\section{\Cs completion}
\label{app:Cstar}

In this appendix, we briefly recall the definitions and basic facts used in the foregoing sections (see, for example, \cite{BR, Sakai, Williams}) and describe in some detail the so-called \Cs completion.

In the following, all the vector spaces are assumed to be complex.

If $\fA$ is a vector space, a map $\fA\times\fA\to\fA$, denoted by $(A,B)\mapsto AB$ for all $A,B\in\fA$, is called a multiplication if it is associative, \ie, if $(AB)C=A(BC)$ for all $A,B,C\in\fA$, and if it is bilinear, \ie, if $(A+B)C=AC+BC$ and $A(B+C)=AB+AC$ for all $A,B,C\in\fA$ (biadditivity) and if $(aA)B=a(AB)$ and $A(bB)=b(AB)$ for all $a,b\in\C$ and all $A,B\in\fA$ (bihomogeneity). Equipped with a multiplication, $\fA$ is called an algebra. If $\fA$ is an algebra and if there exists an element $1\in\fA$ such that $1A=A1=A$ for all $A\in\fA$, this element is called a (multiplicative) identity of $\fA$ and will sometimes be written as $1_\fA$ (the additive identity of $\fA$ is denoted by $0$ or sometimes by $0_\fA$ if necessary). An algebra equipped with an identity is called a unital  algebra.

If $\fA$ is an algebra, a map $\fA\to\fA$, denoted by $A\mapsto A^\ast$ for all $A\in\fA$, is called an involution if it is involutive, \ie, if $(A^{\ast})^\ast=A$ for all $A\in\fA$, if it is antidistributive, \ie, if  $(AB)^\ast=B^\ast A^\ast$ for all $A,B\in\fA$,  and if it is antilinear, \ie, if $(A+B)^\ast=A^\ast+B^\ast$ for all $A,B\in\fA$ (additivity) and if $(aA)^\ast={\bar a} A^\ast$ for all $a\in\C$ and all $A\in\fA$ (antihomogeneity). An algebra equipped with an involution is called a \str algebra. A vector subspace $\fB$ of the \str algebra $\fA$ is called a \str subalgebra of $\fA$ if $AB, A^\ast\in\fB$ for all $A, B\in\fB$. A vector subspace $\fB$ of the \str algebra $\fA$ is called a 2-sided \str ideal of $\fA$ if $AB, BA, B^\ast\in\fB$ for all $A\in\fA$ and all $B\in\fB$.

If $\fA$ and $\fB$ are \str algebras, a map $\pi:\fA\to\fB$ is called a \str homomorphism if, for all $A,B\in\fA$ and all $a\in\C$, it satisfies $\pi(A+B)=\pi(A)+\pi(B)$, $\pi(aA)=a\pi(A)$, $\pi(AB)=\pi(A)\pi(B)$, and $\pi(A^\ast)=\pi(A)^\ast$, and we sometimes say that $\pi$ preserves  the addition, the scalar multiplication, the multiplication, and the involution (of $\fA$), respectively. The set of \str homomorphisms between the \str algebras $\fA$ and $\fB$ is denoted by $\sHom(\fA,\fB)$. If $\fA$ and $\fB$ are unital \str algebras  with identities $1_\fA$ and $1_\fB$, respectively, $\pi\in\sHom(\fA,\fB)$ is called unital if $\pi(1_\fA)=1_\fB$. If $\pi\in\sHom(\fA,\fB)$ is injective or bijective, it is called a \str monomorphism or \str isomorphism, respectively, and we denote by $\sMon(\fA,\fB)$ or $\sIso(\fA,\fB)$ the set of the corresponding \str homomorphisms.  If there exists $\pi\in\sIso(\fA,\fB)$, we write $\fA\cong\fB$. Moreover,  if $\fA=\fB$, the elements of the set $\sAut(\fA):=\sIso(\fA,\fA)$ are called \str automorphisms of $\fA$ and $\sAut(\fA)$ is a group with respect to the usual composition.

 If $\fA$ is a \str algebra and if the map $\|\cdot\|:\fA\to\R$ is a norm on $\fA$ which is submultiplicative, \ie, which satisfies $\|AB\|\le\|A\|\|B\|$ for all $A, B\in\fA$, with respect to which $\fA$ is complete, and which has what we call the \Bs property, \ie, which satisfies,  for all $A\in\fA$, 
\ba
\|A^\ast\|
=\|A\|,
 \ea
the \str algebra $\fA$ is called a Banach \str algebra. If the submultiplicative norm with respect to which $\fA$ is complete has  the so-called \Cs property, 
 \ie, if, for all $A\in\fA$, 
\ba
\|A^\ast A\|
=\|A\|^2,
\ea
the norm $\|\cdot\|$ is called a \Cs norm on $\fA$ and $\fA$ is called a \Cs algebra (recall that all the norms are denoted by $\|\cdot\|$ unless there are several norms on the same vector space). A \str subalgebra $\fB$ of the \Cs algebra $\fA$ is called a \Cs subalgebra of $\fA$ if $\fB$ is closed with respect to the \Cs norm of $\fA$. 

If $\fA$ is a Banach \str algebra, $\fB$ a \Cs algebra, and $\pi\in\sHom(\fA,\fB)$, we know that, for all $A\in\fA$,
\ba
\label{cnt}
\|\pi(A)\|
\le\|A\|,
\ea
\ie, in particular, $\pi\in\mL(\fA,\fB)$,  where, for all normed vector spaces $\mV$ and all Banach spaces $\mB$, we denote by $\mL(\mV,\mB)$ the vector space (with respect to the usual pointwise addition and scalar multiplication) of bounded linear operators from $\mV$ to $\mB$  (\ie, by definition, any $T\in \mL(\mV,\mB)$ is a vector space homomorphism between $\mV$ and $\mB$ for which there exists $C>0$ such that $\|Tv\|\le C\|v\|$ for all $v\in\mV$) and recall that $\mL(\mV,\mB)$ is a Banach space with respect to the so-called operator norm defined by $\|T\|:=\sup_{v\in\mV, \|v\|=1}\|Tv\|$ for all $T\in\mL(\mV,\mB)$. If, in addition, $\fA$ is a \Cs algebra and $\pi\in\sMon(\fA,\fB)$, we know that, for all $A\in\fA$,
\ba
\label{piIso}
\|\pi(A)\|
=\|A\|.
\ea

Also recall that, for any normed vector space $\mV$, the vector space completion of $\mV$ is the couple $(\mV', E)$, where $E\in\mL(\mV,\mV^{\ast\ast})$ is the  isometry defined by $E(v)(\eta):=\eta(v)$ for all $v\in\mV$ and all $\eta\in\mV^\ast$ (and $\mV^\ast:=\mL(\mV,\C)$ is the dual and $\mV^{\ast\ast}:=(\mV^\ast)^\ast$ the double dual equipped with the operator norm) and where
\ba
\label{V'}
\mV'
:=\clo_{\mV^{\ast\ast}}(\ran(E)),
\ea
\ie, $\mV'$ is a Banach space and $E$ an isometry whose range $\ran(E)\subseteq\mV'\subseteq\mV^{\ast\ast}$ is dense in $\mV'$, where, for all normed vector spaces $\mV$ and subsets $\mW\subseteq\mV$, we denote by $\clo_\mV(\mW)$ the closure of $\mW$ with the respect to the norm of $\mV$. 
If $(\wt \mV, F)$ is another couple such that $\wt\mV$ is a Banach space and $F\in\mL(\mV,\wt\mV)$ an isometry whose range is dense in $\wt\mV$, then there exists an isometric isomorphism, \ie, a surjective isometry, from  $\mV'$ to $\wt\mV$ (hence, a vector space completion, when defined as a couple with these three properties, is unique up to isometric isomorphism, only, see Remark \ref{rem:cmpl}). 

If a normed vector space also carries a multiplication and an involution with respect to which the norm is submultiplicative and has the \Cs property, then the vector space completion has the following additional properties.

\bl[\Cs completion]
\label{lem:CsCmpl}
Let $\fA$ be a \str algebra  equipped with a submultiplicative norm which has the \Cs property. Then:
\bn[label=(\alph*), ref={\it (\alph*)}]
\setlength{\itemsep}{0mm}
\item
\label{CsCmpl}
There exist a natural multiplication and involution on $\fA'$ with respect to which the norm of $\fA'$ is submultiplicative and has the \Cs property, \ie, which make $\fA'$ into a \Cs algebra. 

\item
\label{CsE}
$E\in\sMon(\fA,\fA')$ 

\item
\label{CsU}
If $\fA$ is unital, so is $\fA'$.
\en
The couple $(\fA',E)$ is called a \Cs completion of $\fA$. 
\el

\bprf
\ref{CsCmpl}\, 
Let $(\fA', E)$ be the vector space completion of $\fA$ from \eqref{V'} and define the map $\fA'\times\fA'\to\fA'$, for all $A', B'\in\fA'$, by
\ba
\label{CsMult}
A'B'
:=\lim_{n\to\infty} E(A_nB_n),
\ea
where the sequences $(A_n)_{n\in\N}$ and $(B_n)_{n\in\N}$ in $\fA$ are such that, for $n\to\infty$, we have $\|A'-E(A_n)\|\to 0$ and $\|B'-E(B_n)\|\to 0$. We first note that the limit on the right hand side of \eqref{CsMult} exists  since $\|E(A_nB_n)-E(A_mB_m)\|=\|A_nB_n-A_mB_m\|\le \|E(A_n)-E(A_m)\|\|E(B_n)\|+\|E(A_m)\|\|E(B_n)-E(B_m)\|$ for all $n,m\in\N$. We similarly check that it is independent of the choice of the sequences $(A_n)_{n\in\N}$ and $(B_n)_{n\in\N}$ in $\fA$, too, \ie, \eqref{CsMult} is well-defined. Moreover, using the associativity and bilinearity of the multiplication on $\fA$, we see that \eqref{CsMult} indeed defines a multiplication on $\fA'$. Next, we analogously define the involution $\fA'\to\fA'$, for all  $A'\in\fA'$, by
\ba
\label{CsInvo}
A'^ \ast
:=\lim_{n\to\infty}E(A_n^\ast),
\ea
where $(A_n)_{n\in\N}$ is a sequence in $\fA$ such that $\|A'-E(A_n)\|\to 0$ for $n\to\infty$. The limit on the right hand side of \eqref{CsInvo} again exists and is independent of the choice of the sequence $(A_n)_{n\in\N}$ in $\fA$. Moreover, the involutivity, the antidistributivity, and the antilinearity of \eqref{CsInvo} follow from the corresponding properties of the involution of $\fA$. Hence, equipped with the multiplication \eqref{CsMult} and the involution \eqref{CsInvo}, $\fA'$ becomes a \str algebra. The submultiplicativity and the \Cs property of the norm on $\fA'$ directly follow from the corresponding properties of the norm on $\fA$, the continuity of the norm on $\fA'$, and from the fact that $E$ is an isometry. Hence, $\fA'$ is a \Cs algebra. 

\ref{CsE}\, 
Let $A, B\in\fA$ and set $A':=E(A)$, $B':=E(B)$, and, for all $n\in\N$,  $A_n:=A$ and $B_n:=B$. Then, \eqref{CsMult} and \eqref{CsInvo} yield $E\in\sHom(\fA,\fA')$. Since $E$ is an isometry, we get $E\in\sMon(\fA,\fA')$.

\ref{CsU}\, 
If $\fA$ is unital with identity $1_\fA$, the \Cs algebra $\fA'$ is unital with identity $1_{\fA'}:=E(1_\fA)$ since, for all $A'\in\fA'$, we have $\|1_{\fA'} A'-A'\|, \|A' 1_{\fA'}-A'\|\le \|E(1_\fA)\|\|A'-E(A_n)\|+\|E(A_n)-A'\|$ for all $n\in\N$, where $(A_n)_{n\in\N}$ is a sequence in $\fA$ with $\|A'-E(A_n)\|\to 0$ for $n\to\infty$.
\eprf

\br
\label{rem:cmpl}
Let $\mV$ be a normed vector space and $\mC$ the set of Cauchy sequences in $\mV$. Equipped with the addition $\mC\times\mC\to\mC$ and the scalar multiplication $\C\times\mC\to\mC$ defined by $(v_n)_{n\in\N}+(w_n)_{n\in\N}:=(v_n+w_n)_{n\in\N}$ and $\lambda(v_n)_{n\in\N}:=(\lambda v_n)_{n\in\N}$ for all $(v_n)_{n\in\N}, (w_n)_{n\in\N}\in\mC$ and all $\lambda\in\C$, respectively, $\mC$ becomes a vector space. Moreover, $\mC_0:=\{(v_n)_{n\in\N}\in\mC\,|\, \lim_{n\to\infty}\|v_n\|=0\}$ is a vector subspace of $\mC$ and we know that the quotient space $\wt\mV:=\mC/\mC_0$ (see \eqref{V/W}) is complete with respect to the norm $\|\cdot\|:\wt\mV\to\R$ (well-) defined by $\|[(v_n)_{n\in\N}]\|:=\lim_{n\to\infty}\|v_n\|$ for all $[(v_n)_{n\in\N}]\in\wt\mV$. Defining $F:\mV\to\wt\mV$ by $F(v):=[(v)_{n\in\N}]$ for all $v\in\mV$, where $(v)_{n\in\N}$ stands for the constant sequence whose members are all equal to $v$, the couple $(\wt\mV, F)$ is the usual vector space completion of $\mV$.

If  $\fA$ is a \str algebra equipped with a submultiplicative norm having the \Cs property and $(\wt\fA,F)$ the foregoing vector space completion of $\fA$, we easily verify that the Banach space $\wt\fA$ also becomes a \Cs algebra with respect to he multiplication $\wt\fA\times\wt\fA\to\wt\fA$ and the involution $\wt\fA\to\wt\fA$ defined by  $[(v_n)_{n\in\N}][(w_n)_{n\in\N}]:=[(v_n w_n)_{n\in\N}]$ and $[(v_n)_{n\in\N}]^\ast:=[(v_n^\ast)_{n\in\N}]$ for all $[(v_n)_{n\in\N}], [(w_n)_{n\in\N}]\in\wt\fA$, respectively. Moreover, if $(\fA',E)$ is the vector space completion of $\fA$ from \eqref{V'}, we straightforwardly verify that there exists a surjective isometry $\pi\in\mL(\fA',\wt\fA)$ which has the additional property that $\pi\in\sHom(\fA',\wt\fA)$.
\er
\end{appendix}


\end{document}